\documentstyle [floats,preprint,epsf,aps,eqsecnum,table]{revtex}
\advance\textwidth 1mm
\advance\hoffset -0.5mm
\begin{document}
\draft
\global\firstfigfalse
\def\tab{&}
\def\Emax{E_{\rm max}}
\def\Phip{\Phi_{p_e}^*}
\def\rp{r^{\prime}}
\def\phip{\phi_{\bf p}^*({\bf r})}
\def\phipo{\phi_0^*({\vec r})}
\def\psie{\psi ( {\bf r}, t)}

\title{Atomic effects in tritium beta decay}

\preprint{UW/PT-95-16, PURD-TH-95-11}

\author{Lowell S. Brown and Chengxing Zhai\footnote{Present address:{\it
Department of  Physics,Purdue University,West Lafayette, Indiana 47907}}}

\address{ Department of Physics,
    University of Washington,
    Seattle, Washington 98195
    }%

\maketitle

\begin{abstract}

The electron neutrino mass has been measured in several tritium beta
decay experiments. These experiments are sensitive to a small neutrino mass
because the energy release of the decay is small. But the very
smallness of the energy release implies that the Coulomb interactions
of the slowly moving emitted beta electron are relatively large.
Using field theoretic techniques, we derive a systematic
and controlled expansion which accounts for the Coulomb effects,
including the mutual interaction of the beta ray electron and the electron
in the final $^3{\rm He}^+$ ion. In our formulation,
an effective potential which describes the long range Coulomb force
experienced by the beta ray is introduced to ensure that our expansion is
free of infrared divergences. Both the exclusive
differential decay rate to a specific final $^3{\rm He}^+$ state
and the inclusive differential decay rate are calculated to order $\eta^2$,
where $\eta$ is the usual Coulomb parameter.
We analyze the order $\eta^2$ correction to the beta ray spectrum
and estimate how it may affect the neutrino mass squared
parameter and the endpoint energy when this corrected spectrum is used to
compare with the experiments. We find that the effect is small.
\end{abstract}
\newpage
\section{Introduction and summary}
The mass of the electron's anti-neutrino $m_\nu$ has been measured in several
tritium beta decay experiments
\cite{Wilkerson,Robertson,Holzschuh,Kawakami,Stoeffl,Backe}.
These experiments are quite sensitive
to a possible neutrino mass since the end-point energy  $Q$ in this
beta decay is very small ($Q \simeq 18.6$ KeV), and an examination of the
beta ray energy spectra near the end point region can thus reveal a
small neutrino mass. On the other hand, the very smallness of the
decay energy implies that the Coulomb interactions between the outgoing beta
electron and the atomic electron in the final helium ion may play a
significant role in
the interpretation of the experiments. In units where $c=1$
and $\hbar=1$, the strength of this Coulomb
interactions is governed by the parameter
\begin{equation}
    \eta = { \alpha \over v} = {1 \over a_0 p } \,,
\end{equation}
where $ \alpha \simeq 1/137 $ is the fine structure constant and $v$
is the emitted beta electron velocity, or, equivalently,
where $ a_0 $ is the Bohr radius and $ p $ is the momentum of the
outgoing beta ray. Since the end-point energy $ Q $ is small, the beta
electron velocity is small and the Coulomb parameter $ \eta $ is
relatively large. Near the tritium decay end point,
$ \eta \simeq 0.03 $.

To study a small neutrino mass, modern experiments use data
from the beta ray spectrum end point to
approximately $800-1000\,{\rm eV}$ below the end point
\cite{Robertson,Holzschuh,Kawakami,Stoeffl,Backe}.
The actual experiments work with molecular tritium.
The theoretical formula for the spectrum used in the analysis of
the experimental data is the sudden approximation result
for the molecular final states which neglects the Coulomb interaction
between the beta ray electron and the electrons in the daughter
$^3{\rm He}-{\rm T}^+$ molecule
\cite{Robertson,Holzschuh,Kawakami,Stoeffl,Backe,Knapp,Martin}.
In this paper, we shall investigate the Coulomb interaction effect,
hereafter referred to as the atomic effect, and estimate its
size. Instead of molecular tritium, we calculate the effect
for atomic tritium, where the calculation is simpler and brings
out the essential points. The calculation for molecular tritium would be
parallel to that of atomic tritium, but much more complicated.
Currently, tritium beta decay experiments appear to find a squared
electron neutrino mass which is negative,
$ m_\nu^2 \approx - 60\,{\rm eV}^2$
\cite{Wilkerson,Robertson,Holzschuh,Kawakami,Stoeffl,Backe}.
It should be noted that there are two relevant dimensional parameters
which enter here:
The atomic energy scale
${\rm R\!y} \simeq 13.6 \, {\rm eV} $
and the energy range $\Delta E$ where the experimentalists fit their
data. Typically, $\Delta E \simeq 10^3 \, {\rm eV}$. Thus, apart from
potentially large numerical factors, the effect
of the atomic corrections on the determination
of the the neutrino mass squared may be of order $ \eta^2 \Delta E^2
\simeq 900 {\rm eV}^2$
as well as $\eta^2 {\rm R\!y}^2 \simeq 0.2 {\rm eV}^2$.
We shall provide a detailed analysis of how much the atomic effect
influences the neutrino mass squared parameter.

The sudden approximation spectrum used for fitting
the experimental data takes the form
\begin{equation}
    {d \, \Gamma_{\rm exp} \over d E}
        = A \,\, F(2, E) \,\, p \, \sum_{f<f_{\rm max}} W_{fi}
	\,\, (Q_f - E) \, \sqrt{ (Q_f - E)^2 - m_\nu^2} \,,
\label{ExpSpectrum0}
\end{equation}
where $ E $ and $ p $ are the energy and momentum of the outgoing
beta electron and $m_\nu^2$ is the neutrino mass squared.
The spectrum~(\ref{ExpSpectrum0}) needs some explanation.
The first factor $A$ is an overall constant, and  $F(2, E)$ is the usual
Fermi function, which is the square of the beta ray wave function
evaluated at the $^3{\rm He}^+$ nucleus, for the $Z=2$
charged helium nucleus.
The atomic transition probability  $ W_{fi} $ for the initial tritium
state $ | i \rangle$ decaying to the particular final ionic state
$ \langle f | $ is the squared matrix element
\begin{equation}
    W_{fi} = |T_{fi}|^2 \equiv |\langle f | i \rangle|^2 \,.
\end{equation}
The energy release
\begin{equation}
    Q_f = Q + E_i - E_f
\label{Q_f_def}
\end{equation}
is that for the decay process which starts with the initial
tritium atomic energy $ E_i $ and ends
with the final $ ^3{\rm He}^+ $ ion energy $ E_f $.
The upper limit $f_{\rm max}$ on the sum over states is set by
the conservation of energy. It mostly%
\footnote{In a small region
at the very end of the beta ray spectrum, this upper limit corresponds
to a bound state. We ignore these cases since statistically this small
region makes only a small contribution which does not significantly
alter our results.}
corresponds to a state of $^3 {\rm He}^+$
that has become unbound into $^3{\rm He}^{++} + e^-$
with the $e^-$ kinetic energy being
\begin{equation}
    \Emax \equiv K^2 / 2m = Q + E_i - E \,.
\label{EK_def}
\end{equation}
Parameters including the overall strength factor $A$, the end point
energy $Q$, and the neutrino mass squared $m_\nu^2$ are determined
by fitting the spectrum~(\ref{ExpSpectrum0}) to the measured spectrum.

To simplify our work, we shall neglect relativistic corrections
and treat the beta ray non-relativistically. This is valid
since relativistic effects, of order $v^2 \approx 0.06$,
provide only small corrections that can easily be accounted for.
They correct both the sudden approximation result and the atomic
effect. The correction to the sudden approximation result
may be taken care of, to good accuracy, by using an
approximate relativistic Fermi factor $F(Z, E)$ in Eq.~(\ref{ExpSpectrum0})
\cite{Wilkinson,RobertsonRev}. The correction to the atomic effect is
of the order $\eta^2 v^2$, which is negligible\cite{Durand}.

In calculating the beta ray decay rate,
the usual perturbative expansion\cite{Durand,Strikman}
uses the Coulomb potential produced by the final helium nucleus as the
zeroth-order approximation for the potential experienced by the
emitted beta electron. This zeroth-order
approximation does not describe the long distance behavior of
the potential in which the real beta ray moves since
the screening effect of the outside electron has not been taken care of.
Therefore, infrared divergences appear and make the usual perturbative
calculation inconvenient.
Using field-theoretic methods involving the reduction
technique, we shall instead
make use of a conveniently chosen effective potential
as the zeroth-order approximation. Since this effective potential
properly accounts for the long-range character of the screened
Coulomb potential, we can perform a systematic expansion
in powers of the small parameter $ \eta $,
with the expansion free of infrared divergences.
We shall compute the first term in this expansion, which is of
order $ \eta^2 $, for the decays into the individual final ionic
states. The inclusive sum of these individual decay rates agrees with
previous results\cite{Durand,Strikman} under two approximations\footnote
{These approximations are made only for the order $\eta^2$ correction to
the exclusive decay rates caused by the atomic effect.} made there:
1) a ``uniform phase space
factor approximation'' and 2) the closure approximation.
The uniform phase space factor approximation takes the neutrino
phase space factor, which depends on the final helium
ionic energy, to be the same for different
final helium ionic states. The energy of the final ionic state
appearing in the phase space factor is replaced
by the helium ionic ground state energy\cite{Durand}.
Within this uniform phase space factor approximation,
references \cite{Durand,Strikman}
then make the closure approximation, which extends the
upper limit $f_{\rm max}$ of the sum over final helium ionic states to
infinite energy.
The closure approximation over counts contributions from the
final helium ionic states whose energies lie above the limit set by
energy conservation.
In most regions of the beta ray spectrum, this approximation is valid
because the limit determined by energy conservation is
much higher than the atomic energy scale. However,
the closure approximation cannot be used
for the region near the end point of the
beta ray spectrum since here little energy remains to
excite the final helium ionic states, and threshold effects must
be taken into account.
Since the shape of the beta ray energy spectra near the end point region
is crucial for the neutrino mass measurement,
we shall not rely on either the closure approximation
or the uniform phase space factor approximation.

Our result for the case of a vanishing neutrino mass, $ m_\nu = 0 $,
may be expressed in the form
\begin{equation}
    {d \Gamma_{\rm in} \over d E}
	= {m \over 2 \pi^3} F(2, E)
        \, p \, | T_{\beta} |^2 (1{-}2\eta^2) \left [ P \, (Q{-}E)
	+ \eta^2 \, R \, (Q{-}E) \right ] \,,
\label{RawRESULT0}
\end{equation}
where $m$ is the electron mass, the Fermi factor
\begin{equation}
    F(2, E) = {4 \pi \eta \over 1 - e^{- 4 \pi \eta}}
\label{fermi}
\end{equation}
is the value of the squared non-relativistic beta electron wave function
at the $^3{\rm He}^+$ nucleus, and $ | T_\beta |^2 $ is squared
nuclear decay amplitude in the absence of atomic corrections.
Here and in the following we shall make use of the Rydberg energy
\begin{equation}
    {\rm R\!y} = {1 \over 2} \alpha^2 \, m = {\alpha \over 2 a_0 }
        \simeq 13.6 \, {\rm eV} \,.
\end{equation}
In the closure approximation which we have just discussed,
\begin{equation}
\sum_{ {\rm all} \,\, f} |T_{fi}|^2 = 1 \,,
\end{equation}
and a simple calculation, reviewed in Section IV, gives
\begin{equation}
\sum_{ {\rm all} \,\, f} |T_{fi}|^2 (Q_f - E)^2 =
	(Q- E + 2 \, {\rm R\!y})^2 + 4 \, ({\rm R\!y})^2 \,.
\end{equation}
This provides an approximate evaluation of the sum in
Eq.~(\ref{ExpSpectrum0}) when $ m_\nu = 0 $. The finite upper limit on
the sum alters this evaluation and defines the first function that
appears in the curly braces in Eq.~(\ref{RawRESULT0}),
\begin{equation}
    P (Q{-}E) \equiv \sum_{f<f_{\rm max}} |T_{fi}|^2 (Q_f - E)^2 \,.
\label{P_def}
\end{equation}
For later convenience, we take
the independent variable to be $Q-E$, which is related to $\Emax =
K^2/2m $ by Eq.~(\ref{EK_def}).
In Section IV we compute the correction $S(Q{-}E)$ to the closure
approximation such that
\begin{equation}
    P (Q{-}E) = (Q{-}E{+}2 \,{\rm R\!y})^2{+}4 \,{\rm R\!y}^2{+} S (Q{-}E) \,.
\label{P_result}
\end{equation}
The term $\eta^2 R (Q{-}E)$ in Eq.~(\ref{RawRESULT0}) represents
the correction due to the atomic effect.
The major purpose of our work is to compute $R (Q{-}E)$.

In order to describe the effect of our result,
we shall slightly simplify the formula~(\ref{ExpSpectrum0})
used in the experimental data analysis.
Although the experiments measure the
spectra near the end point $ E = Q $, they nonetheless mainly measure
electron energies with $ (E - Q)^2 \gg m_\nu^2 $ since the decay rate
is so small in the region very near the end point.
Hence one may expand the square root in Eq.~(\ref{ExpSpectrum0})
to get a simpler form:
\begin{equation}
    {d \, \Gamma_{\rm exp} \over d E}
	\simeq A \,\, F(2, E) \,\, p \sum_{f<f_{\rm max}} |T_{fi}|^2
        \left[ (Q_f - E)^2 - {1 \over 2} m_\nu^2 \right ] \,.
\end{equation}
The first sum which appears here is just the sum (\ref{P_def}), and
defining
\begin{equation}
    P_1 (Q-E) \equiv \sum_{f<f_{\rm max}} |T_{fi}|^2
\label{P_1_def} \,,
\end{equation}
which is the total probability for the initial tritium to make the
transitions to any
final state below energy $\Emax$,
we have
\begin{equation}
    {d \, \Gamma_{\rm exp} \over d E}
	= A \,\, F(2, E) \,\,p
	\left [P (Q{-}E) - {1 \over 2} m_\nu^2
	P_1 (Q{-}E) \right ] \,,
\label{rightform}
\end{equation}
To see how the atomic correction affects the neutrino mass squared, we
note that Eq.~(\ref{RawRESULT0}) reduces to the sudden
approximation result~(\ref{rightform}) for $m_\nu^2 = 0$
if $R (Q{-}E)$ is ignored.
However, if we include the atomic correction $R (Q{-}E)$,
and require that formula~(\ref{ExpSpectrum0}) fits data described
by spectrum~(\ref{RawRESULT0}) with free parameters $A$, $Q$, and $m_\nu^2$,
we may get a nonzero $m_\nu^2$ for the best fit.
The correction $\eta^2 R(Q{-}E)$ to the spectrum effectively changes
$A$, $Q$, and $m_\nu^2$ by $\Delta A$, $\Delta Q$, and $\Delta m_\nu^2$ in
such a way that the change
\begin{equation}
    \Delta \left ({d \Gamma_{\rm exp} \over d E} \right )
	\approx  A \, F (2, E) \, p \left [{\Delta A \over A} \,
	P (Q{-}E) + \Delta Q \, P'(Q{-}E)
	{-}{1 \over 2} \Delta m_\nu^2 \, P_1 (Q-E) \right ]
\label{mimic}
\end{equation}
to Eq.~(\ref{rightform}) mimics the correction to the spectrum
represented by $\eta^2 R(Q{-}E)$.
Here $P'(Q-E)$ is the first derivative of the function $P$, and we
have ignored terms higher order in the small parameters
$\Delta A$, $\Delta Q$, $\Delta m_\nu^2$, and $m_\nu^2$.
To find the changes $\Delta m_\nu^2$ and $\Delta Q$ due to
our atomic effects of order $ \eta^2 $,
we fit the atomic correction represented by the function\footnote{%
Although $\eta^2$ depends on the beta ray
energy, we shall treat it as a constant in the region near the endpoint
used to determine the neutrino mass.
This is valid because $\eta^2$ varies only by a small amount
$\sim \eta^2 \Delta E/Q$ as the beta ray energy varies from the end point to
$\Delta E$ below the end point.}
$R (Q{-}E)$ to a linear combination of $P (Q-E)$, $P'(Q-E)$, and
$P_1 (Q-E)$ as in the form~(\ref{mimic}).
The neutrino mass measurement is sensitive to energies
from the beta ray spectrum end point down to approximately
$59-74 \, {\rm R\!y}$ ($800-1000\,{\rm eV}$) below the beta ray spectrum
end point\cite{Robertson,Holzschuh,Kawakami,Stoeffl,Backe}.
Fitting in this energy range, we find in Section VI that
\begin{equation}
    R (Q{-}E) \approx 0.94 \, P (Q-E) - 4.6 \, {\rm R\!y} \, P' (Q-E)
        + 6.3 \, {\rm R\!y}^2 \, P_1 (Q-E) \,,
\label{R_fit0}
\end{equation}
which, in view of Eqs.~(\ref{RawRESULT0}) and (\ref{mimic}), gives
\begin{equation}
    \Delta m_\nu^2 \approx - 12.6 \, \eta^2 \, {\rm R\!y}^2
	\simeq - 1.7 \, {\rm eV}^2 \,,
\end{equation}
and
\begin{equation}
    \Delta Q \approx - 4.6 \, \eta^2 \, {\rm R\!y}
	\simeq - 0.047 \, {\rm eV} \,.
\end{equation}
The
fitting formula~(\ref{R_fit0}) depends on the range of energy where
we do the fit. Equation~(\ref{R_fit0}) is done with the energy range
$\Emax$ being $0-64\,{\rm R\!y}$ (0$-$870 eV).
Increasing the range of energy for $\Emax$ from
$0-49 \, {\rm R\!y}$ (0$-$670 eV) to
$0-79 \, {\rm R\!y}$ (0$-$1080 eV)
changes the parameter $\Delta m_\nu^2$ linearly.
For each $5\,{\rm R\!y}$ increase of the energy range, $\Delta m_\nu^2$
decreases by $0.4 \, {\rm eV}^2$. We show this dependence in
Table~\ref{sensitivity}.
The atomic effect causes only a few ${\rm eV}^2$ correction
to the neutrino mass squared and thus does not affect current
experimental bounds on the neutrino mass squared.

\begin{mytable}
\caption{\label{sensitivity}%
The dependence of the neutrino mass squared fitting on
the range of the energy where the fit is made.}
\begin{center}
\begin{tabular}[b]{|c|c|}
\hline
{\rm range of $\Emax$} & {$\Delta m_\nu^2$} \\ \hline
0 --- 49 ${\rm R\!y}$ = 0 --- 670 eV & $-$ 0.80 ${\rm eV}^2$ \\
0 --- 55 ${\rm R\!y}$ = 0 --- 740 eV & $-$ 1.2 ${\rm eV}^2$ \\
0 --- 59 ${\rm R\!y}$ = 0 --- 810 eV & $-$ 1.5 ${\rm eV}^2$ \\
0 --- 64 ${\rm R\!y}$ = 0 --- 870 eV & $-$ 1.7 ${\rm eV}^2$ \\
0 --- 69 ${\rm R\!y}$ = 0 --- 940 eV & $-$ 2.1 ${\rm eV}^2$ \\
0 --- 74 ${\rm R\!y}$ = 0 --- 1010 eV & $-$ 2.5 ${\rm eV}^2$ \\
0 --- 79 ${\rm R\!y}$ = 0 --- 1080 eV & $-$ 2.8 ${\rm eV}^2$ \\
\hline
\end{tabular}
\end{center}
\end{mytable}

The next five sections describe our methods and calculations. In
Section II, we develop, using quantum field theory
techniques, a general formula which is convenient for calculating the
exclusive decay rate to a specific final $^3{\rm He}^+$ ion state.
A comparison potential is introduced to facilitate a systematic and
controlled expansion in the small parameter $\eta$.
With an appropriate choice of the comparison potential, given in
Section III, the exclusive
decay rates are calculated to $\eta^2$ order in Sections IV
and V.
Finally, in Section VI, we examine the atomic
effects in the determination of the neutrino mass squared.
Fitting formulas for the beta ray spectrum are given for the region
near the end point.
In Appendix A, we calculate the value of the wave function of the
beta ray at the origin for the sudden approximation.
Our evaluation of this amplitude confirms an old result \cite{Rose}
and shows, incidentally, that it is accurate through terms of order
$\eta^3$ rather than being valid only through order $\eta^2$.
The details of evaluating
the second order atomic correction are presented in Appendix B.
In Appendix C, we provide the details of the computations for summing
over the final states. Finally, in Appendix D, we investigate the
nature of the exchange corrections. These correspond to the
process in which the electron produced by the weak interaction
is bound in the final
$^3{\rm He}^+$ ion with the initial tritium atomic electron being
ejected. We show that the leading exchange corrections to the
{\it amplitude} are of order $\eta^3$, in contradiction with the
previous order $\eta^4$ estimate that has appeared in the
literature \cite{Bahcall}. The exchange corrections to the
decay {\it rate}, however, are of order $\eta^4$.

We should emphasize that, although the corrections which we have found
for the neutrino mass determination from tritium beta decay are not
significant, the methods which we have developed to treat the
problem may be useful in the examination of other beta decay processes.

\section{General Formulation}
    Since the emitted beta ray has a maximum energy of $18.6$ KeV,
in natural units where $c=1$, the
square of the electron velocity is less than 0.1, it is
valid to treat the electron non-relativistically.
The interactions due to the spins of electrons and nucleons are relativistic
effects. Therefore, the spectrum is not significantly affected by the spins of
these non-relativistic particles except for an overall constant which
accounts for the spin degrees of freedom.
Hence, the effective Hamiltonian for
the beta decay for our purpose may be written in the form
\begin{equation}
    H_{\rm wk}^{\rm eff} = ({\rm const}) \int (d^3 {\bf r}) \,
	\psi_p^{\dagger} ({\bf r}) \psi_n ({\bf r})
	\psi_e^{\dagger} ({\bf r}) \psi_{\nu} ({\bf r})
	+ {\rm h.c.} \,,
\label{Hamdef}
\end{equation}
where the operators $\psi_p$, $\psi_n$, $\psi_e$, and $\psi_{\nu}$ are
spinless field operators, which destroy the proton, neutron, electron, and
neutrino, respectively.

We will calculate the decay rate of the tritium atom expressed as an
energy spectrum integral of the beta ray.
To accomplish this, we assume the following process:
Initially, at time $t = T_i$, there is a tritium atom in its
atomic ground state denoted by $| i \rangle$; at a later time
$t= - T/2$ the weak interaction is turned on and then turned
off at time $t = T/2$; finally the outgoing electron wave packet
travels for certain time and hits the detector far away from the
tritium atom at time $t = T_f$. Here $T$ is the time for the interaction
to act, and the relations $T_i < -T/2$,~~$T/2 < T_f$ are assumed. We shall
then calculate the probability for detecting the outgoing electron,
or the decay rate by dividing this probability by $T$.

To first order in the weak coupling, the amplitude for
the tritium atom decaying to a specific final state is
\begin{equation}
    A_T = - i \langle {\bar \nu}, e, f; T_f | \int_{-T/2}^{T/2}
	dt \, H_{\rm wk}^{\rm eff} (t) \, | i ; T_i \rangle \,,
\label{amplitude}
\end{equation}
where we have denoted the initial tritium atom state at time
$t = T_i$ by $| i ; T_i \rangle$ and
the final state which contains an anti-neutrino, an outgoing electron,
and the helium ion $^3{\rm He}^+$ at time $t = T_f$
by $\langle {\bar \nu}, e, f; T_f |$~.
The time dependence of the Hamiltonian $H_{\rm wk}^{\rm eff} (t)$ is
governed by the Hamiltonian $H$ which contains the particles kinetic
energies and the Coulomb interaction between the charged particles.

In the infinitely heavy nucleus limit, which is appropriate in this
non-relativistic situation, the operator $\psi_p^{\dagger}
({\bf r}, t) \psi_n ({\bf r}, t)$, when evaluated between states
$\langle f |$ and $| i \rangle$, is proportional to $\delta ({\bf r})
Q_+(t)$, where $Q_+(t)$ is the charge raising operator which converts a
neutron into a proton, and the origin is chosen to be located at the nucleus.
Under this infinitely heavy nucleus approximation, the amplitude for the
tritium beta decay reduces to
\begin{equation}
    A_T = ({\rm const}) (-i) \int_{-T/2}^{T/2} dt \,
	e^{i E_\nu t} \langle e, f; T_f |
	\psi_e^{\dagger} ({\bf 0}, t) Q_+ (t) | i; T_i \rangle
	\,,
\end{equation}
where the spatial integral has been performed by exploiting the
$\delta$-function produced from the infinitely heavy nucleus limit,
and the action of the neutrino field operator in the weak decay
effective Hamiltonian has been used,
\begin{equation}
    \langle {\bar \nu}, e, f; T_f | \psi_{\nu} (0, t)
	= e^{i E_\nu t} \langle e, f; T_f | \, .
\end{equation}
Since only the electron non-relativistic field operator is involved
in the following calculations, we shall change notation and replace
$\psi_e$ by $\psi$.

\subsection{Reduction Technique with a Comparison Potential}
We now apply the reduction method to the final outgoing electron.
To achieve this, we first construct an asymptotic wave packet for
the final outgoing electron, which propagates in the Coulomb potential
of the produced $^3{\rm He}^+$. Since the Coulomb potential is a long range
interaction, the wave packet of the final outgoing electron must be
constructed as moving in the Coulomb potential or a comparison
potential $v(r)$ having the same large distance behavior%
\footnote{For the case where the final $^3{\rm He}^+$ is in an unbound state,
besides the anti-neutrino, the final state contains $^3{\rm He}^{++}$
and the two unbound electrons. The electron moving faster is identified
as the beta ray electron. Therefore, the other electron which moves
relatively slowly still plays the role of screening the $^3{\rm He}^{++}$
nucleus.},
\begin{equation}
    v(r) \sim - {e^2 \over 4 \pi r} \quad {\rm as} \quad
	r \to \infty \,.
\end{equation}
Since we only require that the comparison potential $v(r)$ has
the correct long distance behavior, the result~(\ref{correct}) which
we shall obtain should not depend on the specific choice of $v(r)$.
This we shall prove later.
In view of these considerations, the final electron-ion state can be
expressed by the asymptotic state
\begin {equation}
    \langle e, f ; T_f | = \langle f; T_f |
	\left \{ \int (d^3 {\bf r}) \Phi_{p_e}^*({\bf r}, T_f)
	 \psi ({\bf r}, T_f) \right \} \,,
\label{final}
\end{equation}
where the wave function $\Phi_{p_e}^* ({\bf r}, t)$ satisfies the
Schr\"odinger equation
\begin{equation}
    - i {\partial \over \partial t} \Phi_{p_e}^* ( {\bf r}, t)
	= \left \{ - {\nabla^2 \over 2 m} + v(r) \right \}
    	\Phi_{p_e}^* ({\bf r}, t) \,.
\end{equation}
This electron wave packet can be viewed as the superposition
\begin{equation}
    \Phi_{p_e}^* ({\bf r}, t) = \int (d^3 {\bf p}) \, g({\bf p})
	\phi_{\bf p}^*({\bf r}) e^{i E t},
\label{super}
\end{equation}
where $g({\bf p})$ is a probability amplitude peaked at ${\bf p} =
{\bf p}_e$ with a width $\Delta p$, and $\phip$ obeys
\begin{equation}
    \left \{ - {\nabla^2 \over 2 m} + v(r) \right \}
	\phi_{\bf p}^*({\bf r}) = E \, \phi_{\bf p}^*({\bf r}) \,,
\label{comparison}
\end{equation}
with $E = {\bf p}^2 /2m$ and
$\phip$ satisfying the boundary condition that for large $r$
its asymptotic form\footnote{The asymptotic form of this energy
eigenstate also contains an incoming wave which is the time reversal
of the scattered wave.}
contains a plane wave with the momentum ${\bf p}$.

Since the time dependences of the final $^3{\rm He}^+$ state and the initial
tritium atom state are irrelevant phase factors, from now on we
shall replace $\langle f; T_f |$ by $\langle f |$ and $| i; T_i \rangle$
by $| i \rangle$.
Using the construction~(\ref{final}), the usual reduction
method~\cite{Brown} gives
\begin{equation}
    A_T{=}({\rm const}) \! (-i) \! \int (d^3 {\bf x}) \!
	\int_{T_i}^{T_f} d x^0 \!
	\int_{-T/2}^{T/2} dt \, e^{i E_{\nu} t}
	{\partial \over \partial x^0}
        \langle f | {\rm T} \left (\Phi_{p_e}^* (x) \psi (x)
        \psi^{\dagger} ({\bf 0}, t) Q_+ (t) \right ) | i \rangle
        \,,
\label{reduction}
\end{equation}
where we have used
\begin{equation}
    \left \{ \int (d^3 {\bf r}) \Phi_{p_e}^*({\bf r}, T_i)
         \psi ({\bf r}, T_i) \right \} | i \rangle = 0 \,,
\label{discon}
\end{equation}
since, at the very early time $t = T_i$, the wave function $\Phip$
represents an incoming electron wave-packet which has no overlap
with the initial localized electron state in the tritium atom.

Using the superposition~(\ref{super}) and shifting the time integral
variable $x^0$ to $x^0 - t$ in Eq.~(\ref{reduction}), the Heisenberg
equation of motion then displays the $t$-dependence in the form
$\exp\{-i H t\}
\cdots \exp\{i H t\}$, with the left exponential factor acting directly
on the final state and the right factor acting directly on the initial
state. Hence
\begin{eqnarray}
    A_T &=& ({\rm const} ) (-i) \int (d^4 x) \int (d^3 {\bf p})
	\, g({\bf p})
        \int_{-T/2}^{T/2} dt \, e^{i (E_f + E + E_{\nu}
	 - E_i - Q) t}
\nonumber\\
        && \times {\partial \over \partial x^0} \left [
        \phi_p^*({\bf x}) e^{i E x^0} \langle f | {\rm T} \left ( \psi (x)
        \psi^{\dagger} ({\bf 0}, 0) Q_+ (0) \right ) | i \rangle \right ]
	\,,
\end{eqnarray}
where $Q$ is the total energy released which equals the tritium-helium
nucleus mass difference minus the electron mass, $E_f$ is the atomic
energy eigenvalue of the produced $^3{\rm He}^+$ ion, and $E_i$ is the
energy of the initial $^3{\rm H}$ atom.
The time integration now yields
\begin{equation}
        A_T = ({\rm const}) (-i) \int (d^3 {\bf p})
	\,g({\bf p})\, 2 \pi \,\delta_{T}
        (E_f{+}E{+}E_{\nu}{-}E_i{-}Q) {\cal T}
        ({\bf p}) \,,
\label{engyconsv}
\end{equation}
where the function $\delta_{T} (x)$ is defined by
\begin{equation}
    \delta_{T} (x) = {\sin (x T/2) \over \pi x} \,,
\end{equation}
which is approximately a ``$\delta$-function'' but with a
width $1/T$, and the atomic matrix element is defined as
\begin{equation}
    {\cal T} ({\bf p}) = \int (d^4 x) {\partial \over \partial x^0}
	\left \{ \phi_p^*({\bf x}) e^{i E x^0} \langle f |
        {\rm T} \left (\psi (x) \psi^{\dagger}(0) Q_+ (0) \right )
	| i \rangle \right \} \,.
\end{equation}

To calculate the atomic matrix element ${\cal T}({\bf p})$, we write the
time-ordered product in terms of step functions and take
the partial derivative with respect to $x^0$ to get
\begin{equation}
    {\cal T} ({\bf p}) = \phi_{\bf p} ^* (0)
	\langle f | Q_+ | i \rangle + \int (d^4 x) \,
	\langle f | {\rm T} {\partial \over \partial x^0}
        \left ( \phi_{\bf p}^* ({\bf x}) e^{i E x^0} \psi (x) \right )
	\psi^{\dagger}(0) Q_+ (0)| i \rangle \,,
\end{equation}
where we have used the equal-time anticommutator of the electron creation and
annihilation operators,
\begin{equation}
    \left \{ \psi ({\bf r}), \psi^{\dagger} (0)
	\right \} = \delta ({\bf r}) \,.
\label{anticom}
\end{equation}
The time evolution of the electron destruction operator $\psi$ is governed
by the Heisenberg equation of motion
\begin {equation}
    i {\partial \over \partial t} \psi ({\bf r}, t) =
        \left \{ - {\nabla^2 \over 2 m}
        + V ({\bf r}, t) \right \} \psie \,,
\label{Heisenberg}
\end{equation}
where
\begin{equation}
    V({\bf r}, t) = \int (d^3 {\bf r}^{\prime})
        {e^2 \over 4 \pi |{\bf r} - {\bf r}^{\prime}|} \,
         \rho ({\bf r}^{\prime},t) - {\hat Z e^2 \over 4 \pi r},
\label{potential}
\end{equation}
with
\begin{equation}
    \rho ({\bf r}^{\prime},t) =
	\psi^{\dagger} ({\bf r}^{\prime},t) \psi
	({\bf r}^{\prime}, t) \,,
\end{equation}
and the charge operator $\hat Z$ having the properties
\begin{equation}
   {\hat Z} Q_+ = 2 Q_+, \quad Q_+ {\hat Z} = Q_+ \,,
\end{equation}
since the charge of the tritium nucleus is unity and $Q_+$
is the charge raising operator.

Exploiting the equation of motion~(\ref{Heisenberg}) of the operator
$\psi (x)$ and the differential equation~(\ref{comparison}) satisfied by
$\phi_{\bf p}^* ({\bf x})$ and integrating by parts yields\footnote{
The surface integral can be omitted since when the atomic matrix
element ${\cal T}({\bf p})$ is inserted into Eq.~(\ref{engyconsv})
to evaluate
the amplitude $A_T$, the integral over the momentum ${\bf p}$ makes the
integrand vanish on the surface by the Riemann-Lebesgue lemma if the
surface is sufficiently large.}
\begin{equation}
    {\cal T} ({\bf p}){=}\phi_{\bf p}^* (0)
	\langle f | Q_+ | i \rangle
        {-}i \! \int_{T_i}^{T_f}\!\! dt \!\int \!(d^3 {\bf r})
	\langle f | {\rm T} \phi_{\bf p}^* ({\bf r}) e^{- i E t}
         [V ({\bf r}, t){-}v (r)] \psi ({\bf r}, t)
          \psi^{\dagger} (0) Q_+ (0)| i \rangle \,.
\end{equation}
Defining an operator $N_{\bf p} (t)$ by
\begin{equation}
    N_{\bf p} (t) \equiv \int (d^3 {\bf r}) \, \phi_{\bf p}^* ({\bf r})
        [V({\bf r}, t) - v (r)] \psi ({\bf r}, t) \,,
\label{Ndef}
\end{equation}
which accounts for the difference between the real interaction
experienced by the beta ray and the comparison potential,
and utilizing
\begin{equation}
    N_{\bf p} (t) = e^{i H t} N_{\bf p} (0) e^{-i H t} \,,
\end{equation}
we can do the time integral and obtain
\begin{eqnarray}
    {\cal T({\bf p})} &=& \phi_{\bf p}^* (0) \langle f | Q_+ | i \rangle
	- \langle f | N_{\bf p} (0) \, {1{-}e^{i(E_f + E - H)T_f}
	\over H - E - E_f} \, \psi^{\dagger} (0) Q_+ (0) | i \rangle
\nonumber\\
    &&+ \langle f | \psi^{\dagger} (0) Q_+ (0) \,
        {1 - e^{i(E - E_i + H)T_i} \over H + E - E_i} \,
        N_{\bf p} (0) | i \rangle \,.
\label{picture}
\end{eqnarray}
Since we require that the potentials $V({\bf r})$ and $v(r)$ have
the same asymptotic behavior at large distance, the operator $N_{\bf p}(0)$
is localized near the origin. The facts that the final state $\langle f |$
is localized and that the operator $N_{\bf p}(0)$ has a compact support,
enables us to drop the highly oscillating
terms in Eq.~(\ref{picture}).\footnote{Again, if we put the
expression~(\ref{picture}) for the atomic matrix element in
Eq.~(\ref{engyconsv}), by the Riemann-Lebesgue lemma, the integral
over momentum ${\bf p}$ makes the highly oscillating terms
vanish.}
Thus we obtain
\begin{eqnarray}
    {\cal T} ({\bf p}) &=& \phi_{\bf p}^*(0)
	\langle f | Q_+ | i \rangle
	- \langle f | N_{\bf p} (0)
	{1 \over H{-}E{-}E_f{-}i \epsilon}
	\psi^{\dagger} (0) Q_+ (0) | i \rangle
\nonumber\\
	&&+ \langle f | \psi^{\dagger} (0) Q_+ (0)
	{1 \over H + E - E_i - i \epsilon}
	N_{\bf p} (0) | i \rangle \,,
\label{correct}
\end{eqnarray}
where $\epsilon$ is an infinitesimal positive number.\footnote{
This $\epsilon$ specifies the correct way to avoid the
zero in the denominator. Before dropping the highly oscillating term,
while doing the ${\bf p}$ integral, the contour can go on the top of
the zero of the denominator
since the numerator is also zero and thus the integrand is finite.
To apply the Riemann-Lebesgue lemma, we can deform the contour in the
$p$ plane slightly to avoid the zero in the denominator.
As $T_f \to \infty$ and $T_i \to - \infty$, we can drop the highly
oscillating terms with the contour properly deformed so as to make
these large time limits well defined.}
This is our general formula for calculating the atomic matrix element
${\cal T} ({\bf p})$ which exploits the comparison potential.
For later reference, we shall denote the three terms in
Eq.~(\ref{correct}) by $T_1$, $T_2$, and $T_3$, so that this result
is expressed as
\begin{equation}
    {\cal T} ({\bf p}) = T_1 + T_2 + T_3 \,.
\label{calTresult2}
\end{equation}

\subsection{Comparison Potential Invariance Theorem}
We now prove explicitly that the atomic matrix element ${\cal T}
({\bf p})$ does not depend on the short range behavior of the
comparison potential
$v (r)$. To accomplish this, we consider a small variation $\delta
v (r)$ in the comparison potential and
show that ${\cal T} ({\bf p})$ evaluated by Eq.~(\ref{correct}) is unchanged
under this small variation.
Varying $v(r)$ in Eq.~(\ref{comparison}) yields
\begin{equation}
    \delta \left (\phi_{\bf p}^* v \right ) =
	\left \{ {\nabla^2 \over 2m} + E \right \}
	 \delta \phi_{\bf p}^* \,.
\label{varied}
\end{equation}
Inserting the variation~(\ref{varied}) into the definition~(\ref{Ndef}) of
$N_{\bf p}$ gives
\begin{eqnarray}
    \delta N_{\bf p} (0) &=& \int (d^3 {\bf r}) \, \delta
	\phi_{\bf p}^* ({\bf r}) \left \{ V({\bf r}) -
	{{\stackrel {\leftarrow} {\nabla}}^2 \over 2m}
	- E \right \} \psi ({\bf r})
\nonumber\\
    &=& \int (d^3 {\bf r}) \, \delta \phi_{\bf p}^* ({\bf r})
	\left \{ V({\bf r}) -
	{{\stackrel {\rightarrow} {\nabla}}^2 \over 2m}
	- E \right \} \psi ({\bf r}) \,,
\end{eqnarray}
where we have performed two partial integrations and dropped the surface
terms to make the differential operator $\nabla$ acting to the right.
The dropped surface terms vanish because $\delta v(r)$ must be a
localized quantity so as to keep the correct long distance behavior in
the comparison potential $v(r)$. Noticing that
\begin{equation}
    \left \{ - {{\stackrel {\rightarrow} {\nabla}}^2 \over 2m}
	+ V({\bf r}) \right \} \psi ({\bf r}) = [\psi, H] \,,
\end{equation}
we have
\begin{equation}
    \delta N_{\bf p} (0) = \left [ \int (d^3 {\bf r}) \, \delta
	\phi_{\bf p}^*({\bf r}) \psi ({\bf r}), H \right ]
	- E \int (d^3 {\bf r}) \, \delta \phi_{\bf p}^* ({\bf r})
	 \psi ({\bf r}) \,.
\label{Nvary}
\end{equation}
Utilizing this variation of the operator $N_{\bf p} (0)$, we indeed find
a vanishing variation of the atomic amplitude,
\begin{eqnarray}
    \delta {\cal T} ({\bf p}) &=& \delta \phi_{\bf p}^* (0)
	\langle f | Q_+ | i \rangle
	- \langle f | \int (d^3 {\bf r})
	\delta \phi_{\bf p}^*({\bf r})
	\psi ({\bf r}) \psi^{\dagger} (0) Q_+ (0) | i \rangle
\nonumber\\
	&& \quad - \langle f | \psi^{\dagger} (0) Q_+ (0)
	\int (d^3 {\bf r}) \delta \phi_{\bf p}^*({\bf r})
	 \psi ({\bf r}) | i \rangle
\nonumber\\
	&=& 0 \,,
\end{eqnarray}
upon using the anticommutation relation~(\ref{anticom}).

\subsection{Decay Rate}
We now take the plane wave limit to compute the
decay rate. Taking the limit where the distribution amplitude
$g({\bf p})$ becomes a delta function, completing the ${\bf p}$
integral in the
expression~(\ref{engyconsv}), and summing over all the possible
final electron states expresses the decay rate as
\begin{equation}
    \Gamma = \int {(d^3 {\bf p}_{\nu}) \over (2 \pi)^3}
	\int {(d^3 {\bf p}) \over (2 \pi)^3}
	{| ({\rm const}) \, 2 \pi
	\delta_T (E_f{+}E{+}E_{\nu}{-}E_i{-}Q)\,
	{\cal T} ({\bf p}) |^2 \over T} \,.
\end{equation}
Noting that
\begin{equation}
    \lim_{T \to \infty} 2\pi \,
	{[\delta_T (E_f{+}E{+}E_{\nu}{-}E_i{-}Q)]^2 \over T}
	= \delta (E_f{+}E{+}E_{\nu}{-}E_i{-}Q)
\end{equation}
produces the decay rate
\begin{equation}
    \Gamma = \int {(d^3 {\bf p}_{\nu}) \over (2 \pi)^3}
	\int {(d^3 {\bf p}) \over (2 \pi)^3} \, 2 \pi \,
	\delta (E_f{+}E{+}E_{\nu}{-}E_i{-}Q)
	|T_{\beta}|^2 |{\cal T} ({\bf p})|^2 \,.
\label{rateresult}
\end{equation}
Here we have compensated for our previous neglect of the real
structure of the beta decay interaction by inserting a factor
of the true squared amplitude $|T_{\beta}|^2$
in the absence of the atomic effects.\footnote{Summing over the final
polarization of $^3{\rm He}$ nucleus, this squared amplitude is simply
\begin{displaymath}
    |T_{\beta}|^2 = G_F^2 \cos^2 \theta_c
	(|G_V|^2 + 3 \,|G_A|^2) \,,
\end{displaymath}
where $\theta_c$ is the Cabibbo angle and
$G_V \simeq 1$ and $G_A \simeq 1.25$ are two coupling constants
for the charged vector-current and axial-vector-current
nuclear matrix elements which are nearly equal to the corresponding
values for free neutron decay.}
(In the absence of atomic effects, ${\cal T} ({\bf p})
\to 1$.) Performing the momentum integrals yields the differential
rate for the decay to a specific final ion state. Assuming that the
neutrino mass vanishes, this gives
\begin{equation}
    {d \Gamma_{fi} \over d E} = {m \over 2 \pi^3} \, p \, (Q_f - E)^2
	|T_{\beta}|^2 \overline{|{\cal T} ({\bf p})|^2} \,,
\label{diffrate}
\end{equation}
where $m$ is the electron mass, and $Q_f$ is the effective reaction
energy for a specific final atomic state $\langle f |$ defined by
\begin{equation}
    Q_f = Q + E_i - E_f \,.
\label{Qfdef}
\end{equation}
When the decay takes place to an excited $^3{\rm He}^+$ ion with
non-vanishing angular momentum, one must average over the spatial
orientation of this state; this average is indicated by writing
$\overline{|{\cal T} ({\bf p}) |^2}$.

\section{Second Order Evaluation}
We turn now to study each term in our new expression~(\ref{calTresult2})
for the atomic amplitude and estimate their order in terms of the
Coulomb parameter $\eta$ defined by
\begin{equation}
    \eta = { 1 \over pa_0} = {e^2 m \over 4 \pi p}
	= {\alpha m \over p} \,,
\end{equation}
where $a_0$ is the Bohr radius and $\alpha$ is the fine structure constant.
The first term,
\begin{equation}
    T_1 = \phi_{\bf p}^* (0) \, \langle f | Q_+ | i \rangle
	\equiv \phi_{\bf p}^* (0) \, T_{fi} \,,
\end{equation}
which is the result of the sudden approximation, dominates
the amplitude. This term is of order $1$.
It can be calculated by explicitly choosing a comparison
potential $v(r)$ and solving an eigenvalue problem.

The second term, $T_2$, cannot be calculated
analytically without doing any approximation.
We shall find an appropriate approximation to calculate it for the case
where the final outgoing electron has an energy much larger than the
typical atomic energy scale so that the Coulomb parameter $\eta$
is small. We shall, in fact, compute the atomic amplitude through order
$\eta^2$. It is convenient to rewrite the field theoretic expression of
$T_2$ in Eq.~(\ref{correct}) in terms of ordinary quantum mechanics notation.
Using this notation, the two electrons are labeled by the
subscripts $1$ and $2$ and the action of the field operators produces
states that are explicitly antisymmetrized.
For the bra and ket, the first quantum
number refers to electron $1$ while the second one refers to
electron $2$.
With this notation in hand, $T_2$ may be expressed as
\begin{equation}
    T_2 = - \langle {\bf p}, f |
	\left [ {e^2 \over 4 \pi |{\bf r}_1 - {\bf r}_2|}
	- {2 e^2 \over 4 \pi r_1} - v(r_1)
	 \right ] {1 \over H - E - E_f - i \epsilon}
	\left [| {\bf r}_1{=}{\bf 0}, i \rangle
	- | i, {\bf r}_2{=}{\bf 0} \rangle \right ] \,,
\label{QM}
\end{equation}
where
\begin{equation}
    \langle {\bf p}, f | \equiv \int (d^3 {\bf r}_1) \int (d^3 {\bf r}_2)
	\phi_{\bf p}^* ({\bf r}_1) \phi_f^* ({\bf r}_2)
	\langle {\bf r}_1, {\bf r}_2 | \,,
	\quad | {\bf r}_1{=}{\bf 0}, i \rangle \equiv \int (d^3 {\bf r}_2)
	| {\bf r}_1{=}{\bf 0}, {\bf r}_2 \rangle \phi_i ({\bf r}_2)
	\,,
\end{equation}
and $| i, {\bf r}_2{=}{\bf 0} \rangle$ is simply
$| {\bf r}_1{=}{\bf 0}, i \rangle$ with $1$ and $2$ exchanged.
Equation (\ref{QM}) naturally defines the direct term $T_{\rm d}$
as the part associated with the ket $| {\bf r}_1{=}{\bf 0}, i \rangle$
and the change term $T_{\rm e}$ as the part associated with
$| i, {\bf r}_2{=}{\bf 0} \rangle$ so that
\begin{equation}
    T_2 = T_{\rm d} + T_{\rm e} \,.
\label{decomp}
\end{equation}
The Hamiltonian in Eq.~(\ref{QM}) may be obtained by using
the definition~(\ref{potential}) of $V(r)$. It is conveniently partitioned
as
\begin{equation}
    H = H_1 + H_2 + H_I \,,
\end{equation}
with
\begin{eqnarray}
    H_1 &=& {{\bf p}_1^2 \over 2m}- {2 e^2 \over 4 \pi r_1} \,,
\label{H1def}
\\
    H_2 &=& {{\bf p}_2^2 \over 2m} - {e^2 \over 4 \pi r_2} \,,
\label{H2def}
\\
    H_I &=& {e^2 \over 4 \pi |{\bf r}_1 - {\bf r}_2|}
	- {e^2 \over 4 \pi r_2} \,.
\label{interdef}
\end{eqnarray}
We note here that $\phi_i ({\bf r}_2)$ is the ground state
wave function for the Hamiltonian $H_2$ defined in~(\ref{H2def}). This is
the reason why we split the total Hamiltonian in the manner shown above.
We shall find this choice makes the evaluation of the direct term $T_{\rm d}$
easier.  When we consider the exchange term in appendix D,
the definitions of $H_1$ and $H_2$ are switched since there ${\bf r}_1$
and ${\bf r}_2$ are interchanged.

\subsection{Direct Terms; Comparison Potential Choice}
We now examine the direct term $T_{\rm d}$.
Expanding Eq.~(\ref{QM}) in powers of $H_I$, we get the leading
order term for $T_{\rm d}$,
\begin{eqnarray}
    T_{\rm d}^0 &=&\! - \langle {\bf p}, f |
        \left [{e^2 \over 4 \pi |{\bf r}_1{-}{\bf r}_2|}
	{-}{2 e^2 \over 4 \pi r_1}{-}v(r_1) \right ]
        {1 \over H_1 + H_2 - E - E_f - i \epsilon}
        | {\bf r}_1{=}{\bf 0}, i \rangle
\nonumber\\
	&=&\!{-}\langle {\bf p} | \! \int (d^3 {\bf r}_2)
	\phi_f^* ({\bf r}_2)\phi_i ({\bf r}_2) \!
        \left [ {e^2 \over 4 \pi |{\bf r}_1{-}{\bf r}_2|}
	{-}{2 e^2 \over 4 \pi r_1}{-}v(r_1) \right ] \!
	{1 \over H_1{+}E_i{-}E{-}E_f{-}i \epsilon}
	|{\bf r}_1{=}{\bf 0} \rangle \,,
\label{dirlead}
\end{eqnarray}
where we have used $ H_2 | i \rangle = E | i \rangle$.
Since the final ion S-states contribute the major part of
the total decay rate,
we shall first consider only the cases
where the final ionic state is in S-state.
We shall choose the comparison potential to be given by
\begin{equation}
    v (r) = {e^2 \over 4 \pi}
        \int (d^3 {\bf r}^{\prime}) \rho_{fi} (r^{\prime})
        {1 \over |{\bf r} - {\bf r}^{\prime}|} - {2 e^2 \over 4 \pi r} \,,
\label{effpot}
\end{equation}
where\footnote{This density is always real because the
wave functions for the initial and final states can be written as
a real function multiplied by a constant phase factor which is canceled
by the normalization factor $1/T_{fi}$.}
\begin{equation}
    \rho_{fi} (r) = {\langle f | \rho ({\bf r}) Q_+
	| i \rangle \over \langle f | Q_+ | i \rangle}
	= {1 \over T_{fi} } \phi_f^* ({\bf r}) \phi_i ({\bf r}) \,.
\label{rhodef}
\end{equation}
It is easy to see that with this choice the leading order
term~(\ref{dirlead}) vanishes.

With the choice~(\ref{effpot}) of the comparison potential, we show
in Appendix A that, including terms up to order $\eta^3$,
\begin{equation}
    \phi_{\bf p}^* (0) = \sqrt{{p^{\prime} \over p}}
	e^{\pi \eta^{\prime}}
	\Gamma (1 - 2 i \eta^{\prime}) e^{-i\Theta} \,,
\label{Oval}
\end{equation}
which is a modification of the result in reference~\cite{Rose}.
The phase $\Theta$, which is irrelevant in the beta decay
problem, is defined in Appendix A. The shifted momentum
$p^{\prime}$ is related to
\begin{equation}
    V_{fi} = \int (d^3 {\bf r}) \rho_{fi} (r) {e^2 \over 4 \pi r}
	\equiv {\alpha \over a_0} {\tilde v}_{fi}
\end{equation}
via
\begin{equation}
    p^{\prime} = \sqrt {p^2 - 2 m V_{fi}}
	= p \sqrt{1 - 2 \eta^2 {\tilde v}_{fi}}\,,
\label{ppdef}
\end{equation}
and
\begin{equation}
    \eta^{\prime} = \alpha m / p^{\prime} \,.
\end{equation}
Note that $V_{fi}$ is the potential energy at the origin produced
by the charge distribution~(\ref{rhodef}),
which involves both the initial and final states,
while ${\tilde v}_{fi}$ is a dimensionless potential.
This result differs from the correction
derived in the literature \cite{Rose} which uses the charge
density of the initial state, not our ``transition density''
$\rho_{fi}$.

\subsection{Exchange Terms}
The third term $T_3$ in Eq.~(\ref{calTresult2}) and the term
$T_{\rm e}$ in Eq.~(\ref{decomp}) are both exchange terms. They
are examined in Appendix D, where the result~(\ref{exresult})
shows that the exchange amplitudes are of order $\eta^3$.
This contradicts a result appearing in the literature
\cite{Bahcall}, where the leading exchange amplitudes
are claimed to be of order $\eta^4$.
Though the exchange amplitudes are of order $\eta^3$, they only
contribute to the decay rates at order $\eta^4$ since they are
relatively imaginary to the leading sudden approximation amplitude.
The details are given in Appendix D.

\subsection{Order $\eta^2$ Corrections}
The first correction to this leading order direct term $T_{\rm d}$
is obtained by keeping one more term while expanding
Eq.~(\ref{QM}) in powers of $H_I$:
\begin{eqnarray}
    T_{\rm d}^1 &=& \langle {\bf p}, f |
	\left [ {e^2 \over 4 \pi |{\bf r}_1 - {\bf r}_2|}
	- {2 e^2 \over 4 \pi r_1} - v(r_1)
         \right ]
\nonumber\\
        && \qquad \times
	{1 \over H_1{+}H_2{-}E {-}E_f {-}i \epsilon} \;
	H_I \; {1 \over H_1{+}H_2{-}E{-}E_f{-}i \epsilon}
	|{\bf r}_1{=}{\bf 0}, i \rangle \,.
\label{second_direct}
\end{eqnarray}
This leading term of $T_{\rm d}^1$ is of order $\eta^2$.
Therefore, to the second order of $\eta$, we may do
following approximations. First, it is valid to drop $E_f$ and
$H_2$ in the denominators relative to energy $E$ since they are
of order\footnote{We provide here more justification for dropping
the Hamiltonian $H_2$.
Since the wave function $\phi_i ({\bf r}_2')$ is the ground state
eigenfunction of $H_2$, one can replace $H_2$ in the second denominator
by $E_i$. For the first denominator, we can imagine
inserting a complete set of eigenstates of $H_2$ just before
$H_I$ in the second line of Eq.~(\ref{second_direct}).
Since the wave functions $\phi_i ({\bf r}_2')$ and $\phi_f ({\bf r}_2)$
are slowly varying functions, even after multiplied by the Coulomb
interaction factor $1/|{\bf r}_1-{\bf r}_2|$, they have little
overlap with the eigenfunctions of $H_2$ with
energies much higher than the atomic energy due to the highly oscillating
feature of these high energy eigenfunctions. Therefore,
in the sum over the eigenstates of $H_2$,
the main contribution comes from summing over the excitations
with energy being of the order of the atomic energy.
Thus, in both denominators, $H_2$ is of the order of the atomic
energy.} $\eta^2 E$.
Secondly, we can approximate the beta ray wave function
$\phi_{\bf p}^* ({\bf r}_1)$ by the plane wave
$\exp ( -i {\bf p} \cdot {\bf r}_1 )$ or equivalently
treat $\langle {\bf p} |$ as a free particle state with
momentum ${\bf p}$ since their difference is
of order $\eta$. Finally, the Hamiltonian of electron $1$ $H_1$
may be replaced by its free Hamiltonian $H_0$. This is justified
because $H_1$ describes the motion of the beta ray,
thus the Coulomb interaction provides only higher order corrections
in the $\eta$ parameter to its propagation.
With these approximations, to order $\eta^2$, we get on using
Eq.~(\ref{interdef}),
\begin{eqnarray}
    T_{\rm d}^1 &\simeq& \int (d^3 {\bf r}_2)
	\phi_f^* ({\bf r}_2) \phi_i ({\bf r}_2)
	\langle {\bf p} |
        \left [ {e^2 \over 4 \pi |{\bf r}_1 - {\bf r}_2|}
	- {2 e^2 \over 4 \pi r_1} - v(r_1) \right ]
\nonumber\\
        && \qquad \qquad \times {1 \over H_0{-}E{-}i \epsilon}
        \left [{e^2 \over 4 \pi} \left (
	{1 \over |{\bf r}_1 - {\bf r}_2|} - {1 \over r_2} \right )
	\right ]
	{1 \over H_0{-}E{-}i \epsilon}
        |{\bf 0} \rangle
\nonumber\\
	&\equiv& \eta^2 J_{fi} T_{fi} \,.
\label{J_fi_def}
\end{eqnarray}

In Appendix B, we find that $J_{fi}$ is given by
\begin{equation}
    J_{fi} = {1 \over 2}{\tilde v}_{fi} - 1 - {\tilde K}_{fi} \,,
\label{J_fi_result}
\end{equation}
with
\begin{equation}
   {\tilde K}_{fi} =  {1 \over 4}
	\int (d^3 {\bf r}_1) \rho_{fi} (r_1)
        \int (d^3 {\bf r}_2) \rho_{fi} (r_2)
        \ln^2 \left ( {r_2 \over r_1} \right ) \,.
\label{Corrections}
\end{equation}
Combining these second order corrections~(\ref{Corrections}) with
value of the wave function $\phip$ at the origin~(\ref{Oval})
gives the the atomic matrix element ${\cal T} ({\bf p})$
to order $\eta^2$:
\begin{equation}
    {\cal T} ({\bf p}) = T_{fi} \sqrt{{p^{\prime} \over p}}
        e^{\pi \eta} \Gamma (1 - 2 i \eta) e^{-i\Theta}
	\left [1 - \eta^2 \left ( 1 - {1 \over 2}{\tilde v}_{fi}
	+ {\tilde K}_{fi} \right )  \right ] \,.
\label{tobecancel}
\end{equation}
Taking the square of this matrix element, using the
definition~(\ref{ppdef}) of $p^{\prime}$, dropping higher
order terms in $\eta$, and recalling the differential decay
rate formula~(\ref{diffrate}) yields
the exclusive differential decay rate to an ionic S-state:
\begin{equation}
    {d \Gamma_{f0} \over d E} = {m \over 2 \pi^3} \,
	F(2,E) \, p \, | T_{\beta} |^2 \,
	(Q_f - E )^2 \, |T_{fi}|^2 \,
	\left [ 1 - 2 \eta^2 (1 + {\tilde K}_{fi}) \right ] \,,
\label{diffex}
\end{equation}
where $F(2,E)$ is the Fermi function defined in Eq.~(\ref{fermi}).
Note that the term in Eq.~(\ref{tobecancel})
involving ${\tilde v}_{fi}$ cancels the term
from expanding the momentum ratio $p'/p$.

Using the wave functions listed at the end of Appendix B, it is
easy to compute
\begin{eqnarray}
    |T_{1i}|^2 &=& \left ({16 \sqrt{2} \over 27} \right )^2
	= {512 \over 729} \simeq 0.7023 \,,
\\
    |T_{2i}|^2 &=& \left ( - {1 \over 2} \right )^2 = {1 \over 4}
	= 0.25 \,,
\end{eqnarray}
and
\begin{equation}
    |T_{3i}|^2 = \left (- {144 \sqrt{6} \over 5^5} \right )^2
	= {2^9 \, 3^5 \over 5^{10}} \simeq 0.0127 \,,
\end{equation}
which demonstrates the dominance of the first two exclusive decay rates.
It is also straight forward to compute ${\tilde K}_{fi}$ for a transition
to any particular final state. For example, we show in Appendix B that
for the transitions to the first three low energy S-states,
\begin{eqnarray}
    {\tilde K}_{1i} &=& \left ( {\pi^2 \over 12} -
	{5 \over 8} \right ) \simeq 0.1975 \,,
\label{SK1i}
\\
    {\tilde K}_{2i} &=& \left ( {\pi^2 \over 12}
	- {9 \over 8}\right ) \simeq - 0.3025 \,,
\label{SK2i}
\\
    {\tilde K}_{3i} &=& \left ({\pi^2 \over 12}
	- {1045 \over 648}\right ) \simeq - 0.7902 \,.
\label{SK3i}
\end{eqnarray}

\section{Summing over atomic S-wave states}
We shall denote by $\Gamma_0$ the contribution to the inclusive
decay rate from the cases where the final ion is in an S-wave.
According to Eq.~(\ref{diffex}), this is given by
\begin{equation}
    {d \Gamma_0 \over d E} =
	{m \over 2 \pi^3} F(2,E)
	p \, | T_{\beta} |^2
	\sum_{f<f_{\rm max}} (Q_f - E )^2 |T_{fi}|^2
	\left [ 1 - 2 \eta^2 (1 + {\tilde K}_{fi}) \right ] \,,
\label{Sdiffrate}
\end{equation}
where the upper limit $f_{\rm max}$ of the summation corresponds
to the final ionic states with energy
$E_f$ being $\Emax \equiv Q + E_i - E$ since the total energy must be
conserved. In view of Eq.~(\ref{Qfdef}),
\begin{equation}
    (Q_f - E)^2 = (Q - E)^2 + 2 \, (Q - E) (E_i - E_f)
	+ (E_i - E_f)^2 \,,
\label{squareexp}
\end{equation}
we need to do sums weighted by $(E_i-E_f)^n$, with $n = 0, 1, 2\,.$
The result will be expressed as the sudden approximation spectrum
plus the $\eta^2$ order correction:
\begin{equation}
    {d \Gamma_0 \over d E} =
	{m \over 2 \pi^3} F(2,E) \,
	p \, | T_{\beta} |^2 \Biggr [ (1 - 2 \eta^2)
	P(Q-E)
	- {\eta^2 \over 2} C (Q{-}E) \Biggr ] \,,
\label{Sresult}
\end{equation}
We shall first examine the sums corresponding to the sudden
approximation which defines $P(Q-E)$ and then
the sums involving ${\tilde K}_{fi}$ which defines $C(Q{-}E)$.

\subsection{Sudden Approximation Terms}
To do the sum over final states which do not include
energies higher than
\begin{equation}
    {K^2 \over 2m} \equiv \Emax = Q + E_i - E \,,
\end{equation}
we write
\begin{equation}
    \sum_{f<f_{\rm max}} | f \rangle \langle f| = 1 - \int_{k > K}
	{(d^3 {\bf k}) \over (2 \pi)^3} | {\bf k} \rangle
	\langle {\bf k} | \,.
\label{Incomplete}
\end{equation}
Using the squared matrix element calculated in Appendix C,
Eq.~(\ref{ki2result}),
\begin{equation}
    | \langle {\bf k} | i \rangle |^2 = {256 \pi^2
	\over 1 - e^{-4 \pi \gamma}} {1 \over k^3}
	{\gamma^6 \over (1 + \gamma^2)^4} e^{- 8 \gamma \cot^{-1} \gamma} \,,
\label{kisquared}
\end{equation}
where $\gamma = 1/ka_0$,
we can evaluate the needed sums appearing in
the sudden approximation~(\ref{Sdiffrate}) as
\begin{eqnarray}
    \sum_{f<f_{\rm max}} |T_{fi}|^2 (E_i - E_f)^n
	\tab=\tab \langle i | (H_i - H_f)^n
	| i \rangle - \int_{k > K} {(d^3 {\bf k}) \over (2 \pi)^3}
	| \langle {\bf k} | i \rangle |^2
	\left ( E_i - E_k \right )^n
\nonumber\\
	\tab=\tab \langle i | (H_i - H_f)^n | i \rangle -
	{32 \over \pi} \int_0^{1 / Ka_0} d \gamma
\nonumber\\
	\tab\tab \quad \times {4 \pi \gamma \over 1{-}e^{-4 \pi \gamma}}
	{\gamma^4 \over (1{+}\gamma^2)^4}
	e^{- 8 \gamma \cot^{-1} \gamma}
	\left ( - {{\rm R\!y} \over \gamma^2 } \right )^n
	\left (1{+}\gamma^2 \right )^n \,,
\label{suddensums}
\end{eqnarray}
where $E_k$ is the energy of the ionized electron,
\begin{equation}
    E_k = {k^2 \over 2m} \,,
\end{equation}
and ${\rm R\!y}$ is the Rydberg constant,
\begin{equation}
    {\rm R\!y} = {e^2 \over 8 \pi a_0}
	= {1 \over 2 m a_0^2} \simeq 13.6 \, {\rm eV} \,.
\end{equation}
The first term in Eq.~(\ref{suddensums}) is the closure approximation
result, and it can be calculated easily by using the ground state wave
function of hydrogen atom:
\begin{eqnarray}
    \langle i | (H_i - H_f)^n | i \rangle =
	\left \langle i \left | \left ({e^2 \over 4 \pi r} \right )^n
	\right | i \right \rangle =  \left \{
\begin{array}{r}
	1 \,, \quad n=0 \,,
\\
	2 \; {\rm R\!y} \,, \quad n=1 \,,
\\
	8 \; {\rm R\!y}^2 \,, \quad n=2 \,.
\end{array}
\right.
\label{suddenclosure}
\end{eqnarray}
The sudden approximation spectrum is proportional to the sum
\begin{equation}
    P (Q-E) \equiv \sum_{f<f_{\rm max}} |T_{fi}|^2 (Q_f - E)^2 \,,
\label{Sudden_def}
\end{equation}
which may be expressed in terms of the closure
approximation result plus the correction to the closure part due
to the fact that the summation does not include the final states with
energy higher than $\Emax = K^2 / 2m$.
Since $K$ is related to $Q-E$ by the relation $\Emax
= K^2/2m = Q-E+E_i$,
we write $P$ as a function of variable $Q-E$ for the convenience of
later usage.
Explicitly, using the closure results~(\ref{suddenclosure}) above,
we have
\begin{equation}
    P (Q-E) = (Q - E + 2 {\rm R\!y})^2 + 4 {\rm R\!y}^2 + S (Q{-}E) \,,
\label{SKdef}
\end{equation}
where $S (Q{-}E)$ denotes the correction to this closure approximation
result. In view of Eq.~(\ref{suddensums}), this correction comes
from the $\gamma$ integrals, which can be evaluated numerically.

The function $P_1 (E-Q)$ defined by Eq.~(\ref{P_1_def}) in Section I
for the purpose of estimating the atomic effects in neutrino mass
measurements may be obtained immediately by setting $n=0$
in Eq.~(\ref{suddensums}) :
\begin{equation}
    P_1 (Q-E) = 1 - {32 \over \pi} \int_0^{1 / Ka_0} d \gamma
	{4 \pi \gamma \over 1 - e^{-4 \pi \gamma}}
        {\gamma^4 \over (1 + \gamma^2)^4}
        e^{- 8 \gamma \cot^{-1} \gamma} \,.
\end{equation}
This expression enables a numerical evaluation of the function
$P_1 (Q-E)$.

Before displaying the numerical results, we consider the
asymptotic behavior of $S (Q{-}E)$ and $P_1 (Q-E)$ as $Ka_0$ becomes large.
For $Ka_0 \gg 1$, we study the asymptotic expansion of the $\gamma$
integrals in Eq.~(\ref{suddensums}).
This may be accomplished by expanding
the integrands in powers of $\gamma$ since the integral takes the value
of the integrand in the interval $(0, 1/Ka_0)$. One can then do these
integrals easily. Keeping only the three leading
terms in the expansion, the large
$Ka_0$ asymptotic form for $S (Q{-}E)$ is
\begin{equation}
    S (Q{-}E) \simeq {32 \over 15} \left [ - {8 \over \pi} {1 \over Ka_0}
	+ {5 \over (Ka_0)^2}
	- \left ({32 \pi \over 21}
	+ {32 \over 7 \pi} \right ) {1 \over (Ka_0)^3} \right ]
	{\rm R\!y}^2 \,.
\label{sdnasy}
\end{equation}
As one of the intermediate steps above, the leading large $Ka_0$
behavior of $P_1 (Q-E)$ is
\begin{equation}
    P_1 (Q-E) \simeq 1 - {32 \over 5 \pi} {1 \over (Ka_0)^5} \,.
\label{P_1_largeasy}
\end{equation}

\begin {figure}[htbp]
        \epsfysize 5in
        \leavevmode {\hfill \epsfbox {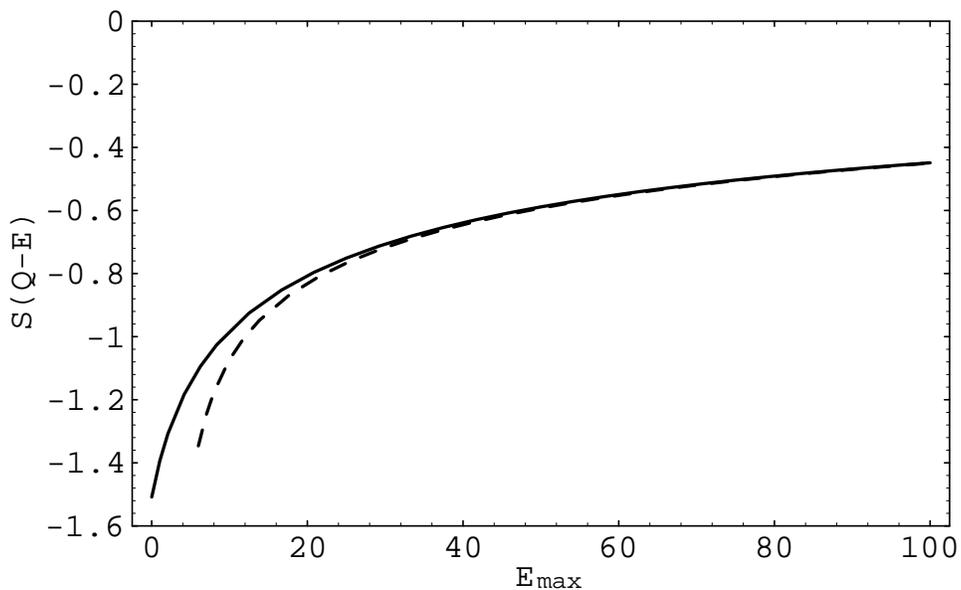} \hfill}
    \protect\caption
        {%
        \advance\baselineskip by -8pt
        Curves for $S (Q{-}E)$ --- the correction to the closure part of the
	sudden approximation --- as a function of the energy threshold
	of the final ionic state, $\Emax = K^2/2m=Q+E_i-E$ and its
	asymptotic behavior for large $Q{-}E$. The solid curve
	is the plot of numerical result of $S (Q{-}E)$, the lower and
	dashed curve is the high energy asymptotic behavior given
	in Eq.~(\protect{\ref{sdnasy}}). The units are
	${\rm R\!y}^2$ for the vertical axis and ${\rm R\!y}$
	for the horizontal axis.
        }%
    \label {fig:one}
\end {figure}

\begin {figure}[htbp]
        \epsfysize 5in
        \leavevmode {\hfill \epsfbox {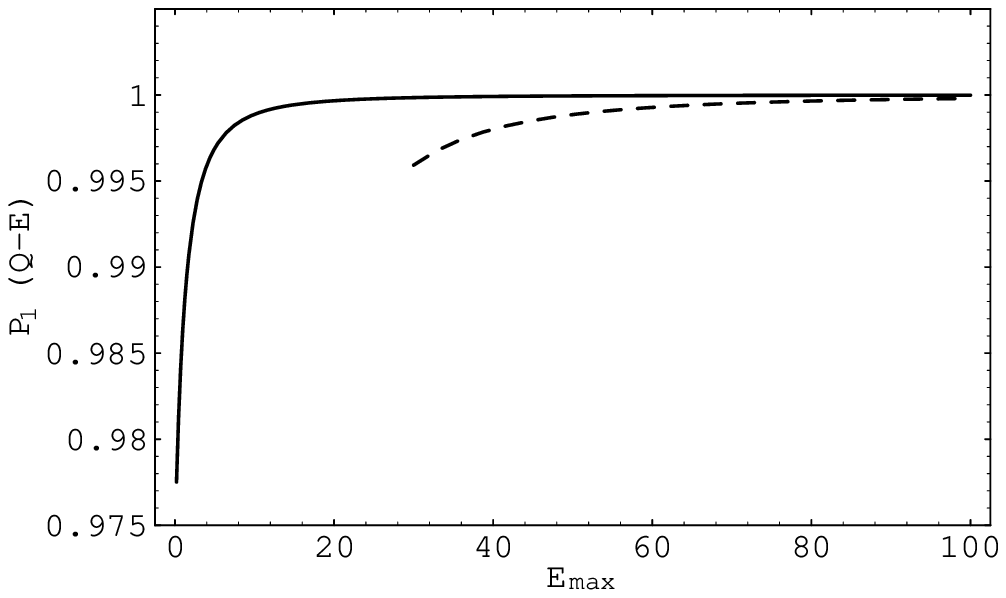} \hfill}
   \protect\caption
        {%
        \advance\baselineskip by -8pt
        Curves for $P_1 (Q-E)$ --- a function measuring the change of
	the spectrum due to per unit change of the neutrino mass squared
	--- as a function of the energy threshold of the final ionic state,
	$\Emax = K^2/2m=Q+E_i-E$ and its asymptotic behavior for large
	$Q{-}E$. The solid curve
	is the plot of numerical result of $P_1 (Q-E)$, the lower and
	dashed curve is the high energy asymptotic behavior given
	in Eq.~(\protect{\ref{P_1_largeasy}}). The units are
	$1$ for the vertical axis and ${\rm R\!y}$ for the
	horizontal axis.
        }%
    \label {fig:two}
\end {figure}

We show in Fig.~\ref{fig:one} the numerical result of $S (Q{-}E)$ as
a function of $\Emax = K^2/2m = Q + E_i - E$.
The asymptotic form~(\ref{sdnasy}) is also shown in Fig.~\ref{fig:one},
where one finds that it describes $S(Q{-}E)$ with a good accuracy until
$Ka_0$ is less than $5$, which is expected by observing that the coefficients
of the three terms in the asymptotic form~(\ref{sdnasy}) are all
of order $1$. The numerical result of function
$P_1 (Q-E)$ is shown in Fig.~\ref{fig:two}
together with its asymptotic form~(\ref{P_1_largeasy})
as functions of $\Emax$.

\goodbreak

\goodbreak

\protect\subsection{${\protect\tilde K}_{fi}$ Corrections}

In this section, we shall deal with the contribution to
the inclusive differential decay rate involving ${\tilde K}_{fi}$.
Recalling the differential decay rate formula~(\ref{Sdiffrate}), the sums
which we need to consider involve the combination $|T_{fi}|^2 {\tilde K}_{fi}$.
Therefore, it is convenient to define
\begin{equation}
    K (\epsilon) \equiv
	\langle i | \, r^{-\epsilon} | f \rangle
	\langle f | \, r^{\epsilon} | i \rangle \,,
\label{cal_def}
\end{equation}
which is related to ${\tilde K}_{fi}$ via
\begin{equation}
    |T_{fi}|^2 \, {\tilde K}_{fi} = {1 \over 4} \, K'' (0) \,,
\label{tilde_cal}
\end{equation}
in view of Eq.~(\ref{Corrections}).
Correspondingly, the sum
\begin{equation}
    K_n (\epsilon) \equiv \sum_{f<f_{\rm max}} \;
	\langle i | \, r^{-\epsilon} | f \rangle
        \langle f | \, r^{\epsilon} | i \rangle \, (E_i - E_f)^n \,,
\label{K_ndef}
\end{equation}
can produce the desired sums through
\begin{equation}
    \sum_{f<f_{\rm max}} \, |T_{fi}|^2 \, {\tilde K}_{fi} \, (E_i - E_f)^n
        = {1 \over 4} \, K_n'' (0) \,.
\label{summations}
\end{equation}
Since the spectrum is corrected by the combination
$\sum_{f<f_{\rm max}} |T_{fi}|^2 {\tilde K}_{fi} (Q_f - E)^2 $, in view
of the relation~(\ref{summations}), the contribution to the
spectrum due to ${\tilde K}_{fi}$ can be expressed in terms of the
quantity $C (Q{-}E)$ defined by
\begin{equation}
    C (Q{-}E) \equiv (Q - E)^2 K_0 '' (0) + 2 (Q - E) K_1 ''(0)
		+ K_2 ''(0) \,.
\label{CKdef}
\end{equation}
The goal of this section is to show the main steps of calculating
$C (Q{-}E)$ and display the numerical results of $C (Q{-}E)$.

For $n=0$, the completeness relation
\begin{equation}
    K_0 (\epsilon) = 1 - \int_{k>K} {(d^3 {\bf k}) \over (2 \pi)^3}
	K (\epsilon)
\end{equation}
and the $\epsilon$ derivatives give
\begin{equation}
    K_0 '' (0) = - \int_{k>K}
	{(d^3 {\bf k}) \over (2 \pi)^3} K'' (0)
	= - {1 \over 2 \pi^2 a_0^3} \int_0^{1/(Ka_0)}
	d \gamma \, \gamma^{-4} K''(0) \,.
\label{K_0result}
\end{equation}
With this, $K_0$ may be then directly evaluated by exploiting the
numerical results for $K''(0)$ obtained in Appendix C.

For $n=1$, we may write similarly
\begin{equation}
    K_1 (\epsilon) = \langle i | (H_i r^{\epsilon}
	- r^{\epsilon} H_f ) r^{-\epsilon} | i \rangle
	- \int_{k>K} {(d^3 {\bf k}) \over (2 \pi)^3}
	K (\epsilon)
	\left (E_i - E_k \right ) \,.
\end{equation}
The first, closure-approximation term is readily evaluated:
\begin{eqnarray}
    \langle i | (H_i r^{\epsilon}
        - r^{\epsilon} H_f ) r^{-\epsilon} | i \rangle
	&=& \langle i | \left [{e^2 \over 4 \pi r}
	+ {\epsilon^2 \over 2 m r^2} + {i \epsilon
	\over 2m} \left ({\bf p} \cdot {{\bf r}
	\over r^2} + {{\bf r} \over r^2} \cdot {\bf p}
	\right ) \right ] | i \rangle
\nonumber\\
	&=& 2 (1 + \epsilon^2) {\rm R\!y} \,.
\end{eqnarray}
Therefore,
\begin{eqnarray}
    K_1 '' (0) &=& \left [ 4 +
	\int_{k>K} {(d^3 {\bf k}) \over (2 \pi)^3}
	K'' (0) (1 + \gamma^{-2}) \right ] {\rm R\!y}
\nonumber\\
	&=& \left [4 + {1 \over 2 \pi^2 a_0^3} \int_0^{1/(Ka_0)}
	    d \gamma \gamma^{-6} (1 + \gamma^2) \,
	    K'' (0)  \right ] {\rm R\!y} \,,
\label{K_1result}
\end{eqnarray}
which enables again a simple numerical calculation of $K_1''(0)$
upon using the result of $K''(0)$.

For the case $n = 2$, since the closure approximation result diverges,
we shall use a different formalism.
It is convenient to separate the sum in the definition~(\ref{K_ndef}),
and thus the definition for $K_2'' (0)$, into two parts:
the contribution from summing the final ionic bound states denoted by
\begin{equation}
    K_{\rm 2b} = \sum_b K''(0) (E_i - E_f)^2 \,,
\label{K_2bdef}
\end{equation}
and the contribution from the continuum with energy under the threshold
$K^2/2m$ denoted by
\begin{equation}
    K_{\rm 2c} = \int_{k<K} {(d^3 {\bf k}) \over (2 \pi)^3}
        K''(0) (E_i - E_k)^2
	= {1 \over 2 \pi^2} \int_{1/Ka_0}^{\infty} d \gamma \,
	  \gamma^{-8} (1 + \gamma^2) \, K''(0) \, {\rm R\!y}^2 \,.
\end{equation}
The sum of $K_{\rm 2b}$ and $K_{\rm 2c}$ produces $K_2''(0)$.
The second part $K_{\rm 2c}$ can again be readily evaluated by using the
numerical results for $K''(0)$ given in Appendix C, and this Appendix
also provides the evaluation
\begin{equation}
    K_{\rm 2b} = 32.26 \, {\rm R\!y}^2 \,.
\end{equation}

\begin {figure}[htbp]
        \epsfysize 5in
        \leavevmode {\hfill \epsfbox {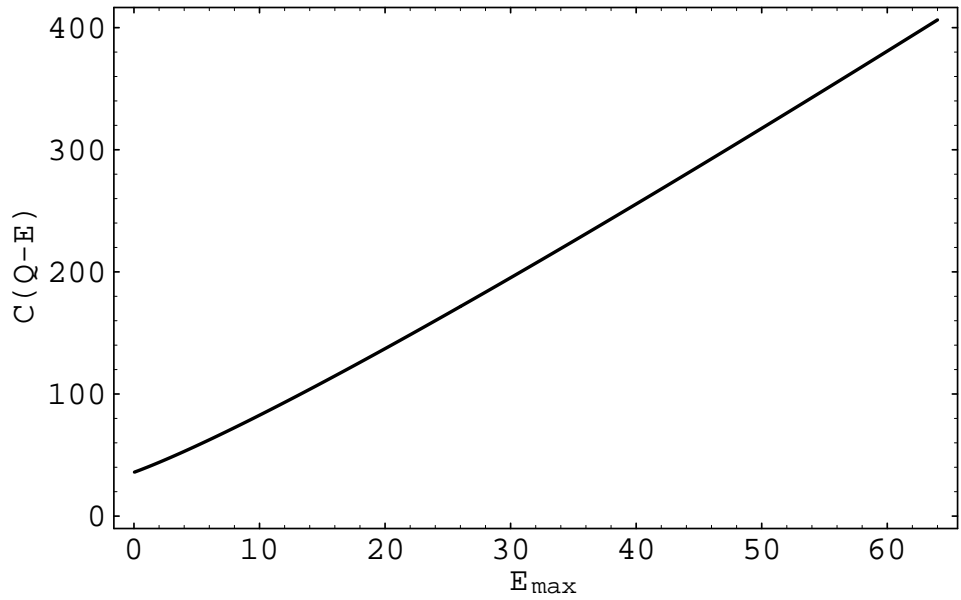} \hfill}
    \caption
        {%
        \advance\baselineskip by -8pt
        Curve for $C (Q{-}E)$ --- the correction to the spectrum due to
	${\protect\tilde K}_{fi}$ as a function of the threshold energy
	of the final ionic state $\Emax = K^2/2m=Q+E_i-E$; the units are
	${\rm R\!y}^2$ for the vertical axis and ${\rm R\!y}$ for the
	horizontal axis.
        }%
    \label {fig:three}
\end {figure}

\begin {figure}[htbp]
        \epsfysize 5in
        \leavevmode {\hfill \epsfbox {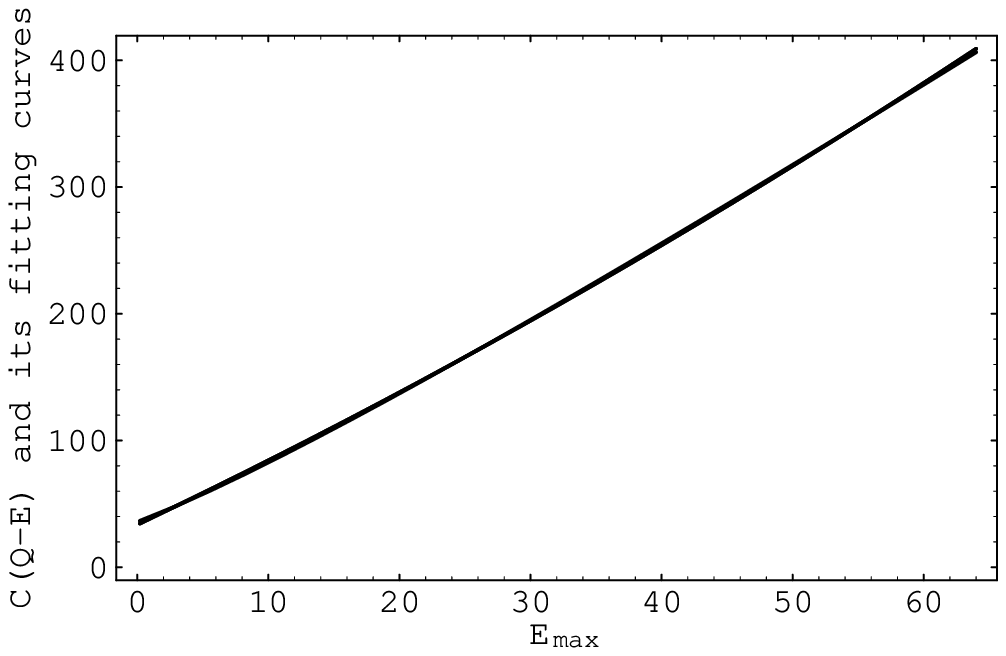} \hfill}
    \caption
        {%
        \advance\baselineskip by -8pt
        Curves for $C (Q{-}E)$ and its two fitting forms $C_{f1} (Q{-}E)$ and
	$C_{f2} (Q{-}E)$ which can barely
	be distinguished. The units are the same as
	in previous figure, ${\rm R\!y}^2$ for the vertical axis
	and ${\rm R\!y}$ for the horizontal axis.
        }%
    \label {fig:four}
\end {figure}

\begin {figure}[htbp]
        \epsfysize 5in
        \leavevmode {\hfill \epsfbox {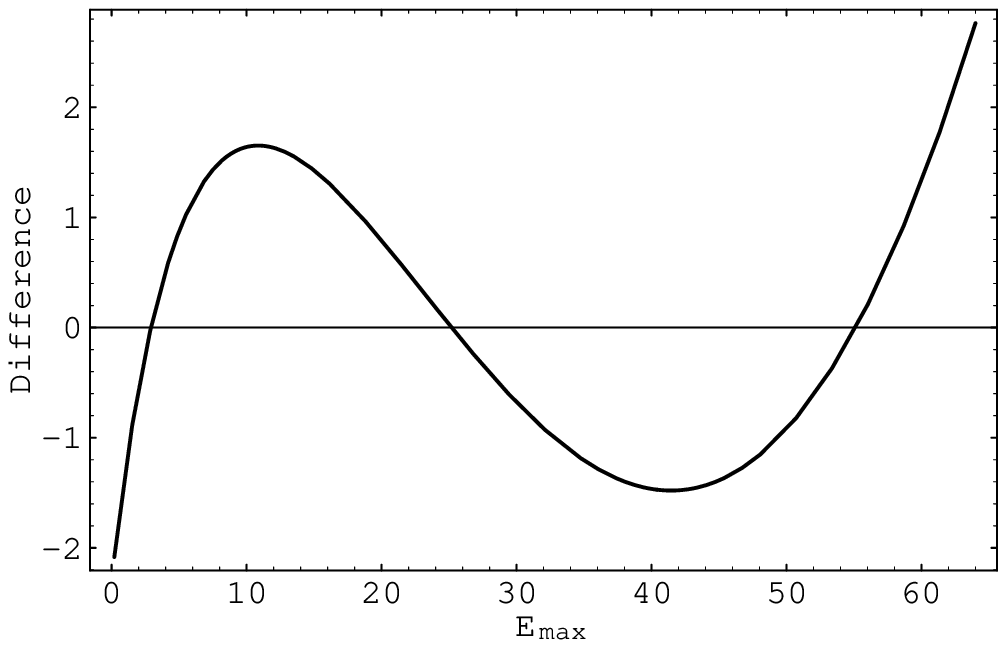} \hfill}
    \caption
        {%
        \advance\baselineskip by -8pt
        Curve for the
	difference $C_{f1} (Q{-}E)-C (Q{-}E)$ as a function of $\Emax
	= Q+E_i-E$.  The units are the same as
	in previous figure, ${\rm R\!y}^2$ for the vertical axis
	and ${\rm R\!y}$ for the horizontal axis.
        }%
    \label {fig:five}
\end {figure}

\begin {figure}[htbp]
        \epsfysize 5in
        \leavevmode {\hfill \epsfbox {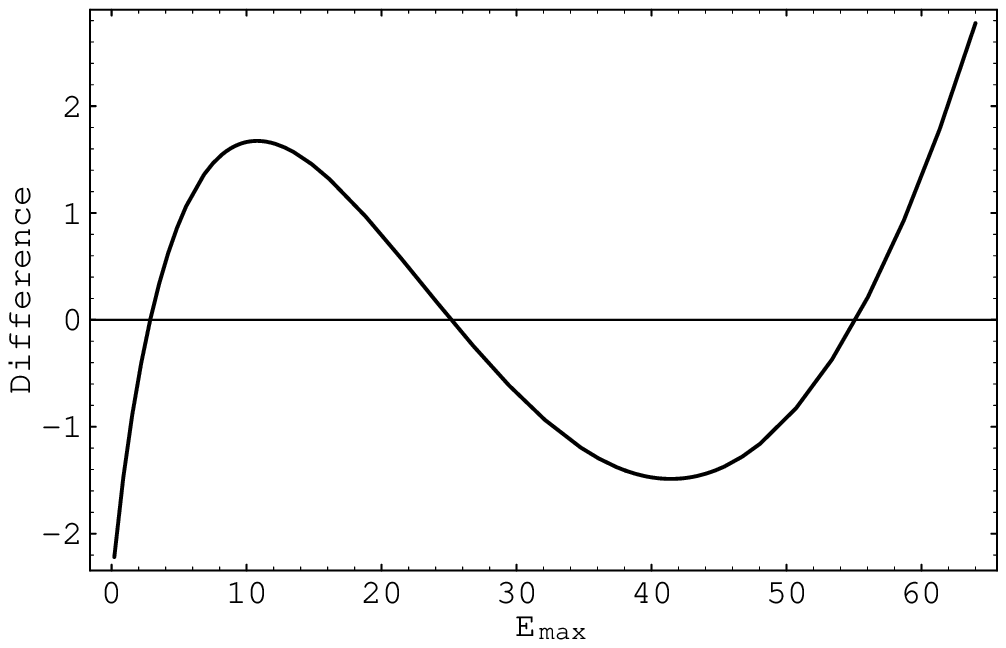} \hfill}
    \caption
        {%
        \advance\baselineskip by -8pt
        Curve for the
	difference $C_{f2} (Q{-}E)-C (Q{-}E)$ as a function of $\Emax
	= Q + E_i - E$. The units are the same as
	in previous figure, ${\rm R\!y}^2$ for the vertical axis
	and ${\rm R\!y}$ for the horizontal axis.
        }%
    \label {fig:six}
\end {figure}

The curve for $C (Q{-}E)$ as a function of the
threshold energy $\Emax$ is shown in Fig.~\ref{fig:three}.
\goodbreak
We find that a quadratic function $C_{f1} (Q{-}E)$,
\begin{eqnarray}
    C_{f1} (Q{-}E) &=& 33.6 \, {\rm R\!y}^2 + 4.92 \, {\rm R\!y} \, \Emax
		+ 0.0148 \, \Emax^2
\nonumber\\
	&=& 28.7 \, {\rm R\!y}^2 + 4.89 \, {\rm R\!y} \, (Q - E)
		+ 0.0148 \, (Q - E)^2 \,,
\label{Cfitting}
\end{eqnarray}
describes $C(Q{-}E)$ for $\Emax$ in the range
$0-64 \, {\rm R\!y}$ ($0 - 870$ eV) with a
good accuracy. To analyze how the atomic effects change the
experimental data fitting, we fit $C (Q{-}E)$ to a linear combination
of the functions $P (Q{-}E)$, $P ' (Q{-}E)$, and $ P_1 (Q{-}E)$ in the
same range of the spectrum and get
\begin{equation}
    C (Q{-}E) \approx C_{f2} (Q{-}E) \equiv 0.0148 \, P (Q{-}E) +
	2.42 \, {\rm R\!y} \, P ' (Q{-}E) + 18.9 \, {\rm R\!y}^2
	P_1 (Q{-}E) \,.
\label{Cfitting2}
\end{equation}
We show $C(Q{-}E)$ and its two fitting formula $C_{f1} (Q{-}E)$ and
$C_{f2} (Q{-}E)$ simultaneously in Fig.~\ref{fig:four}. The three
curves can barely be distinguished. The errors due to the
fitting are shown in Fig.~\ref{fig:five} for $C_{f1} (Q{-}E)$ and
Fig.~\ref{fig:six} for $C_{f2} (Q{-}E)$ respectively.

\goodbreak

\goodbreak
\goodbreak

\section{Final atomic states with non-zero angular momentum}
So far, we have considered only the final ion state $\langle f |$
being in an S-state.
To obtain the inclusive decay rate, we also need to consider the
case where the final ion states are not S-states.
For these cases, the first term $T_1$, the amplitude of the sudden
approximation, vanishes.
Therefore, to order $\eta^2$, as far as the decay rate is concerned,
it is sufficient to calculate the amplitude to order $\eta$,
which comes from the leading term of $T_{\rm d}$.
Recalling expression~(\ref{dirlead}) for this leading term $T_{\rm d}^0$,
we have, to order $\eta$,
\begin{equation}
    T_{\rm d}^0 \simeq
	{-}\langle {\bf p} | \! \int (d^3 {\bf r}_2)
	\phi_f^* ({\bf r}_2)\phi_i ({\bf r}_2) \!
        \left [ {e^2 \over 4 \pi |{\bf r}_1{-}{\bf r}_2|}
	{-}{2 e^2 \over 4 \pi r_1}{-}v(r_1) \right ] \!
	{1 \over H_1{+}E_i{-}E{-}E_f{-}i \epsilon}
	|{\bf r}_1{=}{\bf 0} \rangle \,.
\end{equation}
The last two terms in the square brackets vanish upon
integrating over the solid angle of ${\bf r}_2$ since now
$\phi_f ({\bf r}_2)$ contains only higher partial waves with $l \ge 1$.
Hence
\begin{equation}
    T_{\rm d}^0 \simeq  -  \int (d^3 {\bf r}_2)
        \phi_f^* ({\bf r}_2)\phi_i ({\bf r}_2) \langle {\bf p} |
	{e^2 \over 4\pi |{\bf r}_1 - {\bf r}_2|}
	{1 \over H_0 - E - i \epsilon} |{\bf 0} \rangle \,,
\label{otherlead}
\end{equation}
where we have neglected $E_i$, $E_f$ compared with $E$ and replaced
$H_1$ by the free Hamiltonian $H_0$.
Suppose that final state $ \langle f|$ has the angular momentum
quantum number $(l, m)$, or equivalently
\begin{equation}
    \phi_f^* ({\bf r}) = R_{fl}^* (r) Y_{lm}^* ({\hat {\bf r}}) \,,
\end{equation}
with $R_{fl} (r)$ the radial wave function.
In Appendix B, we find that, to the leading order,
\begin{equation}
    T_{\rm d}^0 \simeq i {4 \pi \over l (l+1)} \eta
	 Y_{lm}^* ({\hat {\bf p}})
	\int_0^{\infty} d r_2 r_2^2 R_{fl}^* (r_2) \phi_i (r_2) \,.
\label{lresult1}
\end{equation}

The differential decay rate involves the angular average
of the square of
the spherical harmonic function which appears in Eq.~(\ref{lresult1}),
\begin{equation}
    \int {d \Omega_{\bf p} \over 4 \pi}
	\left | Y_{lm} ({\hat {\bf p}}) \right |^2
	= {1 \over 4 \pi} \,,
\end{equation}
and thus, in view of Eq.~(\ref{diffrate}), the differential decay
rate to states with specific energy and
angular momentum $l$ is given by
\begin{equation}
    {d \Gamma_{fl} \over d E} = {m \over 2 \pi^3} p \,
        |T_{\beta}|^2 (Q_f - E)^2 \,
	\eta^2 \, {4 \pi (2l + 1) \over l^2 (l+1)^2}
	\left |\int_0^{\infty} d r_2 r_2^2 R_{fl}^* (r_2) \phi_i (r_2)
	\right |^2 \,,
\label{3.9}
\end{equation}
where we have done the summation over the magnetic quantum number $m$
which generates the factor $2l+1$.

Summing over $l$ and $f$ produces the differential decay rate for
the final $^3{\rm He}^+$ ion having nonzero angular momentum.
Using the expansion~(\ref{squareexp}) we encounter the
sums\footnote{Here the sum over $f$ should still be understood as the
sum with the upper bound $f_{\rm max}$ as before.}
\begin{equation}
    M_n = \sum_{fl} {4 \pi (2l + 1) \over l^2 (l+1)^2}
	\left |\int_0^{\infty} d r \, r^2 R_{fl}^* (r) \phi_i (r)
	\right |^2 (E_i - E_f)^n \,,
\end{equation}
with $n=0,1,2$.
To facilitate the calculation, we define
\begin{equation}
    M (k) = \left (E_i - E_k \right )^2
	\sum_{l=1}^{\infty} {4 \pi (2l + 1) \over l^2 (l+1)^2}
	\left |\int_0^{\infty} d r \, r^2 R_{kl}^* (r)
	\phi_i (r) \right |^2 \,,
\label{Mdef}
\end{equation}
where $R_{kl} (r)$ is the radial wave function of the final unbound
ion state with energy $E_k = k^2/(2m)$ and angular momentum $l$.
$M(k)$ is calculated in Appendix C.
We can consequently calculate $M_n$ for $n=0,1,2$.

For $n=0$, exploiting the completeness relation
\begin{equation}
    \sum_f R_{fl}^* (r^{\prime}) \, R_{fl} (r) ={1 \over r^2} \,
	\delta (r^{\prime} - r) \,,
\label{comp}
\end{equation}
we can write
\begin{eqnarray}
    M_0 &=& \sum_l {2 l{+}1 \over l^2 (l{+}1)^2}
	\int_0^{\infty}  4 \pi r^2 dr |\phi_i (r)|^2
	- \sum_l {4 \pi (2l{+}1) \over l^2 (l{+}1)^2}
	\int_K^{\infty}  dk {2k^2 \over \pi}
	\left | \int_0^{\infty} dr \, r^2
	R_{kl} (r) \phi_i (r) \right |^2
\nonumber\\
	&=&  1 - \int_K^{\infty} dk {2 k^2 \over \pi}
	M (k) \left (E_i - E_k \right )^{-2} \,,
\label{S0exp}
\end{eqnarray}
where the sum
\begin{equation}
    \sum_{l=1}^{\infty} {2l+1 \over l^2 (l+1)^2}
	= \sum_{l=1}^{\infty} \left [{1 \over l^2}
	- {1 \over (l+1)^2} \right ] = 1
\label{cancelsum}
\end{equation}
and the unit norm of the initial wave function $\phi_i (r)$ have been used.
This expression enables a numerical evaluation of $M_0$.

\begin {figure}[htbp]
        \epsfysize 5in
        \leavevmode {\hfill \epsfbox {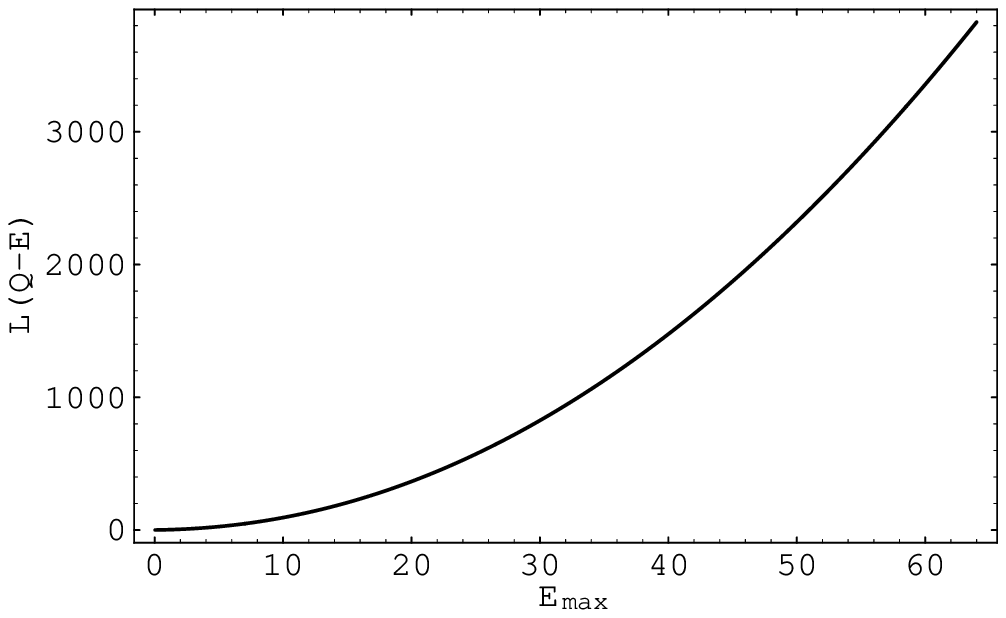} \hfill}
    \caption
        {%
        \advance\baselineskip by -8pt
        Numerical curve for $L (Q{-}E)$ ---  the correction to the spectrum
	due to atomic effect for non-S-wave final ionic states --- as a
	function of
	$\Emax = Q + E_i - E$; vertical axis has the unit ${\rm R\!y}^2$
	and horizontal axis has the unit ${\rm R\!y}$.
        }%
    \label {fig:seven}
\end {figure}

\begin {figure}[htbp]
        \epsfysize 5in
        \leavevmode {\hfill \epsfbox {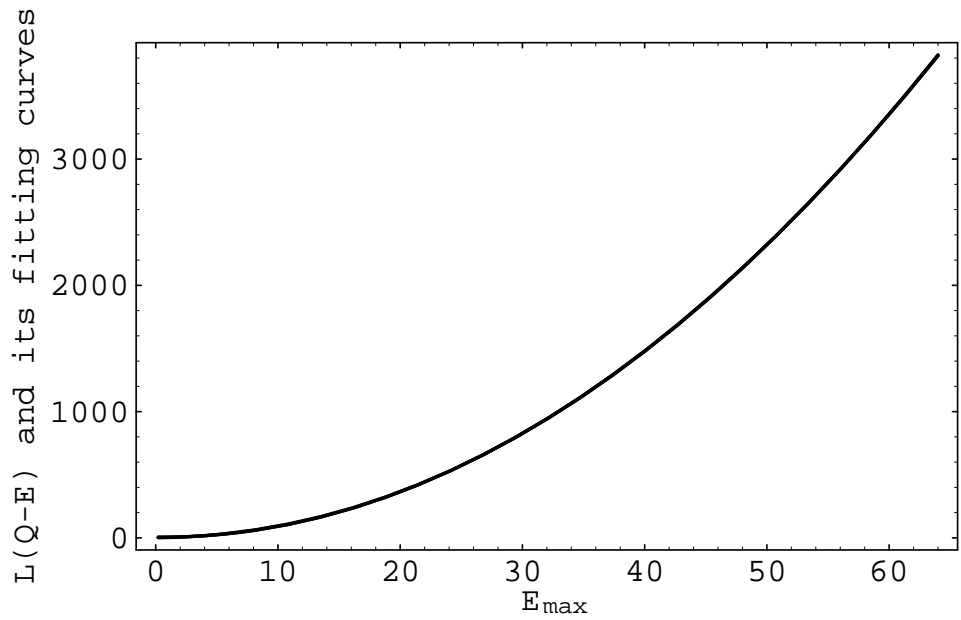} \hfill}
    \caption
        {%
        \advance\baselineskip by -8pt
        Numerical curve for $L (Q{-}E)$ and its two fitting formula
	$L_{f1} (Q{-}E)$ and $L_{f2} (Q{-}E)$ as functions of $\Emax
	=Q+E_i-E$; vertical axis has the unit ${\rm R\!y}^2 a_0^3$
	and horizontal axis has the unit ${\rm R\!y}$.
        }%
    \label {fig:eight}
\end {figure}

\begin {figure}[htbp]
        \epsfysize 5in
        \leavevmode {\hfill \epsfbox {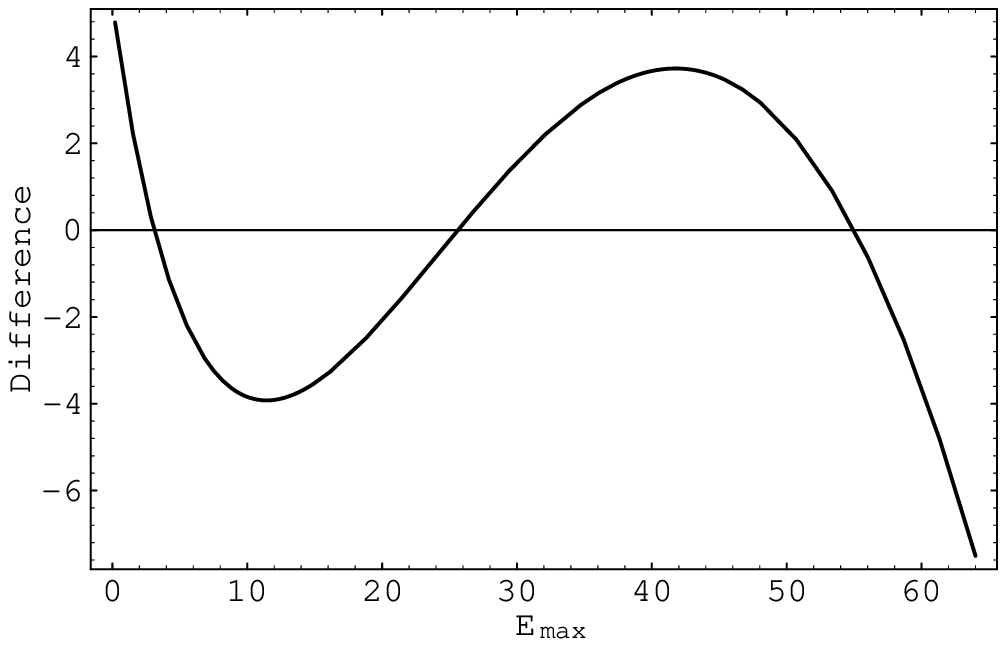} \hfill}
    \caption
        {%
        \advance\baselineskip by -8pt
        Numerical curve
        for the difference between $L_{f1} (Q{-}E)-L (Q{-}E)$
	as a function of $\Emax=Q+E_i-E$,
	vertical axis has the unit ${\rm R\!y}^2$
	and horizontal axis has the unit ${\rm R\!y}$.
        }%
    \label {fig:nine}
\end {figure}

\begin {figure}[htbp]
        \epsfysize 5in
        \leavevmode {\hfill \epsfbox {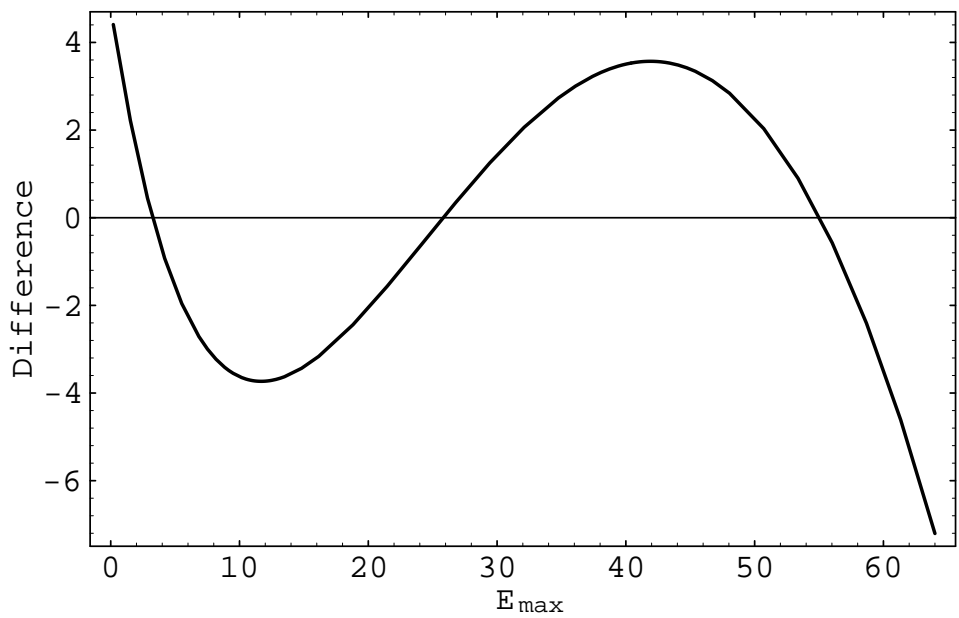} \hfill}
    \caption
        {%
        \advance\baselineskip by -8pt
        Numerical curve
        for the difference between $L_{f2} (Q{-}E)-L (Q{-}E)$
	as a function of $\Emax=Q+E_i-E$,
	vertical axis has the unit ${\rm R\!y}^2$
	and horizontal axis has the unit ${\rm R\!y}$.
        }%
    \label {fig:ten}
\end {figure}

To calculate $M_1$ and $M_2$, we write
$M_n$ as the sum of two parts --- $M_{n{\rm b}}$
the part coming from summing over the final bound ionic states and
$M_{n{\rm c}}$ the part coming from summing over the final continuous
ionic states,
\begin{equation}
    M_n = M_{n{\rm b}} + M_{n{\rm c}} \,.
\end{equation}
Following the definition of $M_{n{\rm c}}$, we have
\begin{equation}
    M_{n{\rm c}} = \int_0^K d k {2 k^2 \over \pi} M (k)
		(E_i - E_k)^{n-2} \,,
\label{Sncexp}
\end{equation}
which may be easily calculated numerically.
For $M_{n{\rm b}}$, we shall not investigate it numerically;
instead, we estimate it to a good accuracy.
To facilitate the notation, we define $L (Q{-}E)$ by
\begin{equation}
    L (Q{-}E) \equiv M_0 \,\, (Q - E)^2 + 2 M_{\rm 1c} \,\,
	(Q - E) + M_{\rm 2c} \,,
\label{LK_def}
\end{equation}
which enables us to write the differential inclusive decay rate
for the final ion state having nonzero angular momentum as
\begin{equation}
    {d \Gamma^{\prime} \over d E} \equiv
        \sum_{fl} {d \Gamma_{fl} \over d E} =
	{m \over 2 \pi^3} \, p \, |T_{\beta}|^2 \, \eta^2 \,
	\left [ L (Q{-}E){+}2 M_{\rm 1b} (Q{-}E){+}M_{\rm 2b} \right ] \,.
\label{Nonzero}
\end{equation}
Upon using the numerical result of $M (k)$ obtained in Appendix C,
one may obtain $M_0$, $M_{\rm 1c}$, and $M_{\rm 2c}$ by
numerically performing the integrals over $k$ in
expressions~(\ref{S0exp}) and (\ref{Sncexp}).
Definition~(\ref{LK_def}) thus gives a numerical result for
$L (Q{-}E)$ which we display
as a function of $\Emax = K^2/(2m)$ in Fig.~\ref{fig:seven}.
We find also a quadratic function $L_{f1} (Q{-}E)$
\begin{eqnarray}
    L_{f1} (Q{-}E) &=& 5.90 \, {\rm R\!y}^2 -
	1.03 \, {\rm R\!y} \, \Emax +
	0.947 \, \Emax^2
\nonumber\\
	&=& 7.88 \, {\rm R\!y}^2 - 2.93 \, {\rm R\!y} \, (Q - E) -
	0.947 \, (Q - E)^2
\label{Lfitting}
\end{eqnarray}
which fits $L (Q{-}E)$ for $\Emax$ in the range
$0-64 \, {\rm R\!y}$ ($0 - 870$ eV)
with a fairly good accuracy. Like in previous section, we fit
$L (Q{-}E)$ to a linear combination of functions $P (Q{-}E)$,
$P'(Q{-}E)$, and $P_1 (Q{-}E)$ to generate the fitting formula
\begin{equation}
    L (Q{-}E) \approx L_{f2} (Q{-}E) \equiv 0.948 P (Q{-}E)
	- 3.39 {\rm R\!y} P' (Q{-}E) + 15.7 {\rm R\!y}^2 P_1 (Q{-}E) \,.
\label{Lfitting2}
\end{equation}
We show $L (Q{-}E)$ and its two fitting functions $L_{f1} (Q{-}E)$ and
$L_{f2} (Q{-}E)$ together in Fig.~\ref{fig:eight}. They can hardly be
distinguished. The errors caused by the fitting are shown
in Fig.~\ref{fig:nine} for $L_{f1} (Q{-}E)$ and Fig.~\ref{fig:ten} for
$L_{f2} (Q{-}E)$ respectively.

We now turn to provide numerical bounds for $M_{\rm 1b}$ and $M_{\rm 2b}$.
The energy of an ionic state with principal quantum number $n_f$ is
given by
\begin{equation}
    E_f = {4 \over n_f^2} E_i = - {4 \over n_f^2} {\rm R\!y} \,.
\end{equation}
Since the lowest energy state with nonzero angular momentum ($n_f=2$) has
the energy $E_i$, it does not contribute to the summation defining
$M_{n{\rm b}}$ for $n=1,2$ due to the factor $(E_i - E_f)^n$.
For the other states, since $n_f > 2$, $E_f \ge 4 E_i/9$, and hence
\begin{equation}
    (E_f - E_i)^n \ge \left ({5 \over 9} \, {\rm R\!y} \right )^n \,.
\end{equation}
Therefore, according the definition of $M_{n{\rm b}}$, we have
\begin{eqnarray}
    - M'_{\rm 0b} \,\, {5 \over 9} \,\, {\rm R\!y}
	> &M_{\rm 1b}& > - M'_{\rm 0b} \,\, {\rm R\!y} \,,
\nonumber\\
    M'_{\rm 0b} \,\, {25 \over 81} \,\, {\rm R\!y}^2
	< &M_{\rm 2b}& < M'_{\rm 0b} \,\, {\rm R\!y}^2 \,,
\end{eqnarray}
where $M'_{\rm 0b}$ is defined by excluding the lowest energy state
($n_f{=}2,l{=}1$)
in the $(n_f, l)$ summation of the definition of $M_{\rm 0b}$, {\it i.e.},
\begin{equation}
    M'_{\rm 0b} \equiv M_{\rm 0b} - {3 \over 4} \, 4 \pi \left |
        \int_0^{\infty} dr r^2 R_{21} (r) \phi_i (r) \right |^2 \,.
\end{equation}
Evaluating the matrix element,
\begin{equation}
    4 \pi \left |
	\int_0^{\infty} dr r^2 R_{21} (r) \phi_i (r) \right |^2
	= {3 \over 4} \,,
\end{equation}
and using the numerical result
\begin{equation}
    M_{\rm 0b} = 0.659 \,,
\end{equation}
gives the explicit bounds:
\begin{equation}
    - 0.05 \, {\rm R\!y} > M_{\rm 1b} > - 0.1 \, {\rm R\!y} \,,
	\qquad 0.03 \, {\rm R\!y}^2 < M_{\rm 2b} < 0.1 \, {\rm R\!y}^2 \,.
\label{bounds}
\end{equation}

\section{Inclusive Decay Rate}

Adding the differential decay rates~(\ref{Sresult})
and~(\ref{Nonzero}), we find that
the inclusive differential decay rate is given by
\begin{eqnarray}
    {d \Gamma_{\rm in} \over d E}
	&=&{d \Gamma_0 \over d E}
	+ {d \Gamma^{\prime} \over d E}
\nonumber\\
	&\simeq&{m \over 2 \pi^3}
	F(2,E) \, p \, | T_{\beta} |^2 (1{-}2\eta^2)
	\left [ P(Q-E) + \eta^2 R (Q{-}E)] \right ] \,,
\label{RawRESULT}
\end{eqnarray}
where we have defined the correction term $ R (Q{-}E) $ by
\begin{equation}
    R (Q{-}E) \equiv - {C (Q{-}E) \over 2} + L (Q{-}E)
	+ 2 M_{\rm 1b} (Q - E) + M_{\rm 2b} \,,
\label{R_def}
\end{equation}
and dropped higher order terms.
Here $C (Q{-}E)$ and $L (Q{-}E)$ are evaluated numerically;
$M_{\rm 1b}$ and $M_{\rm 2b}$ are very small contributions
bounded by Eq.~(\ref{bounds}).
The term containing $ R (Q{-}E)$ in the second line of Eq.~(\ref{RawRESULT})
represents the correction to the sudden approximation
result due to the atomic effect.

\subsection{Comparing with previous results}
We now compare our result~(\ref{RawRESULT}) with previous results
\cite{Durand,Strikman}.
To reproduce the previous results, two approximations must be made,
as mentioned in Section I.
Under the ``uniform phase space factor approximation'', we have
\begin{eqnarray}
    && C (Q{-}E) \sim (Q - {\bar E})^2 K_0'' (0) \,,
\nonumber\\
    && L (Q{-}E) \sim (Q - {\bar E})^2 M_0 \,,
\end{eqnarray}
where ${\bar E}$ is some average beta ray energy
which differs from the beta ray energy $E$ by an amount
of order ${\rm R\!y}$.%
\footnote{For example, in reference\protect\cite{Durand}, ${\bar E}$ is
chosen to be $E-3\,{\rm R\!y}$. The details of the choice of ${\bar E}$
does not matter, since there the order $\eta^2 {\rm R\!y}/(Q-E)$ terms
are omitted.}
Under the closure approximation,
Eq.~(\ref{K_0result}) and Eq.~(\ref{S0exp}) read $K_0'' (0) \simeq 0$ and
$M_0 \simeq 1$ respectively.
With these approximations, the definition~(\ref{R_def}) gives
$ R (Q{-}E) \sim (Q-{\bar E})^2 $, which implies that
the result~(\ref{RawRESULT}) above reduces to a modified sudden
approximation result:
\begin{equation}
    {d \Gamma_{\rm in} \over d E} =
        {m \over 2 \pi^3} F(2,E)
        {p'} \, | T_{\beta} |^2
        P(Q-E) \,,
\label{ComResult}
\end{equation}
where terms of order $\eta^2 S (Q{-}E)$ and $\eta^2 {\rm R\!y}^2$ have been
discarded with the approximations we are considering.
The result~(\ref{ComResult}) agrees with previous results
\cite{Durand,Strikman},
with the modified momentum ${ p'}$ being defined by
\begin{equation}
    {{ p'}^2 \over 2m } + 2 \, {\rm R\!y} = { p^2 \over 2m } \,.
\end{equation}
The momentum ${ p'}$ equals
the momentum of an emitted beta electron
whose energy at short distances is
modified by the repulsive Coulomb interaction energy with the original
electron bound in the tritium atom.
This modification which changes $p$ to ${p'}$ was found by Rose
a long time ago \cite{Rose}.

\subsection{Estimation of the atomic effects in the neutrino mass
determination}

To estimate how the atomic effects change the neutrino mass squared
parameter, we first recall that the spectrum used in the experimental
data analysis is the sudden approximation result
\cite{Robertson,Holzschuh,Kawakami,Stoeffl,Backe,Knapp}
\begin{equation}
    {d \Gamma_{\rm exp} \over dE}
	= A \,\, F (2, E) \,\, p \sum_{f<f_{\rm max}}
	W_{fi} (Q_f - E) \,\sqrt{ (Q_f - E)^2 - m_\nu^2} \,.
\label{sudden_molecule}
\end{equation}
The terms in the right hand side of Eq.~(\ref{sudden_molecule}) require
some explanation.
The first factor $F (2, E)$ is the usual Fermi function\footnote{Usually a
relativistic Fermi function is used to take care of the dominant
relativistic correction.} with the nucleus charge $Z=2$, and
$ p $ is the momentum of the beta ray. $W_{fi}$ is the transition
probability for the initial tritium state decaying to the state
$f$ of the final $^3{\rm He}^+$ ion.
In real experiments, molecular tritium is used. Therefore, $W_{fi}$ is
the square of the matrix element for the molecular state.
Since only atomic tritium is considered in this article,
we have made the replacement $W_{fi} = |T_{fi}|^2$.
Finally, $m_\nu^2$ is the neutrino mass squared.
Though one focuses on measuring the beta ray spectrum near the
end point to probe the neutrino mass, most of the data obtained
in the experiments are in the range where $ (E - Q)^2 \gg m_\nu^2 $
since the decay rate is tiny at the very end of the spectrum.
Expanding the square root, we get
\begin{eqnarray}
    {d \, \Gamma_{\rm exp} \over d E} &\simeq& A \,\,
	F (2, E) \,\, p \sum_{f<f_{\rm max}} |T_{fi}|^2
        \left[ (Q_f - E)^2 - {1 \over 2} m_\nu^2 \right ]
\nonumber\\
        &=& A \,\, F (2, E) \,\, p \,
	\left [P (Q{-}E) - {1 \over 2} m_\nu^2
        P_1 (Q{-}E) \right ] \,,
\label{RightForm}
\end{eqnarray}
where we have used the definitions~(\ref{P_def}) and (\ref{P_1_def})
for the functions $P (Q{-}E)$ and $P_1 (Q{-}E)$ which are evaluated
numerically in section IV.

In real experimental data analysis,
the parameters $A$, $Q$, and $m_\nu^2$
are determined by comparing the spectrum~(\ref{RightForm})
with the measured spectrum. Therefore, any theoretical correction to the
spectrum~(\ref{RightForm}) has the effect of changing the parameters
$A \to {\bar A} = A+\Delta A$,
$Q \to {\bar Q} = Q+\Delta Q$, and
$m_\nu^2 \to  \Delta m_\nu^2$ so that
the theoretical correction to the spectrum may be included by using
an effective spectrum described by the same form~(\ref{RightForm}) but with
the effective parameters ${\bar A}$, ${\bar Q}$, and $\Delta m_\nu^2$.
The change of the spectrum~(\ref{RightForm})
\begin{equation}
    \Delta \left ({d \Gamma_{\rm exp} \over dE} \right )
	\approx A \,\, F(2, E) \,\, p \left [
	{\Delta A \over A} \, P (Q{-}E)
	+ \Delta Q P'(Q{-}E)
	- {1 \over 2} \Delta m_\nu^2 P_1 (Q{-}E) \right ]
\label{mimic2}
\end{equation}
mimics the corresponding theoretical correction to the spectrum.

We now estimate $\Delta Q$, $\Delta m_\nu^2$ corresponding to the correction
due to the atomic effect which accounts for the interaction between
the beta ray and the electron of the $^3{\rm He}^+$ ion.
This we shall do by requiring the correction $R (Q{-}E)$ in
Eq.~(\ref{RawRESULT}) be mimicked by a linear combination of
$P (Q{-}E)$, $P ' (Q{-}E)$, and $P_1 (Q{-}E)$ as appearing
in the right hand side of Eq.~(\ref{mimic2}).
We can neglect the small parameters $M_{\rm 1b}$ and $M_{\rm 2b}$ since
they are bound by Eq.~(\ref{bounds}).
Since the region important for the neutrino mass measurement goes from the
the beta ray end point to approximately
$59-74 \, {\rm R\!y}$ ($800-1000\, {\rm eV}$)
below end point\cite{Robertson,Holzschuh,Kawakami,Stoeffl,Backe},
the previous fitting formula $C_{f2} (Q{-}E)$ and $L_{f2} (Q{-}E)$ for the
energy
$\Emax$ range $0-64 {\rm R\!y}$ may be used here.
We shall discuss the sensitivity to the range of the energy of the result
later.
Replacing $C (Q{-}E)$ and $L (Q{-}E)$ with their fitting
formulas~(\ref{Cfitting2}) and
(\ref{Lfitting2}), the definition~(\ref{R_def}) reads
\begin{equation}
    R (Q{-}E) \approx 0.94 \, P (Q{-}E) - 4.6 \, {\rm R\!y} \, P ' (Q{-}E)
	+ 6.3 \, {\rm R\!y}^2 \, P_1 (Q{-}E) \,.
\label{R_fit}
\end{equation}
Inserting this fitting form of $R (Q{-}E)$ into Eq.~(\ref{RawRESULT}) yields
\begin{eqnarray}
    {d \Gamma_{\rm in} \over d E}
	&\approx&
	{m \over 2 \pi^3} F(2,E)
	p \, | T_{\beta} |^2 (1{-}1.06 \, \eta^2)
\nonumber\\
	&& \qquad \times \left [P (Q{-}E)
	- 4.6 \, \eta^2 \, {\rm R\!y} \, P ' (Q{-}E)
	+ 6.3 \, \eta^2 \, {\rm R\!y}^2 \, P_1 (Q{-}E) \right ] \,,
\label{RESULT}
\end{eqnarray}
where we have shifted the argument of $P$ to absorb the second term
which causes an negligible error of order $O (\eta^4)$.
In view of the argument above, specifically Eq.~(\ref{mimic2}),
Eq.~(\ref{RESULT}) shows that the atomic effect
gives a correction to neutrino mass squared of
\begin{equation}
    \Delta m_\nu^2 \approx - 12.6 \,\eta^2 {\rm R\!y}^2
	\simeq - 1.7 \, {\rm eV}^2 \,,
\label{mass2result}
\end{equation}
and the endpoint changes by
\begin{equation}
    \Delta Q \approx - 4.6 \, \eta^2 {\rm R\!y}
	\simeq - 0.047 \, {\rm eV} \,.
\label{endpointresult}
\end{equation}
We now examine the sensitivity of the result
to the energy range $\Emax$ used in the fit.
Since $\Delta Q$ is tiny, we shall only consider how $\Delta m_\nu^2$
depends on the range in which we do the fit.
We fit $R (Q{-}E)$ in various energy $\Emax$ ranges and display
the corresponding $m_\nu^2$ in Table~\ref{sensitivity} shown in
the introduction.
The neutrino mass squared has basically a linear dependence on the
energy range where we do the fit. Increasing the energy range by
each $5 \, {\rm R\!y}$ causes $\Delta m_\nu^2$ to decrease by
$-0.4 \, {\rm eV}^2$.
The atomic effect changes the neutrino mass squared parameter on the order
of a few ${\rm eV}^2$. It is not a big effect.

\section{Conclusion}
We have developed a systematic expansion for the tritium
beta decay amplitude in the Coulomb parameter $\eta$. By choosing
a convenient comparison potential, one can avoid the infrared divergences
due to the long range characteristic of the Coulomb force.
Both the exclusive and the inclusive decaying rates are calculated to
order $\eta^2$. The inclusive decay rate agrees with previous results.
The estimation on how this order $\eta^2$ correction affects
the neutrino mass squared parameter is provided. We find that the effect
is small and does not suffice to explain the mysterious negative
electron anti-neutrino mass squared obtained in modern experimental
data analysis\cite{Wilkerson,Robertson,Holzschuh,Kawakami,Stoeffl,Backe}.
We also remark that the order $\eta^2$ correction
to the spectrum is small and can not provide any explanation for
the anomalous structure in the beta decay spectrum in the last
55 eV closest to the end point presented in the last article of
reference\cite{Stoeffl}.

\acknowledgments
We need to thank several people. D.G. Boulware collaborated in our
initial formulation. L. Durand, III provided encouragement and advise.
Discussions with P. B. Arnold were helpful.
The work was supported, in part,
by the U. S. Department of Energy under grant DE-AS06-88ER40423.

\newpage
\appendix
\section{Wave function calculation}
We shall find the value of the wave function $\phip$
at the origin up to and including terms of order $\eta^3$.
Since $\phi_{\bf p}^* (0)$ involves only the S-wave component,
we may replace
\begin{equation}
    \phi_{\bf p} ({\bf r}) \to {1 \over pr} u_p (r) \,,
\label{A1}
\end{equation}
where $u_p (r)$ obeys the S-wave radial Schr\"odinger equation
\begin{equation}
    \left \{ - {1 \over 2m} {d^2 \over dr^2} + v(r) - E \right \}
	u_p (r) = 0 \,.
\label{Schrdger}
\end{equation}
The radial wave function $u_p (r)$ vanishes at the origin and obeys
the asymptotic boundary condition
\begin{equation}
    r \to \infty : \quad u_p (r) \sim {1 \over 2i}
	\left [e^{ipr + i\eta \ln (2pr)}
	- e^{-ipr - i \eta \ln (2pr) - 2i \delta_0} \right ] \,,
\label{inftyasy}
\end{equation}
where $\delta_0$ is the S-wave phase shift.
This boundary condition contains a non-trivial phase structure
because, at large distances, the particle moves in the long-range
Coulomb field of unit charge.

It is convenient to divide space into two regions: $0 < r < r_c$ and
$r_c < r < \infty$, with $r_c$ being an intermediate distance between
the Bohr radius $a_0$ and the de Broglie wave length $1/p = \eta a_0$.
We take $r_c$ to be of the order
\begin{equation}
    \left ({r_c / a_0} \right )^2 \sim O(\eta^{1 + \zeta}) \,,
\label{r_cdef}
\end{equation}
with $0 < \zeta < 1$ ({\it e.~g.} $r_c = \eta^{2/3} a_0$).
As we shall see, there are two reasons for this choice.
One reason is that in the region where $0 < r < r_c$,
we have $r \ll a_0$ which enables us to expand
the expression (\ref{effpot}) for $v(r)$ as
\begin{equation}
    v(r) = - {2 e^2 \over 4 \pi r} +
	V_{fi} + O (e^2 \eta^3 p^3 r^2) \,,
\label{shortrange}
\end{equation}
where
\begin{equation}
    V_{fi} = \lim_{r \to 0}
	\left [v (r) + {2 e^2 \over 4 \pi r } \right ]
	= \int (d^3 {\bf r}) \rho_{fi} (r) {e^2 \over 4 \pi r} \,.
\end{equation}
The expansion~(\ref{shortrange}) of $\Delta v(r)$ does not contain
a term linear in $r$. Such a term would correspond to a charge
distribution $- \nabla^2 r \sim 1/r$ which
has an unphysical singularity at the origin.
The correction to the first two terms in the expansion~(\ref{shortrange})
comes from the consideration that, for any reasonable comparison potential,
the characteristic length scale for $\Delta v(r)$ to vary is the
Bohr radius $a_0$, and so the leading correction is of the order
\begin{equation}
    {e^2 \over r} \left ({r \over a_0} \right )^3 \sim
	e^2 \eta^3 p^3 r^2 \,.
\end{equation}
Inserting the expansion~(\ref{shortrange}) of the potential
into the radial Schr\"odinger equation~(\ref{Schrdger}),
we see that the ``interior'' solution $u_p (r)$ in the region
$0 < r < r_c$ is a constant $C$ times the
Coulomb S-state radial wave function with charge $2$ and energy
$E_{p^{\prime}} = E - V_{fi}$.
This radial wave function involves the shifted momentum
\begin{equation}
    p^{\prime} = \sqrt{p^2 - 2 m V_{fi}}
\end{equation}
and the correspondingly altered Coulomb parameter
\begin{equation}
    2 \eta^{\prime} = 2 \alpha m / p^{\prime} \,.
\end{equation}
In the region $r < r_c$, the higher order terms in the
expansion~(\ref{shortrange}) give rise to corrections,
compared with the energy $E$ of the beta ray, of order
\begin{equation}
    {e^2 \eta^3 p^3 r^2 \over E} \sim \eta^4 (p r)^2
	\le \eta^4 (p r_c)^2 \sim o (\eta^3) \,,
\end{equation}
for our choice~(\ref{r_cdef}) of $r_c$.
Therefore, this Coulomb wave function with the shifted momentum
$p^{\prime}$ obeys the Schr\"odinger equation~(\ref{Schrdger})
with an error which is less than order $\eta^3$.
As is well known~\cite{Gottfried}, this wave function gives the
limits,
\begin{equation}
    r \to 0: \quad u_p (r) \to C p^{\prime} r
	e^{\pi \eta^{\prime}} \Gamma ( 1 + 2 i \eta^{\prime} ) \,,
\label{0val}
\end{equation}
and
\begin{equation}
    r \to \infty : \quad u_p (r) \sim C
	\left [ 1 + {4 \eta^{\prime} \over p^{\prime} r}
	 \right ]^{-1/4} {1 \over 2i}
	{\Biggl \{}e^{ ip^{\prime} r + 2 i \eta^{\prime}
	\ln p^{ \prime} r} + \cdots {\Biggr \}} \,,
\label{shortasy}
\end{equation}
where the ellipsis $\cdots$ stands for the incoming wave contribution
which involves $\exp \{-i (p^{\prime} r + 2 \eta^{\prime}
\ln 2 p^{\prime} r) \}$.
Note that, for $r \sim r_c$, the
asymptotic expansion~(\ref{shortasy}) is valid
because $p r \gg 1$ in this region. This is the other reason for
the choice~(\ref{r_cdef}) of $r_c$.

In the region where $r > r_c$, we can get the asymptotic
form of the wave function $u_p (r)$ by iterating Eq.~(\ref{Schrdger})
starting with the limiting behavior~(\ref{inftyasy})
of $u_p (r)$ for $r \to \infty$.
Matching this ``exterior'' solution with the previous ``interior''
solution at $r \sim r_c$ determines the constant $C$.
The two linearly independent solutions at large radius $r$ correspond
to outgoing and incoming waves,
\begin{equation}
    u_p (r) = u_p^{(+)} (r) + u_p^{(-)} (r) \,.
\end{equation}
Since we have only one constant $C$ to be determined,
it is sufficient to match the outgoing wave part $u_p ^{(+)} (r)$ to
connect the solutions in the two regions.
We write this part of the exterior solution as the W.K.B. approximate
solution times an arbitrary function,
\begin{equation}
    u_p^{(+)} (r) = {1 \over 2i}
	\left [{E \over E - v(r)} \right ]
	^{1 /4} e^{i S (r)} \, w_p^{(+)} (r) \,.
\label{WKB}
\end{equation}
Here $S(r)$ is defined by
\begin{equation}
    {d S (r) \over dr} = \sqrt{2m [E - v (r)]} \,,
\label{Sdef}
\end{equation}
with the boundary condition that
\begin{equation}
    S(r) \to pr + \eta \ln 2pr \,, \quad {\rm as} \, r \to \infty \,,
\label{Slimit}
\end{equation}
which is consistent with the long-range limit of $v(r)$
in Eq.~(\ref{Sdef}).
Thus, the boundary condition~(\ref{inftyasy}) is obeyed by requiring
that $w_p^{(+)} (r) \to 1$ as $r \to \infty$.
Iterating the solution~(\ref{WKB}) in the Schr\"odinger
equation~(\ref{Schrdger})
yields the asymptotic expansion of $w_p ^{(+)} (r)$.
For example, the first iteration gives
\begin{equation}
    w_p^{(+)} (r) \simeq 1 - {im \over 4 p^3} {dv(r) \over dr} \,.
\label{itecor}
\end{equation}
When $r \sim r_c$, which is in the asymptotic region, the form~(\ref{WKB})
for $u_p^{(+)} (r)$ with $w_p^{(+)} (r) = 1$ is appropriate for our discussion.
In the region $r \sim r_c$,
the first factor containing the power $1/4$ in the
solution~(\ref{WKB}), by using the expansion~(\ref{shortrange}),
can be written as
\begin{equation}
    \left [{E \over E - v (r)} \right ]^{1/4}
	\simeq \sqrt{p \over p^{\prime}} \left [
	1 + {4 \eta^{\prime} \over p^{\prime} r}
	\right ]^{-1/4} \,,
\end{equation}
which matches the corresponding factor in the asymptotic
expansion~(\ref{shortasy}) of the interior solution.
The phase $S(r)$, in this region, to order $\eta^3$, is
given by
\begin{equation}
    S (r) \approx p^{\prime} r + 2 \eta^{\prime}
	\ln 2 p^{\prime} r + \Theta \,,
\label{longasy}
\end{equation}
where $\Theta$ is a constant phase.
The fact that $\Theta$ is a constant may be justified
by observing that
\begin{eqnarray}
    {d \over dr} \left [ S (r) - p^{\prime} r - 2 \eta^{\prime}
	\ln 2 p^{\prime} r \right ] \tab=\tab
	\sqrt{2m (E - v(r))} - {m^2 v(r)^2 \over 2 p^3}
	 - p^{\prime} - {2i \eta^{\prime} \over r}
\nonumber\\
 	\tab\approx\tab  p - {m v (r) \over p}
	- {m^2 v(r)^2 \over 2 p^3}
	 - p^{\prime} - {2i \eta^{\prime} \over r}
	= o ({\eta^3 \over r_c}) \,,
\end{eqnarray}
where we have used the definition of $S (r)$, expanded the
square root in powers of $v (r)$, and replaced $v(r)$ by its
expansion~(\ref{shortrange}).\footnote{The phase $\Theta$ may be
explicitly determined to $O(\eta^3)$
by integrating Eq.~(\ref{Sdef}) inward from the limit~(\ref{Slimit}).}
Therefore, in the region where $r \sim r_c$, the leading terms in the
solutions~(\ref{WKB})~and~(\ref{shortasy}) match to the order $\eta^3$.
We did not include,
in the asymptotic expansions~(\ref{shortasy}) and (\ref{WKB}),
sub-leading terms in $1/r$ such as $\eta/(pr)^2$ and $\eta / (pr)^3$
which are of the order $O(\eta^2)$ and $O (\eta^3)$ in the matching
region.
These terms must also match to the order $\eta^3$ because we are
solving the same differential equation by using different approaches
with both having an accuracy of order $\eta^3$.\footnote{
This remark shows that we have presented more detail of how the matching
works than is actually necessary. This we did for
the sake of clarity.}
For example, the first correction in Eq.~(\ref{itecor}) shows
the matching of terms of order $\eta/(pr)^2$, if we recall the leading
behavior of $v(r)$ at the region $r \sim r_c$ and include the term
of order $\eta/(pr)^2$ in the asymptotic behavior~(\ref{shortasy}).

Requiring that the outgoing wave in the asymptotic
form~(\ref{shortasy}) to be the same as that of $u_p^{(+)} (r)$ in
Eq.~(\ref{WKB}) thus gives
\begin{equation}
    C \simeq \sqrt{{p \over p^{\prime}}} \, e^{i \Theta } \,.
\label{Cval}
\end{equation}
Consequently, inserting this expression for $C$ into
Eq.~(\ref{0val}) and using Eq.~(\ref{A1})
gives, including terms up to order $\eta^3$,
\begin{equation}
    \phi_{\bf p} (0) = e^{\pi \eta^{\prime}} \Gamma (1 + 2 i \eta^{\prime})
	\sqrt{{p^{\prime} \over p}} \, e^{i \Theta} \,.
\end{equation}
The phase $\Theta$ is, of course, irrelevant since only the absolute value
of the reduced matrix element occurs in the decay rate.

\section{ Second Order Corrections}

\subsection{Evaluation of $J_{fi}$}
We now calculate $J_{fi}$ defined by Eq.~(\ref{J_fi_def}).
Defining
\begin{equation}
    L_{\bf p} ({\bf r}) \equiv \langle {\bf p} |
	\left [ {1 \over |{\bf r}_1{-}{\bf r}|}
        {-} {1 \over r_1}{-}\Delta {\bar v} (r_1) \right ]
        {1 \over H_0{-}E {-}i \epsilon}
        \left ({1 \over  |{\bf r}_1{-}{\bf r}|}
	{-}{1 \over r} \right )
        {1 \over H_0{-}E{-}i \epsilon}
        |{\bf r}_1{=}{\bf 0} \rangle
\label{isotro}
\end{equation}
with $\Delta {\bar v} (r)$ being given by
\begin{equation}
    {e^2 \over 4 \pi} \Delta {\bar v} (r)
	\equiv v(r) + {e^2 \over 4 \pi r} \,,
\label{delta_v_def}
\end{equation}
expresses $J_{fi}$ as
\begin{equation}
    \eta^2 J_{fi} =  \left ({e^2 \over 4 \pi} \right )^2
         \int (d^3 {\bf r}) \rho_{fi} (r) L_{\bf p} ({\bf r}) \,.
\label{clear}
\end{equation}
The representation
\begin{equation}
    {1 \over H_0 - E - i \epsilon} =
	i \int_0^\infty dt e^{-i t(H_0 - E)} \,,
\end{equation}
expresses
\begin{equation}
    L_{\bf p}({\bf r}) = - \int_0^\infty \!dt\! \int_0^\infty\! dt'
	\langle {\bf p} |
	\left ( {1 \over |{\bf r}_1{-}{\bf r}|}
        {-} {1 \over r_1}{-}\Delta {\bar v} (r_1) \right )
	e^{-it (H_0{-}E)}
	\left ({1 \over  |{\bf r}_1{-}{\bf r}|}
        {-}{1 \over r} \right ) e^{-it' (H_0{-}E)}
        |{\bf r}_1{=}{\bf 0} \rangle \,.
\end{equation}
By introducing the Heisenberg picture operator
\begin{equation}
    {\bf r}_1 (t) \equiv e^{i H_0 t} {\bf r}_1 e^{-i H_0 t} \,,
\end{equation}
$L_{\bf p} ({\bf r})$ may be simply expressed as
\begin{eqnarray}
    L_{\bf p}({\bf r}) &=& - \int_0^\infty dt \int_0^\infty dt' \,
	\langle {\bf p} | \left ( {1 \over |{\bf r}_1 (t+t') - {\bf r}|}
        - {1 \over r_1 (t+t')} - \Delta {\bar v} [r_1(t{+}t')] \right )
\nonumber\\
	&& \qquad\qquad\qquad \qquad \qquad
	\times \left ({1 \over  |{\bf r}_1(t') - {\bf r}|}
        - {1 \over r} \right )|{\bf r}_1{=}{\bf 0} \rangle \,,
\label{qmatrix}
\end{eqnarray}
where we have used $\langle {\bf p} | \; (H_0 - E) = 0$.
Since the Hamiltonian $H_0$ describes a free particle,
the equation of motion for ${\bf r}_1 (t)$ gives
\begin{equation}
    {\bf r}_1 (t) = {\bf r}_1 (0) + {{\bf p}_1 \over m} t \,.
\end{equation}
We shall need only the leading order term of $L_{\bf p}({\bf r})$
for $p$ being much larger that $1/a_0$.
This is the term that gives the leading correction of order
$\eta^2 = 1 /(p a_0)^2$. It is then valid to
treat the momentum operator ${\bf p}_1$ as commuting with the
coordinate operator ${\bf r}_1$ since
$[{\bf r}_1 , {\bf p}_1] \sim 1  \ll p a_0$.
Thus, the leading order in $\eta$ evaluation of the matrix element in
Eq.~(\ref{qmatrix}) is given by its classical limit with the
operator ${\bf p}_1$
replaced by its eigenvalue ${\bf p}$ and ${\bf r}_1$ set to zero and with
$\langle {\bf p} | {\bf r}_1{=}{\bf 0} \rangle$=1:
\begin{equation}
    L_{\bf p}({\bf r}) = - \int_0^\infty dt \int_0^\infty dt'
	\left \{ {1 \over | {{\bf p} \over m} (t{+}t') - {\bf r}|}
	- {1 \over {p \over m} (t{+}t')}
	- \Delta {\bar v} \left [{p \over m} (t{+}t') \right ] \right \}
	\left ({1 \over  |{{\bf p} \over m} t' - {\bf r}|}
	- {1 \over r} \right ) \,.
\label{cmatrix}
\end{equation}
Performing a change of variables
\begin{equation}
    s = {m r \over p} (t + t') \,, \qquad \lambda = {t' \over t+t'} \,,
\end{equation}
puts this in the form
\begin{equation}
    L_{\bf p}({\bf r}) =
	 - {m^2 \over p^2} \int_0^\infty ds s \int_0^1 d \lambda
	\left ( {1 \over | {\hat {\bf p}} s - {\hat {\bf r}}|} -
	{1 \over s} - r \, \Delta {\bar v} (s r) \right )
	\left ( {1 \over | {\hat {\bf p}} \lambda s - {\hat {\bf r}}|}
	-1 \right ) \,.
\label{classical}
\end{equation}

Since only the angular average in ${\bf r}$ contributes, it is
convenient to use the spherical harmonic expansion
\begin{equation}
    {1 \over 4 \pi |{\bf r}_1 - {\bf r}_2|}
        = \sum_{l=0}^{\infty} \sum_{m=-l}^l
        {1 \over 2l+1} {r_<^l \over r_>^{l+1}} \,
        Y_{lm}^* ({\hat {\bf r}}_1) \,
        Y_{lm} ({\hat {\bf r}}_2 )
\label{LapGreen2}
\end{equation}
to obtain
\begin{eqnarray}
    L_{\bf p}({\bf r}) &=& - {m^2 \over p^2}
	\int_0^\infty ds s \int_0^1 d \lambda
	\left [\sum_{l=1}^\infty \sum_{m=-l}^l
	{4 \pi \over 2l{+}1} {s_<^l \over s_>^{l{+}1}}
	Y_{lm}^* ({\hat {\bf r}}) Y_{lm} ({\hat {\bf p}})
	+ \left (1{-}{1 \over s} \right ) \theta(1{-}s) -
	\Delta {\bar v} (s r) \right ]
\nonumber\\
	&& \times \left [\sum_{l'=0}^\infty \sum_{m'=l'}^{l'}
	{4 \pi \over 2l'+1} {s_<^{\prime \; l'} \over s_>^{\prime \; l'+1}}
	Y_{l'm'}^* ({\hat {\bf p}}) Y_{l'm'} ({\hat {\bf r}})
	+ \left ({1 \over \lambda s} - 1 \right )
	\theta(\lambda s - 1) \right ] \,,
\end{eqnarray}
where
\begin{eqnarray}
	s_< &=& {\rm min} \{s, 1\}, \qquad s_> = {\rm max} \{s, 1\} \,,
\nonumber\\
	s'_< &=& {\rm min} \{\lambda s, 1\},
	\qquad s'_> = {\rm max} \{\lambda s, 1\} \,.
\end{eqnarray}
The orthonormality of the spherical harmonics and the property
\begin{equation}
    {4 \pi \over 2l+1}
	\sum_m Y^*_{lm} ({\hat {\bf r}}) Y_{lm} ({\hat {\bf r}})
	= P_l (1) = 1
\end{equation}
now yield the angular average
\begin{equation}
    \overline {L_{\bf p}({\bf r})} = - {m^2 \over p^2}
	\int_0^\infty ds s \int_0^1 d \lambda
        \left [\sum_{l=1}^\infty {1 \over 2l+1}
	{s_<^l \over s_>^{l+1}}
	{s_<^{\prime \;l} \over s_>^{\prime \;l+1}}
	- r \Delta {\bar v} (s r) \left ({1 \over \lambda s} -1 \right )
	\theta (\lambda s - 1) \right ] \,.
\label{averaged}
\end{equation}
For the first term in the integrand,
exchanging the order of the integrals and sum,
it is straight forward to evaluate
\begin{eqnarray}
    &&\int_0^1 d \lambda \sum_{l=1}^\infty {1 \over 2l+1}
	\int_0^\infty ds \, s {s_<^l \over s_>^{l+1}}
        {s_<^{\prime l} \over s_>^{\prime l+1}}
\nonumber\\
    && \quad \qquad = \int_0^1 d \lambda \sum_{l=1}^\infty {1 \over 2l+1}
	\left (\int_0^1 ds \, s \, s^{2l+1} \lambda^l
	+ \int_1^{1 \over \lambda} ds \, s \, \lambda^l
	+ \int_{1 \over \lambda}^\infty ds \, s
	{1 \over s^{2l+2} \lambda^{l+1}} \right )
\nonumber\\
    && \quad \qquad = \int_0^1 d \lambda \sum_{l=1}^\infty
	\left ({\lambda^{l-1} \over 2 l}
	- {\lambda^l \over 2l+2} \right)
     = \int_0^1 d \lambda \; {1 \over 2} = {1 \over 2} \,.
\end{eqnarray}
Therefore,
\begin{equation}
    \overline {L_{\bf p}({\bf r})} = - {m^2 \over p^2}
	\left [ {1 \over 2} -
	\int_1^\infty \!\! ds \, (\ln s - s + 1) \, r
	\, \Delta {\bar v} (s r) \right ] \,,
\label{great}
\end{equation}
where we have carried out the remaining $\lambda$ integral.
The expansion~(\ref{LapGreen2}) gives
\begin{equation}
    \Delta {\bar v} (r) = \int (d^3 {\bf r}')
	\rho_{fi} (r') \left ({1 \over |{\bf r} - {\bf r}'|}
	- {1 \over r} \right )
	= \int (d^3 {\bf r}') \rho_{fi} (r') \theta (r'{-}r) \left (
	{1 \over r'} -{1 \over r} \right ) \,.
\end{equation}
Inserting this expression for $\Delta {\bar v} (r)$ into Eq.~(\ref{great})
and performing the $s$ integral produces
\begin{equation}
    \overline {L_{\bf p}({\bf r})} 	= - {m^2 \over 2 p^2}
        \left \{ 1 - \int (d^3 {\bf r}') \rho_{fi} (r')
	\theta (r'{-}r) \left [{r \over r'} - 2 + {r' \over r}
	- \ln^2 \left ({r' \over r} \right ) \right ] \right \} \,.
\end{equation}
Putting this result in to Eq.~(\ref{clear}) gives
\begin{equation}
    J_{fi} = - \left \{1 - \int (d^3 {\bf r}) \rho_{fi} (r)
        \int (d^3 {\bf r}') \rho_{fi} (r')
	\left [{r \over 2 r'} - {1 \over 4}
	\ln^2 \left ({r' \over r} \right ) \right ] \right \} \,,
\end{equation}
where the fact that the integrand is symmetric in $r$ and $r'$ has
been used to replace $\theta(r{-}r')$ by $1/2$.

To simplify this result, we note that
\begin{equation}
    {1 \over r} = {4 \pi \over e^2} (H_i - H_f)
\end{equation}
to obtain
\begin{eqnarray}
    \langle i | {1 \over r} | f \rangle
    	\langle f | r | i \rangle &=& {4 \pi \over e^2}
	\langle i | f \rangle (E_i - E_f)
	\langle f | r | i \rangle
\nonumber\\
	&=& {4 \pi \over e^2}
	\langle i | f \rangle
	\langle f | r H_i - H_f r | i \rangle
\nonumber\\
	&=& {a_0 \over 2} \langle i | f \rangle
	\langle f | \left [r, {\bf p}^2 \right ] | i \rangle
	+ \langle i | f \rangle \langle f | i \rangle
\nonumber\\
	&=& a_0 \langle i | f \rangle
	\langle f | \left (i {\hat {\bf r}} \cdot {\bf p}
	+ {1 \over r} \right ) | i \rangle +
	\langle i | f \rangle \langle f | i \rangle \,.
\end{eqnarray}
Since $\langle {\bf r} | i \rangle \sim \exp\{-r/a_0\}$,
\begin{equation}
    i a_0 {\hat {\bf r}} \cdot {\bf p} | i \rangle
	= - | i \rangle \,,
\end{equation}
and so
\begin{eqnarray}
    \langle i | {1 \over r} | f \rangle
    	\langle f | r | i \rangle &=&
	\langle i | f \rangle
	\langle f | {a_0 \over r} | i \rangle
\nonumber\\
	&=& a_0^2 m | T_{fi} |^2 V_{fi} \,,
\end{eqnarray}
or
\begin{equation}
    \int (d^3 {\bf r}_1) \rho_{fi} (r_1)
	\int (d^3 {\bf r}_2) \rho_{fi} (r_2)
	\left ({r_2 \over r_1} \right )
	= a_0^2 m V_{fi} \,.
\end{equation}
Therefore, $J_{fi}$ can be repackaged as
\begin{equation}
    J_{fi} = {1 \over 2} {\tilde v}_{fi} - 1 - {\tilde K}_{fi} \,,
\end{equation}
with
\begin{equation}
    {\tilde K}_{fi} = {1 \over 4}
	\int (d^3 {\bf r}_1) \rho_{fi} (r_1)
        \int (d^3 {\bf r}_2) \rho_{fi} (r_2)
	\ln^2 \left ( {r_2 \over r_1} \right ) \,.
\end{equation}
This is the result~(\ref{Corrections}) quoted in the text.

We shall calculate ${\tilde K}_{fi}$ for the final
$^3{\rm He}^+$ state being in $1s$, $2s$, and $3s$ states.
To evaluate ${\tilde K}_{fi}$, it is convenient to consider the
integral
\begin{equation}
    I (\epsilon) = \int (d^3 {\bf r}_1) \rho_{fi} (r_1)
        \int (d^3 {\bf r}_2) \rho_{fi} (r_2)
	\left ({r_2 \over r_1} \right )^{\epsilon} \,,
\end{equation}
which enables us to express ${\tilde K}_{fi}$ in terms of $I(\epsilon)$ as
\begin{equation}
    {\tilde K}_{fi} =  {1 \over 4}
	\left . {d^2 \over d \epsilon^2}
	I (\epsilon) \right |_{\epsilon = 0}  \,.
\label{KinI}
\end{equation}
For the case where the final $^3{\rm He}^+$ is in its ground state,
the $1s$ state, using the wave functions
\begin{equation}
    \phi_f (r) = \phi_{1s} (r) = {1 \over \sqrt{\pi}}
	\left ({2 \over a_0} \right )^{3/2} e^{-2r/a_0} \,,
	\quad \phi_i (r) = {1 \over \sqrt{\pi}} a_0^{-3/2}
	e^{-r/a_0}
\label{fiwavefunctions}
\end{equation}
we have $\rho_{fi} (r) \propto \phi_f^* (r) \phi_i (r) \propto e^{-sr}$
with $s = 3/a_0$.
Using this form of $\rho_{fi} (r)$ and performing the
changes of variables $x_1 = s r_1$ and $x_2 = s r_2$ yields
\begin{equation}
    I (\epsilon) = {1 \over 4}
        \int_0^{\infty} d x_1 x_1^{2 - \epsilon} e^{x_1}
	\int_0^{\infty} d x_2
	x_2^{2+\epsilon} e^{-x_2}
	= {1 \over 4} \Gamma (3 + \epsilon)
	\Gamma (3 - \epsilon) \,,
\end{equation}
where we have used $I(0) = 1$ to get the correct normalization factor.
Using
\begin{equation}
    \Gamma (1 + \epsilon) \Gamma (1 - \epsilon) =
	{\pi \epsilon \over \sin \pi \epsilon} \,,
\end{equation}
expression~(\ref{KinI}) now reads
\begin{equation}
    {\tilde K}_{1i} = {1 \over 16}
        \left . {d^2 \over d \epsilon^2} \left [
	{\pi \epsilon (4 - \epsilon^2) (1 - \epsilon^2)
	\over \sin \pi \epsilon} \right ]
	\right |_{\epsilon = 0}
	= \left ( {\pi^2 \over 12} - {5 \over 8} \right )  \,.
\label{Kval}
\end{equation}
Similarly, exploiting the wave functions for the final $2s$ and $3s$
ion states
\begin{eqnarray}
    \phi_{2s} (r) &=& {1 \over \sqrt{\pi}}
	\left ({1 \over a_0} \right )^{3/2}
	\left [1 - {r \over a_0} \right ]
	\exp \left (-{r \over a_0} \right ) \,,
\\
    \phi_{3s} (r) &=& {1 \over 81 \sqrt{3 \pi}}
	\left ({2 \over a_0} \right )^{3/2}
	\left [27 - 36 \left ({r \over a_0}
	\right ) + 8 \left ({r \over a_0} \right )^2 \right ]
	\exp \left ( - {2 r \over 3 a_0} \right ) \,,
\end{eqnarray}
we find the corresponding $K_{2i}$ and $K_{3i}$:
\begin{equation}
    {\tilde K}_{2i} = \left ( {\pi^2 \over 12} - {9 \over 8} \right ) \,,
	\quad {\tilde K}_{3i} = \left ( {\pi^2 \over 12} - {1045 \over 648}
	 \right ) \,.
\end{equation}

\subsection{Evaluation of $T_{\rm d}^0$ for final atomic states with
non-zero angular momentum}
Here we provide the intermediate steps for deriving the result~(\ref{lresult1})
from the expression~(\ref{otherlead}) for $T_d^0$.
Following steps similar to those that lead to Eq.~(\ref{classical}) in the
previous subsection, we can rewrite $T_d^0$ to the leading order as
\begin{equation}
    T_{\rm d}^0 \simeq  - i \eta
	\int (d^3 {\bf r}_2) \phi_f^* ({\bf r}_2)\phi_i ({\bf r}_2)
	\int_0^\infty d s {1 \over |{\hat {\bf p}} s - {\hat {\bf r}}_2|} \,.
\end{equation}
For the final state $\langle f |$ with angular momentum $(l,m)$ and thus
the wave function
\begin{equation}
    \phi_f^* ({\bf r}) = R_{fl}^* (r) Y_{lm}^* ({\hat {\bf r}}) \,,
\end{equation}
we use the expansion~(\ref{LapGreen2}) to get
\begin{eqnarray}
    T_{\rm d}^0 &\simeq&  - i \eta
        \int (d^3 {\bf r}_2) R_{fl}^* (r_2) Y_{lm}^* ({\hat {\bf r}_2})
	\phi_i (r_2) \int_0^\infty d s \sum_{l'm'} {4 \pi \over 2l'+1}
	Y_{l'm'}^*({\hat {\bf p}})
	Y_{l'm'}({\hat {\bf r}}_2) {s_<^{l'} \over s_>^{l'+1}}
\nonumber\\
	&=& - i \eta \int_0^\infty d r_2 \, r_2^2 R_{fl}^* (r_2) \phi_i (r_2)
	{4 \pi \over 2 l+1} Y_{lm}^* ({\hat {\bf p}})
	\int_0^\infty ds \left [ \theta(1{-}s)s^l +
	\theta(s{-}1) s^{-l-1} \right ]
\nonumber\\
	&=& -i \eta {4 \pi \over l(l+1)}
	Y_{lm}^* ({\hat {\bf p}})
	\int_0^\infty d r_2 \, r_2^2 R_{fl}^* (r_2) \phi_i (r_2) \,.
\label{lresult2}
\end{eqnarray}

\section{Inclusive Details}
\subsection{Coulomb Wave Functions}
To facilitate the calculations below, we display here the Coulomb
wave function for the scattering state on the final $^3{\rm He}^{++}$
nucleus of charge two \cite{Gottfried}
\begin{equation}
    \psi_{\bf k} ({\bf r}) = \langle {\bf r} | {\bf k} \rangle
	= e^{\pi \gamma} \Gamma (1 - 2i \gamma)
	e^{ i {\bf k} \cdot {\bf r}} F (2 i \gamma, 1; ikr
	- i {\bf k} \cdot {\bf r}) \,,
\label{Coulombwave}
\end{equation}
where $F(a, c; z)$ is the confluent hypergeometric function, and
the parameter $\gamma$ is defined by
\begin{equation}
    \gamma = {1 \over ka_0} \,.
\end{equation}
The appropriate integral representation of the confluent hypergeometric
function yields
\begin{equation}
    \psi_{\bf k} ({\bf r}) = e^{ \pi \gamma}
        \Gamma (1 - 2i \gamma)
        \oint_C {dt \over 2 \pi i}
        e^{ikrt + i{\bf k} \cdot {\bf r} (1-t)}
	t^{2i\gamma-1} (t - 1)^{ - 2 i \gamma} \,,
\label{psiintrep}
\end{equation}
where the contour $C$ wraps the cut in $t$ plane connecting the
branch points $t=0$ and $t=1$ in a counter-clockwise sense.
This wave function has the partial wave expansion
\begin{equation}
    \psi_{\bf k} ({\bf r}) = \sum_{l=0}^{\infty}
	(2l+1) P_l (\cos \theta) e^{i \pi l/2} R_{kl} (r) \,,
\label{Coulombpartial}
\end{equation}
with
\begin{equation}
    R_{kl} (r) = {e^{\pi \gamma} \Gamma (l + 1 - 2i \gamma)
	\over (2l + 1)!}
	(2 k r)^l e^{i k r} F (l + 1 - 2i \gamma, 2l+2; -2ikr) \,.
\end{equation}
Here $P_l (\cos \theta)$ is the Legendre function of the first kind
and $\theta$ is the angle between ${\bf k}$ and ${\bf r}$.
These radial wave functions can be recast into the following integral
forms by using different integral representations of the confluent
hypergeometric functions \cite{Landau}:
\begin{equation}
    R_{kl} (r) = e^{\pi \gamma} \Gamma (1 + l - 2i \gamma)
	{i \over (-2 kr)^{l+1}} \oint_C {dt \over 2 \pi i}
	e^{2ikrt} \left (t + {1 \over 2} \right )^{-l-1+2 i\gamma}
	\left (t - {1 \over 2} \right )^{-l-1-2i\gamma} \,,
\label{Clintrep1}
\end{equation}
and
\begin{equation}
    R_{kl} (r) = - e^{\pi \gamma} \Gamma ( - l - 2i \gamma)
        (2 kr)^l \oint_C {dt \over 2 \pi i}
        e^{2ikrt} \left (t + {1 \over 2} \right )^{l+2 i\gamma}
        \left (t - {1 \over 2} \right )^{l-2i\gamma} \,,
\label{Clintrep2}
\end{equation}
which we shall use later.
Here the contour $C$ wraps the cut connecting the branch points at
$t = \pm 1/2$ in the counter-clockwise sense.

\subsection{$\protect \langle {\protect\bf k} | i \protect \rangle$}
The matrix element $\langle {\bf k} | i \rangle$ can be evaluated by using
the Coulomb wave function directly. Since the wave function of $| i \rangle$
is spherically symmetric, in view of Eq.~(\ref{Coulombpartial}) only
$R_{k0} (r)$ is needed.
Utilizing Eq.~(\ref{Clintrep1}) for $l=0$
and performing the $r$ integral yields
\begin{eqnarray}
    \langle i | {\bf k} \rangle
	\tab =\tab  \int_0^{\infty} 4 \pi r^2 dr
	\, \phi_i (r) \, R_{k0} (r)
\nonumber\\
	\tab =\tab
	4 \sqrt{\pi} a_0^{-3/2} {e^{\pi \gamma} \Gamma (1{-}2i \gamma)
	\over (2ik)^3} \oint_C {dt \over 2 \pi i}
	{1 \over (t{+}i\gamma/2)^2}
	\left (t + {1 \over 2} \right )^{{-}1{+}2 i\gamma}
        \left (t - {1 \over 2} \right )^{{-}1{-}2i\gamma} .
\end{eqnarray}
Deforming the contour $C$ to wrap the double pole at $t= - i\gamma/2$
enables us to evaluate the contour integral as
\begin{eqnarray}
    \langle i | {\bf k} \rangle &=& - 4 \sqrt{\pi} a_0^{-3/2}
	{e^{\pi \gamma} \Gamma (1 - 2i\gamma)
	\over (2ik)^3} {d \over dt} \left . \left [
	\left (t + {1 \over 2} \right )^{-1+2 i\gamma}
        \left (t - {1 \over 2} \right )^{-1-2i\gamma}
	\right ] \right |_{t = - i\gamma/2}
\nonumber\\
	&=& - {8 \sqrt{\pi} \over k^3 a_0^{3/2}}
	e^{ \pi \gamma} \Gamma (1 - 2i \gamma)
	{\gamma \over (1 + \gamma^2)^2}
	\left ( {i\gamma - 1 \over i\gamma + 1} \right )^{2i\gamma} \,.
\label{kiresult}
\end{eqnarray}
Consequently,
\begin{equation}
    | \langle {\bf k} | i \rangle |^2 =
    | \langle i | {\bf k} \rangle |^2 =
	 {256 \pi^2 \over 1 - e^{-4 \pi \gamma}} {1 \over k^3}
	{\gamma^6 \over (1 + \gamma^2)^4}
	 e^{- 8 \gamma \cot^{-1} \gamma} \,.
\label{ki2result}
\end{equation}

\subsection{$K'' (0)$}
In this subsection, we compute
\begin{equation}
    K'' (0) \equiv \left. { d^2 \over d\epsilon^2 }
	\langle i | \, r^{-\epsilon} | {\bf k} \rangle
\langle {\bf k} | \, r^{\epsilon} | i \rangle \right |_{\epsilon{=}0}
\label{quark}
\end{equation}
for Coulomb states in the continuum $ | {\bf k} \rangle $.
We first express $K'' (0)$ in terms of several
one parameter integrals, then study its asymptotic behavior
as a function of $\gamma$, and finally present numerical results.

Using the representation
\begin{equation}
    \left ({r \over a_0} \right )^{-1 - \epsilon}
	= {1 \over \Gamma(1 + \epsilon)} \int_0^{\infty}
	dx \, x^{\epsilon} \exp \left ( - x {r \over a_0} \right )
\label{brownstrick}
\end{equation}
and the Coulomb wave function~(\ref{Clintrep1}) for $l=0$ and performing the
radial integral, we get
\begin{eqnarray}
	\langle i | \left ({a_0 \over r} \right )^{\epsilon}
	| {\bf k} \rangle \tab=\tab  - {\sqrt{\pi a_0^3} e^{\pi \gamma}
	\Gamma (1 - 2i \gamma) \over 2 \Gamma (1 + \epsilon)}
	\gamma^4 \int_0^{\infty} dx \, x^{\epsilon}
\nonumber\\
	\tab\tab \times \oint_C {dt \over 2 \pi i}
	\left [ t + {i \gamma (1 + x) \over 2} \right ]^{-3}
	\left (t + {1 \over 2} \right )^{-1+2i\gamma} \left (
	t - {1 \over 2} \right )^{-1-2i\gamma} .
\end{eqnarray}
Deforming the contour to wrap the triple pole at $ t = - i\gamma(1 + x)/2$
gives
\begin{eqnarray}
	\langle i | \left ({a_0 \over r} \right )^{\epsilon}
        | {\bf k} \rangle \tab=\tab{\sqrt{\pi a_0^3} e^{\pi \gamma}
        \Gamma (1 - 2i \gamma) \over 2 \Gamma (1 + \epsilon)}
        \gamma^4 \int_0^{\infty} dx \, x^{\epsilon}
\nonumber\\
        \tab\tab \times \left . {1 \over 2} {d^2 \over dt^2}
	\left [ \left (t + {1 \over 2} \right )^{-1+2i\gamma}
	\left (t - {1 \over 2} \right )^{-1-2i\gamma}
	\right ] \right |_{t = - i \gamma (1 + x)/2 } \,,
\end{eqnarray}
which can be also written in form
\begin{equation}
    \langle i | \left ({a_0 \over r} \right )^{\epsilon}
        | {\bf k} \rangle
	={-}{ 4 \sqrt{\pi a_0^3} e^{\pi \gamma}
        \Gamma (1{-}2i \gamma) \over \Gamma (1{+}\epsilon)}
	\gamma^2 \int_0^{\infty} dx \, x^{\epsilon} {d^2 \over dx^2}
	\left [ {1 \over [1{+}\gamma^2 (1{+}x)^2]}
	e^{ - 4 \gamma \cot^{-1} \gamma (1{+}x)} \right ] \,.
\label{clmatrix}
\end{equation}
Putting this into the definition (\ref{quark}) of $K (\epsilon)$ yields
\begin{eqnarray}
	K (\epsilon) \tab = \tab 16 \pi a_0^3
	{\sin \pi \epsilon \over \pi \epsilon}
	{4 \pi \gamma \over 1 - e^{- 4 \pi \gamma}} \gamma^4
	\int_0^{\infty} dx \int_0^{\infty} dy \left ({x \over y }
	\right )^{\epsilon} \qquad \qquad \qquad
\nonumber\\
	\tab \tab  \times {d^2 \over dx^2} \!
	\left [ {1 \over [1{+}\gamma^2 (1{+}x)^2]}
	e^{{-}4 \gamma \cot^{-1}\!\gamma (1{+}x)} \right ] \!
	{d^2 \over dy^2}\!\left [ {1 \over [1{+}\gamma^2 (1{+}y)^2]}
	e^{{-} 4 \gamma \cot^{-1}\!\gamma (1{+}y)} \right ] .
\label{cal_K_epsilon_exp}
\end{eqnarray}
Rescaling the dummy variables $x, y$ by a factor $\gamma$ and
taking the derivative with respect to $\epsilon$ gives
\begin{eqnarray}
    K'' (0) \tab = \tab {-}32 \pi a_0^3
	{4 \pi \gamma \over 1 - e^{- 4 \pi \gamma}} \gamma^6
\nonumber\\
	\tab \tab \times \left \{ {2 \pi^2 \over 3}
	{\gamma^2 \over (1{+}\gamma^2)^4}
	e^{-8 \gamma \cot^{-1} \gamma}
	{+}{2 \gamma \over (1{+}\gamma^2)^2} e^{-4 \gamma \cot^{-1} \gamma}
	I_2 (\gamma){+}I_1 (\gamma)^2 \right \} \,,
\label{cal_K_result}
\end{eqnarray}
where we have defined the integrals
\begin{equation}
    I_k (\gamma) \equiv \int_0^{\infty} dx \, \ln^k x {d^2 \over dx^2}
	\left [ {1 \over 1 + (\gamma + x)^2}
	e^{- 4 \gamma \cot^{-1} (\gamma + x)} \right ] \,,
\label{I_k_def}
\end{equation}
with $k=1,2$.
We have exhausted our analytic power at this point.
This expression is the starting point of the numerical computation.

Before displaying the numerical data of $K'' (0)$, let
us explore its asymptotic behavior as $\gamma \to 0$ and
$\gamma \to \infty$.
For a small $\gamma$, we first study the corresponding asymptotic behavior
of $I_k$.
Noting that
\begin{equation}
    \cot^{-1} (\gamma + x) = {\pi \over 2}
	- \tan^{-1} (\gamma + x) \,,
\end{equation}
expanding the integrand, and keeping only the first two leading
terms in the small $\gamma$ expansion, reduces Eq.~(\ref{I_k_def}) to
\begin{equation}
    I_k (\gamma) \simeq e^{-2 \pi \gamma}
	\int_0^{\infty} dx \ln ^k x
	{d^2 \over dx^2} \left [{1 \over 1 + x^2}
	- { 2 \gamma x \over (1 + x^2)^2}
	+ {4 \gamma \over 1 + x^2} \tan^{-1} x \right ] \,.
\end{equation}
Examining Eq.~(\ref{cal_K_result}), we find that,
to the first two leading
terms in the small $\gamma$ asymptotic expansion, we need to keep only
the first term for $I_2 (\gamma)$ in the equation above.
For $I_2$, the leading term can be calculated by doing the $x$ integral
by parts and evaluating a contour integral. The result is
\begin{equation}
    I_2 (\gamma) \simeq - \pi  \qquad {\rm as} \, \gamma \to 0 \,.
\end{equation}
For $I_1$, the two leading terms must be kept.
It can be expressed as
\begin{equation}
    I_1 (\gamma) \simeq 2 e^{-2 \pi \gamma}
	\int_0^{\infty} dx \left [{1 \over (1 + x^2)^2}
	+ {4 \gamma \over (1 + x^2)^2}
	\tan^{-1} x + \ln x {d^2 \over dx^2}
	{2 \gamma x \over (1 + x^2)^2 } \right ] \,,
\label{I_1_exp}
\end{equation}
where we have split the integrand into two parts and
done the $x$ integral by parts for one of them.
The $x$ integrals in the equation above can be evaluated
either by using their corresponding indefinite integral or
by writing them as derivatives of the Gamma functions.
We shall not go through the details of the derivation here, but
only display the results:
\begin{equation}
    I_1 (\gamma) \simeq {\pi \over 2}
	- \left ({\pi^2 \over 2} + 3 \right ) \gamma
	\qquad {\rm as} \, \gamma \to 0 \,.
\end{equation}
Inserting these into Eq.~(\ref{cal_K_result}) and making
appropriate expansions of the other functions of $\gamma$,
we obtain the asymptotic form of $K'' (0)$:
\begin{equation}
    K'' (0) \simeq - 8 \pi^3 \gamma^6 a_0^3
	+ 160 \pi^2 \gamma^7 a_0^3 \quad {\rm as} \;
	\gamma \to 0 \,.
\label{cal_K_small_asy}
\end{equation}

We now study the asymptotic expansion of $K'' (0)$ for large
$\gamma$.
It is convenient to look at the $\gamma \to \infty$ limit
of Eq.~(\ref{cal_K_epsilon_exp}), which is
\begin{eqnarray}
    K (\epsilon) &=& 64 \pi^2 \gamma a_0^3
	{\sin \pi \epsilon \over \pi \epsilon}
	\int_0^{\infty} dx \int_0^{\infty} dy
	\left ({x \over y} \right )^{\epsilon}
\nonumber\\
	&& \times {d^2 \over dx^2} \left [{1 \over (1 + x)^2}
	e^{-4/ (1 + x)} \right ]
	{d^2 \over dy^2} \left ({1 \over (1 + y)^2}
	e^{-4/ (1 + y)} \right ) \,.
\end{eqnarray}
Taking the derivative with respect to $\epsilon$ produces
\begin{equation}
    K'' (0) = - 128 \pi^2 \gamma a_0^3
	\left [ {2 \pi^2 \over 3} e^{-8}
	+ 2 e^{-4} C_2 + C_1^2 \right ] \,,
\end{equation}
where $C_1, C_2$ are two constants
\begin{equation}
    C_k \equiv \int_0^{\infty} dx \ln^k x
	{d^2 \over dx^2} \left ({e^{-4/(1+x)} \over (1+x)^2}
	\right ) \,.
\end{equation}
Numerically, $C_1 = 0.0525$ and $C_2 = -0.0664$.
Using these values, the asymptotic behavior becomes
\begin{equation}
    K '' (0) \simeq -3.19 \, \gamma \, a_0^3
	\quad {\rm as} \; \gamma \to \infty \,.
\label{cal_K_large_asy}
\end{equation}

\begin {figure}[htbp]
        \epsfysize 5in
        \leavevmode {\hfill \epsfbox {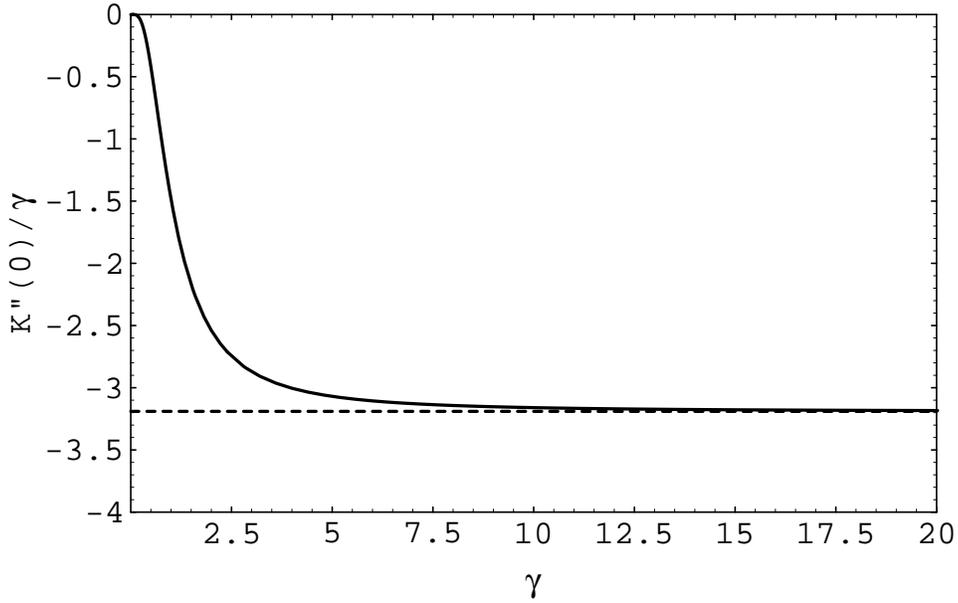} \hfill}
    \caption
        {%
        \advance\baselineskip by -8pt
	Numerical curve
	for $K''(0) / \gamma$ as a function of $\gamma$.
	The dashed horizontal line is its asymptotic behavior
	for large $\gamma$. The unit for the vertical axis
	is $a_0^{-3}$ and the horizontal variable $\gamma$
	is dimensionless.
        }%
    \label {fig:eleven}
\end {figure}
\goodbreak

\begin {figure}[htbp]
        \epsfysize 5in
        \leavevmode {\hfill \epsfbox {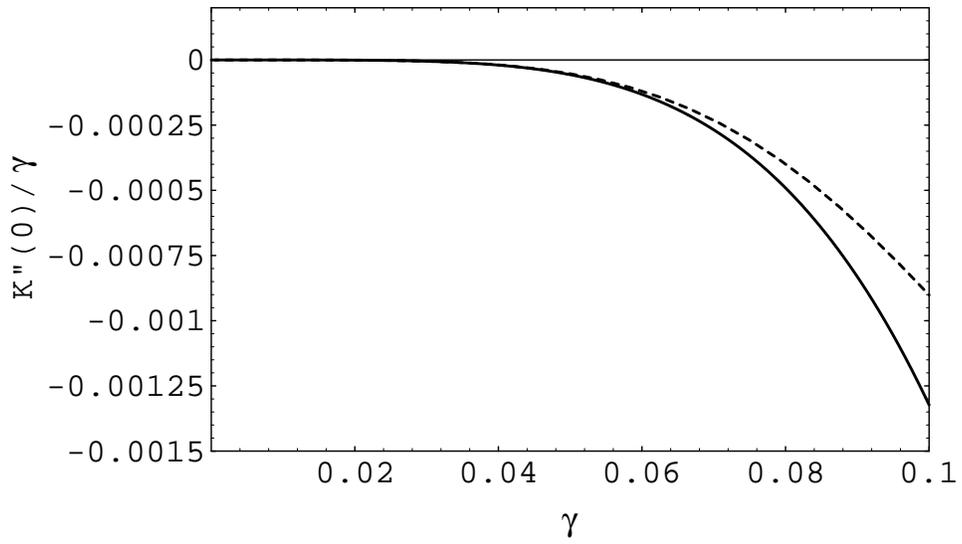} \hfill}
    \caption
        {%
        \advance\baselineskip by -8pt
        Numerical curves
        for $K''(0) / \gamma$ as a function of $\gamma$
	and its asymptotic behavior.
        The lower curve represents $K''(0)$ and
	the upper dashed curve represents its asymptotic behavior
        for small $\gamma$. The unit for the vertical axis
        is $a_0^{-3}$ and the horizontal axis represents the
        dimensionless variable $\gamma$.
        }%
    \label {fig:twelve}
\end {figure}

\begin {figure}[htbp]
        \epsfysize 5in
        \leavevmode {\hfill \epsfbox {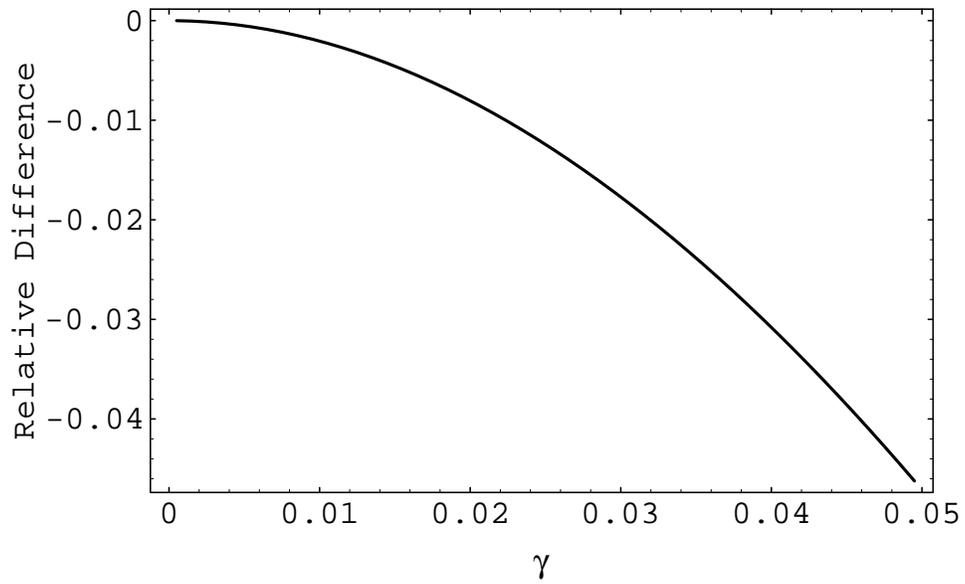} \hfill}
    \caption
        {%
        \advance\baselineskip by -8pt
        Relative accuracy of the asymptotic form
	$[K''(0) + 8 \pi^3 \gamma^6 -160 \pi^2 \gamma^7]/
	(8 \pi^3 \gamma^6)$ for small $\gamma$ as a function of
	$\gamma$; both vertical and horizontal
	axes are dimensionless.
        }%
    \label {fig:thirteen}
\end {figure}

We present the curves of $K''(0)$ as a function of $\gamma$
and its large $\gamma$ asymptotic behavior in Fig.~\ref{fig:eleven}.
Since the small $\gamma$ asymptotic form~(\ref{cal_K_large_asy})
fits $K''(0)$ only at a small region, we display it separately
in Fig.~\ref{fig:twelve} and Fig.~\ref{fig:thirteen}.
Our numerical data shows that at the large $\gamma$ region,
$K''(0)$ quickly approaches its asymptotic
form~(\ref{cal_K_large_asy}).
For $\gamma$ larger than $10$, the asymptotic form has
the accuracy of $0.01$. At the small $\gamma$ region,
one can see from Fig.~\ref{fig:thirteen} that
the asymptotic form~(\ref{cal_K_small_asy}) has the accuracy $0.01$
for $\gamma$ smaller than $0.02$.

\subsection{$K_{\rm 2b}$}
We now turn to evaluate
\begin{eqnarray}
    K_{\rm 2b} &=& \sum_b (E_i - E_f)^2 K''(0)
\nonumber\\
		&=&
 \left . {d^2 \over d \epsilon^2} \sum_b  (E_i - E_f)^2
	\langle i | r^{- \epsilon} | f \rangle
	\langle f | r^\epsilon | i \rangle \right |_{\epsilon=0} \,,
\label{ferrari}
\end{eqnarray}
where the sum runs only over the final bound states.
This was defined by Eq.~(\ref{K_2bdef}) in the text.
We first note that
\begin{eqnarray}
    (E_i - E_f) \langle f | r^{\epsilon} | i \rangle
	&=& \langle f | \left ( r^{\epsilon} H_i
	- H_f r^{\epsilon} \right ) | i \rangle
\nonumber\\
	&=& \langle f | r^{\epsilon} \left [
	{e^2 \over 4 \pi r} + {\epsilon (\epsilon + 1)
	\over 2 m r^2} + {\epsilon \over m r} i {\hat {\bf r}}
	\cdot {\bf p} \right ] | i \rangle \,.
\end{eqnarray}
Since $\phi_i (r) \propto \exp (-r/a_0)$, the operator $i {\hat {\bf r}}
\cdot {\bf p}$ is equivalent to the factor $-1/a_0$. Therefore,
\begin{equation}
    (E_i - E_f) \langle f | r^{\epsilon} | i \rangle
	= {\rm R\!y} \langle f | r^{\epsilon}
	\left [ {2 (1 - \epsilon) a_0 \over r}
	+ {\epsilon (\epsilon + 1) a_0^2 \over r^2}
	\right ] | i \rangle \,.
\label{trick}
\end{equation}
Inserting Eq.~(\ref{trick}) into the definition (\ref{ferrari}) and
carrying out the derivatives is straightforward algebra. The result
may be simplified by two remarks. Since the radial Schr\"odinger
equations are real, so are their regular solutions, except for an
overall phase factor. Hence,
\begin{equation}
\langle i | f(r) | f \rangle \langle f | g(r) | i \rangle =
	\langle i | g(r) | f \rangle \langle f | f(r) | i \rangle \,.
\end{equation}
Moreover, the definition (\ref{ferrari}) is not changed by the
reflection $\epsilon \to - \epsilon$ . Making use of these remarks
puts the result in the form
\begin{eqnarray}
    \sum_b (E_i{-}E_f)^2 K'' (0) \tab = \tab  {\rm R\!y}^2 \sum_b
	\left \{ - 2 \langle i | {a_0^2 \over r^2} | f \rangle
	\langle f | {a_0^2 \over r^2} | i \rangle
	+ 16 \langle i | {a_0 \over r} | f \rangle
	\langle f | {a_0^2 \over r^2} | i \rangle \right .
	- 8 \langle i | {a_0 \over r} | f \rangle
	\langle f | {a_0 \over r}| i \rangle
\nonumber\\
	\tab \tab \qquad \qquad
	+ 8 {d \over d \epsilon} \left .
	\langle i | \left ( {r \over a_0} \right )^{-1 - \epsilon}
	| f \rangle \langle f |
	\left ( {r \over a_0} \right )^{-2 + \epsilon}
	| i \rangle  \right |_{\epsilon = 0}
\nonumber\\ \tab \tab \qquad \qquad
	+ 4 {d^2 \over d \epsilon^2} \left . \left .
	\langle i | \left ( {r \over a_0}
	\right )^{{-}1{-}\epsilon} | f \rangle \langle
	f | \left ( {r \over a_0} \right )^{{-}1{+}\epsilon}
	| i \rangle \right |_{\epsilon = 0}
	\right \} \,.
\label{dderivative}
\end{eqnarray}

The first three terms in the right hand side of Eq.~(\ref{dderivative})
may be readily calculated by  using the
bound state wave functions for the Coulomb potential.
The bound states for Coulomb potential are labeled by the
principal quantum number $n$ and the angular quantum numbers $l,m$.
Since only the final S-wave states give non-vanishing matrix elements,
we can simply use the principal quantum number $n=1,2,\cdots$ to denote the
final states.
The wave function may be expressed in terms of the generalized
Laguerre polynomials $L_n^m (x)$ which may be written in terms of an
integral representation to express
\begin{equation}
    \phi_f ({\bf r}) = {1 \over \sqrt{\pi}}
	\left ({2 \over a_0} \right )^{3/2} {1 \over n^{5/2} }
    	\oint_C {dt \over 2 \pi i}
	\exp\left\{- (1 + 2t) (2r / na_0) \right\}
	\left( 1 + {1 \over t} \right)^n \,,
\end{equation}
with the contour $C$ encircling $t=0$ in the counter-clockwise sense.
Using this and the initial wave function~(\ref{fiwavefunctions}) and
performing the trivial radial integral gives
\begin{equation}
    \langle i | \left ({a_0 \over r} \right )^m | f \rangle
	= {2^{7/2} \, \Gamma (3{-}m) \over n^{5/2}}
	\oint_C {dt \over 2 \pi i} \left (1 + {1 \over t} \right )^n
	\left (1 + {2 \over n} + {4 t \over n} \right )^{m-3} \,.
\end{equation}
The contour can be deformed to wrap the pole at
$t=-n/4-1/2$ to carry out the $t$ integral. For $m=2$, there is
a contribution from the contour at the infinity.
Therefore,
\begin{equation}
    \langle i | \left ({a_0 \over r} \right )^m | f \rangle
	= - {2^{7/2} \, \Gamma (3{-}m) \over n^{5/2} \, (2{-}m)!}
	\left ({n \over 4} \right )^{3-m} \left [
	{d^{(2-m)} \over dt^{(2-m)}} \left . \left (1+{1 \over t} \right )^n
	\right |_{t=-{n+2 \over 4}} - \delta_{m,2} \right ] \,.
\end{equation}
For $m=1,2$, it is straight forward to evaluate
\begin{equation}
    \langle i |{a_0 \over r} | f \rangle
	= {8 \sqrt{2 n} \over (n+2)^2}
	\left ({n-2 \over n+2} \right )^{n-1} \,,
	\qquad \langle i | \left ({a_0 \over r} \right )^2 | f \rangle
	= \left ({2 \over n} \right )^{3/2}
	\left [1 - \left ({n-2 \over n+2} \right )^n \right ] \,.
\end{equation}
Hence, the first three sums in the right hand side of Eq.~(\ref{dderivative})
can be written as
\begin{eqnarray}
    \sum_b \langle i | {a_0^2 \over r^2} | f \rangle
        \langle f | {a_0^2 \over r^2} | i \rangle
	&=& 8 \sum_{n=1}^\infty {1 \over n^3}
	\left [1 - \left ({n-2 \over n+2} \right )^n \right ]^2
	\simeq 15.82 \,,
\nonumber\\
    \sum_b \langle i | {a_0\over r} | f \rangle
        \langle f | {a_0^2 \over r^2} | i \rangle
        &=& 32 \sum_{n=1}^\infty {1 \over n (n+2)^2}
	\left ({n-2 \over n+2} \right )^{n-1}
	\left [1 - \left ({n-2 \over n+2} \right )^n \right ]
	\simeq 4.78 \,,
\nonumber\\
    \sum_b \langle i | {a_0\over r} | f \rangle
        \langle f | {a_0\over r} | i \rangle
        &=& 128 \sum_{n=1}^\infty {n \over (n+2)^4}
	\left ({n-2 \over n+2} \right )^{2(n-1)}
	\simeq 1.58 \,.
\end{eqnarray}

For the fourth and fifth terms in Eq.~(\ref{dderivative}),
we shall compute the bound-state sums by exploiting the completeness
of the the sum over all the intermediate $S$-wave states. We write
the bound-state sum in terms of the identity operator (in the
$S$-wave sector) --- which is equivalent to using the closure result
---  and then subtract the sum over all the continuum $S$-wave Coulomb
scattering states:
\begin{equation}
    \sum_b | f \rangle \langle f | = 1 -
	\int_0^{\infty} {1 \over 2 \pi^2} k^2 dk \,
	| k \rangle \langle k |  \,.
\label{b_sum}
\end{equation}
Since the closure results --- the terms involving the identity operator
--- do not have any $\epsilon$ dependence and thus do not contribute
to the derivative, the identity operator in Eq.~(\ref{b_sum}) may
be omitted in our calculations.

Thus the fourth term in Eq.~(\ref{dderivative}) may be written as
\begin{equation}
   \sum_b {d \over d \epsilon} \left .
	\langle i | \left({a_0 \over r}\right)^{1 + \epsilon}\!\!\!\!
	| f \rangle \langle f | \left({a_0 \over r}\right)^{2 - \epsilon}
	 \!\!\!\! | i \rangle \right |_{\epsilon{=}0} \!\!\!\!
	 = {-}\int_0^{\infty} dk  {k^2 \over 2 \pi^2}
	{d \over d \epsilon} \! \left .
	\langle i | \left({a_0 \over r}\right)^{1{+}\epsilon} \!\!\!\!
	| k \rangle \langle k | \left({a_0 \over r}\right)^{2{-}\epsilon}
	\!\!\!\! | i \rangle \right |_{\epsilon = 0} .
\end{equation}
Using the representation~(\ref{brownstrick})
and taking the derivative with respect to $\epsilon$ gives
\begin{eqnarray}
    \sum_b {d \over d \epsilon} \tab\tab\!\!\!\! \left .
        \langle i | \left({a_0 \over r}\right)^{1 + \epsilon}
        | f \rangle \langle f | \left({a_0 \over r}\right)^{2 - \epsilon}
        | i \rangle \right |_{\epsilon = 0}
\nonumber\\
	\tab=\tab\int_0^{\infty} {1 \over 2 \pi^2} k^2 dk
	{d \over d \epsilon} \left .
	{\sin \pi \epsilon \over \pi \epsilon}
	\int_0^{\infty} dx x^{\epsilon} \int_0^{\infty} dy y^{-\epsilon}
	\langle i | e^{-xr/a_0} | k \rangle
	\langle k | {a_0 \over r} e^{-yr/a_0}
	| i \rangle \right |_{\epsilon{=}0}
\nonumber\\
	\tab=\tab{16 \over \pi} \int_0^{\infty} d \gamma
	\int_0^{\infty} dx \int_0^{\infty} dy \ln
	\left ({x \over y} \right )
	{4 \pi \gamma \over 1 - e^{-4\pi \gamma}}
	{\gamma^2 (1 - x) \over [1{+}\gamma^2 (1{+}x)^2]^2}
	{1 \over 1 + \gamma^2 (1 + y)^2}
\nonumber\\
	\tab\tab \qquad \times e^{-4 \gamma \cot^{-1}
	\gamma (1{+}x){-}4 \gamma \cot^{-1} \gamma (1{+}y) }
\nonumber\\
	\tab=\tab {4 \over \pi} \int_0^{\infty} \! d \gamma \gamma^{-2}
	{4 \pi \gamma \over 1{-}e^{-4 \pi \gamma}} \left [
	{2 \gamma \over 1{+}\gamma^2} e^{-4 \gamma \cot^{-1} \gamma}
	I_3 (\gamma) + (1{-}e^{-4 \gamma \cot^{-1} \gamma} ) I_4 (\gamma)
	\right ] \,,
\end{eqnarray}
where the integrals $I_3$ and $I_4$ are defined by
\begin{equation}
    I_3 (\gamma) \equiv \int_0^{\infty} dx \ln x
	 {\gamma \over 1 + \gamma^2 (1 + x)^2}
	e^{-4 \gamma \cot^{-1} \gamma (1 + x)} \,,
\end{equation}
and
\begin{equation}
    I_4 (\gamma) \equiv \int_0^{\infty} dx \ln x {\gamma^2 (1 - x)
	\over [1 + \gamma^2 (1 + x)^2]^2} e^{-4 \gamma
	\cot^{-1} \gamma (1 + x)} \,.
\end{equation}
We display here the numerical value of this fourth term
\begin{equation}
    \sum_b {d \over d \epsilon} \left .
        \langle i | \left ({a_0 \over r} \right )^{1 + \epsilon}
        | f \rangle \langle f | \left ({a_0 \over r} \right )^{2 - \epsilon}
        | i \rangle \right |_{\epsilon = 0}
	= - 1.00  \,.
\end{equation}

In the same fashion, we can deal with the fifth term in
Eq.~(\ref{dderivative}),
\begin{eqnarray}
	{d^2 \over d \epsilon^2} \!\!\!
	\tab\tab \sum_f \left . \langle i |
	\left ({a_0 \over r} \right )^{1{+}\epsilon}
	| f \rangle \langle f | \left ({a_0 \over r}
	\right )^{1{-}\epsilon} | i \rangle
	\right |_{\epsilon = 0}
\nonumber\\
	\tab=\tab{-}{d^2 \over d \epsilon^2}
	\left . {\sin \pi \epsilon
	\over \pi \epsilon} \int_0^{\infty}  dx x^{\epsilon} \!
	\int_0^{\infty}  dy y^{-\epsilon} \!
	\int_0^{\infty} \! {1 \over 2 \pi^2}
	k^2 dk \langle i | e^{{-}xr/a_0}| k \rangle
	\langle k | e^{{-}yr/a_0} | i \rangle  \right |_{\epsilon{=}0}.
\end{eqnarray}
Evaluating the derivative gives
\begin{eqnarray}
    {d^2 \over d \epsilon^2} \!\!\!
	\tab\tab \sum_b \langle i |
	\left . \left ({a_0 \over r} \right )^{1 + \epsilon}
	| f \rangle \langle f | \left ({a_0 \over r}
	\right )^{1 - \epsilon} | i \rangle
	\right |_{\epsilon = 0}
\nonumber\\
	\tab=\tab
	{\pi^2 \over 3} \int_0^{\infty}  {1 \over 2 \pi^2}
	k^2 dk \left | \langle i | {a_0 \over r} | k \rangle \right |^2
	{-}64 \pi a_0^3 \int_0^{\infty} dx \int_0^{\infty} dy \ln^2
	\left ( {x \over y} \right )
	\int_0^{\infty}  {1 \over 2 \pi^2}
	k^2 dk {4 \pi \gamma \over 1{-}e^{{-}4\pi \gamma}}
\nonumber\\
	\tab\tab \times
	{\gamma^4 (1{-}x) \over [1 + \gamma^2 (1{+}x)^2]^2}
	{\gamma^4 (1{-}y) \over [1 + \gamma^2 (1{+}y)^2]^2}
	e^{-4 \gamma \cot^{-1} \gamma (1{+}x)
	{-} 4 \gamma \cot^{-1} \gamma (1{+}y) }
\nonumber\\
	\tab=\tab{8 \over \pi} \int_0^{\infty}  d \gamma
	{4 \pi \gamma \over 1{-}e^{{-}4 \pi \gamma}}
	\left [{\pi^2 \over 3}
	{1 \over (1{+}\gamma^2)^2} e^{- 8 \gamma \cot^{-1} \gamma}
	{+}4 I_4 (\gamma)^2{+}{2 \over 1{+}\gamma^2}
	e^{{-}4 \gamma \cot^{-1} \gamma} I_5 (\gamma) \right ] ,
\end{eqnarray}
where we have defined the integral $I_5$
\begin{equation}
    I_5 (\gamma) \equiv \int_0^{\infty} dx \ln^2 x{\gamma^2 (1 - x)
        \over [1 + \gamma^2 (1 + x)^2]^2} e^{-4 \gamma
        \cot^{-1} \gamma (1 + x)} \,,
\end{equation}
and used the result
\begin{equation}
    \int_0^{\infty} dx {\gamma^2 (1 - x)
        \over [1 + \gamma^2 (1 + x)^2]^2} e^{-4 \gamma
        \cot^{-1} \gamma (1 + x)} = - {1 \over 2(1{+}\gamma^2)}
	e^{-4 \gamma \cot^{-1} \gamma} \,.
\end{equation}
Numerically, this fifth term has the value
\begin{equation}
    {d^2 \over d \epsilon^2} \sum_b \langle i |
        \left . \left ({a_0 \over r} \right )^{1 + \epsilon}
        | f \rangle \langle f | \left ({a_0 \over r}
        \right )^{1 - \epsilon} | i \rangle
        \right |_{\epsilon = 0}
	= 2.02 \,.
\end{equation}

Putting these numerical values into Eq.~(\ref{dderivative}) yields
finally
\begin{equation}
        K_{\rm 2b} =
	\sum_b (E_i - E_f)^2 K'' (0) \simeq 32.26 \, {\rm R\!y}^2 \,.
\end{equation}

\subsection{M(k)}
We turn now to calculate $M(k)$ defined in Eq.~(\ref{Mdef}),
\begin{equation}
    M (k) = \left (E_i - E_k \right )^2
	\sum_{l=1}^{\infty} {4 \pi (2l + 1) \over l^2 (l+1)^2}
	\left |\int_0^{\infty} d r r^2 R_{kl}^* (r)
	\phi_i (r) \right |^2 \,.
\label{Mdef2}
\end{equation}
We shall express $M(k)$ as a one parameter integral.

To do this, let us consider first the simpler
sum $G(\xi', \xi'')$ defined by
\begin{equation}
    G (\xi', \xi'') \equiv \sum_{l=1}^{\infty}
	(2l + 1) g_{kl} (\xi') g_{kl}^* (\xi'') \,,
\label{kernel_def}
\end{equation}
where
\begin{equation}
     g_{kl} (\xi) \equiv k^{-2} \int_0^{\infty} dr
        R_{kl}^* (r) \phi_i (\xi, r) \,,
\end{equation}
with
\begin{equation}
    \phi_i (\xi, r) =
        {1 \over \sqrt{\pi a_0^3}} e^{-\xi r / a_0} \,,
\end{equation}
which, up to an overall constant,
is the ground state wave function of a
hydrogen-like atom with nuclear charge $+\xi e$.
The reason for considering $G(\xi', \xi'')$ is that,
as we shall see later, $M(k)$ may be expressed in terms of
$G (\xi', \xi'')$ and $G(\xi', \xi'')$ may be evaluated
in a closed form.
We first complete the evaluation of $G(\xi', \xi'')$.
Using the orthogonality relation of the Legendre function,
\begin{equation}
    \int_{-1}^1 dx P_l (x) P_{l'} (x) = {2 \over 2 l + 1}
	\delta_{l,l'} \,,
\end{equation}
the expansion~(\ref{Coulombpartial}) expresses the radial Coulomb
wave function as
\begin{equation}
    R_{kl} (r) = {1 \over 2 \, i^l} \int_{-1}^1 d \cos \theta
	P_l (\cos \theta) \psi_{\bf k} (r, \theta, 0) \,.
\end{equation}
Inserting this into the definition of $g_{kl} (\xi)$
and using the completeness relation
\begin{equation}
    \sum_{l=0}^{\infty} (2l+1) P_l (x) P_l (y)
	= 2 \delta (x - y)
\end{equation}
yields
\begin{eqnarray}
    G (\xi', \xi'') &=& {1 \over 2 k^4} \int_{-1}^1 d \cos \theta
	\int_0^{\infty} dr \psi_{\bf k}^* (r, \theta, 0)
	\phi_i (\xi', r)
	\int_0^{\infty} dr' \psi_{\bf k} (r', \theta, 0)
	\phi_i (\xi'', r')
\nonumber\\
	&& - k^{-4} \int_0^{\infty} dr R_{k0}^* (r) \phi_i (\xi', r)
	\int_0^{\infty} dr' R_{k0} (r') \phi_i (\xi'', r') \,,
\label{Gsummed}
\end{eqnarray}
where the second term accounts for the fact that we do not include
the $l=0$ term in the summation~(\ref{kernel_def}).
Making use of the integral representation~(\ref{psiintrep}) and
the definition of $\phi_i (\xi, r)$,
it is straight forward to first perform the $r$ integration and then,
deforming the contour of the $t$-integration to encircle the pole as
in the evaluation of~(\ref{kiresult}), to compute
\begin{eqnarray}
    \int_0^{\infty} \!dr \psi_{\bf k}^* (r, \theta, 0) \phi_i (\xi', r)
	\tab=\tab\! {e^{\pi \gamma} \Gamma (1{+}2 i \gamma)
	\over \sqrt{\pi a_0}} \gamma \!
	\oint_C \! {dt \over 2 \pi i}
	{1 \over \xi' \gamma{+}i \cos \theta
	{-} i t (1{-}\cos \theta)}
	t^{2i\gamma{-}1} (t{-}1)^{{-}2 i \gamma}
\nonumber\\
	\tab = \tab \! { e^{\pi \gamma} \Gamma (1 + 2 i \gamma)
	\over \sqrt{\pi a_0}} { \gamma
	\over \xi' \gamma + i \cos \theta }
	\left ({\xi' \gamma + i \cos \theta \over \xi' \gamma + i}
	\right )^{- 2 i \gamma} \,.
\label{Gcontour}
\end{eqnarray}
The remaining angular integration in Eq.~(\ref{Gsummed}) may now
be done in closed form:
\begin{eqnarray}
    &&\int_{-1}^1 d \cos \theta {\gamma \over \xi' \gamma + i \cos \theta }
	\left ({\xi' \gamma + i \cos \theta \over \xi' \gamma + i}
        \right )^{-2 i \gamma} {\gamma \over \xi'' \gamma - i \cos \theta }
        \left ({\xi'' \gamma - i \cos \theta \over \xi'' \gamma - i}
	\right )^{2i \gamma}
\nonumber\\
	&& \quad = - {i \gamma \over \xi'{+}\xi''}
	\left ({\xi' \gamma{+}i \over \xi'' \gamma{-}i}
	\right )^{2 i \gamma}
	\int_{\cos \theta = -1}^{\cos \theta = 1}
	d \ln \left ({\xi' \gamma{+}i \cos \theta
	\over \xi'' \gamma{-}i \cos \theta} \right )
	\left ( {\xi' \gamma{+}i \cos \theta
	\over \xi'' \gamma{-}i \cos \theta} \right )^{-2i \gamma}
\nonumber\\
	&& \quad = - {1 \over 2 (\xi' + \xi'')}
	\left ({\xi' \gamma{+}i \over \xi'' \gamma{-}i}
	\right )^{2 i \gamma}
	\left [ \left ( {\xi' \gamma{+}i \over \xi'' \gamma{-}i}
	\right )^{-2i \gamma}
	- \left ( {\xi' \gamma{-}i \over \xi'' \gamma{+}i}
	\right )^{-2i \gamma} \right ]
\nonumber\\
	&& \quad = - {1 \over 2 (\xi' + \xi'')}
	\left (e^{- 4 \gamma (\pi  - \tan^{-1} \xi' \gamma
	- \tan^{-1} \xi'' \gamma)} - 1 \right ) \,.
\end{eqnarray}
With this, it is straight forward to evaluate the first term in
Eq.~(\ref{Gsummed}).
For the second term, using the integral
representation~(\ref{Clintrep2}) for the radial wave function
$R_{kl} (r)$ for $l=0$ and doing the $r$ integral, we find that
\begin{eqnarray}
    \int_0^{\infty} d r \phi_i (\xi, r) R_{k0}(r)
        \tab = \tab {e^{\pi \gamma} \Gamma (1{-}2i \gamma)
        \over 4 \sqrt{ \pi a_0}} \oint {dt \over 2 \pi i}
        {1 \over t{+}i \xi \gamma / 2} \left (
        t{+}{1 \over 2} \right )^{2 i \gamma} \left (
        t{-}{1 \over 2} \right )^{-2i \gamma}
\nonumber\\
        \tab = \tab {e^{\pi \gamma} \Gamma (1{-}2i \gamma)
        \over 4 \sqrt{ \pi a_0}}
        \left (1{-}e^{- 4 \gamma \cot^{-1} \xi \gamma}
        \right ) \,,
\label{r_2kxi}
\end{eqnarray}
where the contour integral has been evaluated in the fashion
as in the previous evaluations in Eqs.~(\ref{kiresult}).
Note that, while deforming the contour, the contour at infinity
contributes the $1$ in the last parentheses.
We have now found that
\begin{eqnarray}
    G (\xi', \xi'') &=& {\gamma^5 a_0^3 \over 1 - e^{-4 \pi \gamma}}
	\Biggr [ {1 \over \xi' + \xi''} \left (1 -
	e^{- 4 \gamma (\pi  - \tan^{-1} \xi' \gamma
	- \tan^{-1} \xi'' \gamma)} \right )
\nonumber\\
	&& \qquad - {1 \over 4} \left (1 - e^{-2 \pi \gamma
	+ 4 \gamma \tan^{-1} \xi' \gamma} \right )
	\left (1 - e^{-2 \pi \gamma
	+ 4 \gamma \tan^{-1} \xi'' \gamma} \right ) \Biggr ] \,.
\label{Gresult}
\end{eqnarray}

As mentioned before, $G (\xi', \xi'')$ is related with $M(k)$,
as we shall now prove.
To facilitate the derivation, it is convenient to define
\begin{equation}
    f_{kl} (\xi) \equiv {a_0 \over l (l+1)}
	\int_0^{\infty} dr r R_{kl}^* (r) \phi_i (\xi, r) \,,
\end{equation}
which generates the desired matrix element involved in the
definition~(\ref{Mdef2}) via
\begin{equation}
    {1 \over l(l+1)} \int_0^{\infty} dr r^2 R^*_{kl} (r) \phi_i (r)
	= - \left . {d \over d \xi} f_{kl} (\xi) \right |_{\xi = 1} \,.
\label{dxi}
\end{equation}
The Schr\"odinger equation satisfied by $R_{kl} (r)$,
\begin{equation}
    \left [ - {1 \over 2m} {d^2 \over dr^2}
	- {2 e^2 \over 4 \pi r} - {k^2 \over 2m} \right ]
	(r R_{kl}^* (r))
	= {l(l+1) \over 2 m r^2} (r R_{kl}^* (r)) \,,
\end{equation}
enables us to relate $g_{kl} (\xi)$ and $f_{kl} (\xi)$ through
\begin{eqnarray}
    g_{kl} (\xi) &=& - {1 \over l(l+1)} \int_0^{\infty} dr
		(r \phi_i (\xi, r))
		\left [{d^2 \over dr^2} + {4 m e^2 \over 4 \pi r}
		+ k^2 \right ] (r R_{kl}^* (r))
\nonumber\\
	&=& (1 + \xi^2 \gamma^2) {d \over d \xi} f_{kl} (\xi)
		+ 2 (\xi - 2) \gamma^2 f_{kl} (\xi) \,,
\label{gf_rel}
\end{eqnarray}
where we have done the integral by parts and used the explicit form of
the wave function $\phi_i (\xi, r)$.
Relation~(\ref{gf_rel}) may be easily inverted:
\begin{equation}
    f_{kl} (\xi) = - {1 \over 1 + \xi^2 \gamma^2}
		e^{4 \gamma \tan^{-1} \xi \gamma}
		\int_{\xi}^{\infty} d \xi'
		e^{-4 \gamma \tan^{-1} \xi' \gamma}
		g_{kl} (\xi') \,,
\label{fg_rel}
\end{equation}
where the boundary condition has been appropriately chosen.
Combining this with Eq.~(\ref{dxi}) yields
\begin{equation}
    {1 \over l(l{+}1)} \int_0^{\infty} dr r^2 R^*_{kl} (r) \phi_i (r)
	= - \left [ {g_{kl} (1) \over 1{+}\gamma^2}
	{-}{2 \gamma^2 \over (1{+}\gamma^2)^2}
	e^{4 \gamma \tan^{-1} \gamma}  \int_1^{\infty} d \xi'
	e^{{-}4 \gamma \tan^{-1} \xi' \gamma}
	g_{kl} (\xi') \right ] \,.
\end{equation}
Consequently, $M(k)$ may be expressed in terms of $G (\xi', \xi'')$ as
\begin{eqnarray}
	M (k) \tab = \tab 4 \pi {\rm R\!y}^2 \gamma^{-4} \Biggr [
	G(1, 1) - {4 \gamma^2 \over 1 + \gamma^2}
	e^{4 \gamma \tan^{-1} \gamma} \int_1^{\infty}
	d \xi \, G (1, \xi) e^{-4 \gamma \tan^{-1} \xi \gamma}
\nonumber\\
	\tab\tab {+}{8 \gamma^4 \over (1{+}\gamma^2)^2}
	e^{8 \gamma \tan^{-1} \gamma}
	\int_1^{\infty} d \xi' \int_{\xi'}^{\infty} d \xi''
	G (\xi', \xi'') e^{-4 \gamma (\tan^{-1} \xi' \gamma
	{+}\tan^{-1} \xi'' \gamma)} \Biggr ] \,,
\label{Mresult0}
\end{eqnarray}
where we have used the fact that $G(\xi', \xi'')$ is symmetric
about its two arguments.

Inserting the explicit form~(\ref{Gresult}) for $G(\xi', \xi'')$ into
the expression above for $M (k)$ and making the changes of the variables
$\xi = 1/t$, $\xi' = 1/t'$, and $\xi''= 1/t''$ yields the lengthy expression
\begin{eqnarray}
	M (k) \tab=\tab
	{\rm R\!y}^2 a_0^3 {4 \pi \gamma \over 1 - e^{-4 \pi \gamma}}
	\Biggr \{ {1 \over 4}
	\left (1 - e^{-4 \gamma \tan^{-1} \gamma^{-1}} \right )
	\left (1 + 3 e^{-4 \gamma \tan^{-1} \gamma^{-1}} \right )
\nonumber\\
	\tab\tab - {4 \gamma^2 \over 1 + \gamma^2}
	e^{- 4 \gamma \tan^{-1} \gamma^{-1}} G_1 (\gamma)
	+ {8 \gamma^4 \over (1 + \gamma^2)^2}
	e^{- 8 \gamma \tan^{-1} \gamma^{-1}} G_2 (\gamma) \Biggr \} \,,
\end{eqnarray}
where the integrals $G_1 (\gamma)$ and $G_2 (\gamma)$ are defined by
\begin{eqnarray}
    G_1 (\gamma) &\!=&\! \int_0^1 {dt \over t} \left [
	{1 \over 1{+}t} \left (e^{-4 \gamma \tan^{-1} (t/\gamma)}
	{-}e^{-4 \gamma \tan^{-1} \gamma^{-1}} \right ) \right .
\nonumber\\
	&&  \qquad \left .
	- {1 \over 4 t} \left (e^{4 \gamma \tan^{-1} (t/\gamma)}
	{-}1 \right ) \left (1{-}e^{-4 \gamma \tan^{-1} \gamma^{-1}}
	\right ) \right ]
\end{eqnarray}
and
\begin{eqnarray}
    G_2 (\gamma) &=& \int_0^1 d t' \int_0^{t'} d t'' {1 \over t' t''}
	\left [{1 \over t' + t''} \left (
	e^{4 \gamma (\tan^{-1} (t'/\gamma) + \tan^{-1} (t''/\gamma))}
	- 1 \right ) \right .
\nonumber\\
	&& \left . - {1 \over 4 t' t''}
	\left (e^{4 \gamma \tan^{-1} (t'/\gamma)} - 1 \right )
	\left (e^{4 \gamma \tan^{-1} (t''/\gamma)} - 1 \right )
	\right ] \,.
\end{eqnarray}
The integral defining $G_2 (\gamma)$ may be simplified to a one
parameter integral.
This can be done by first making the change of variable $t' = t$,
$t''= s t$, which gives
\begin{eqnarray}
    G_2(\gamma) &=& \int_0^1 {dt \over t^2} \int_0^1 {ds \over s}
	\left [ {e^{4 \gamma (\tan^{-1} (t/\gamma)
	+ \tan^{-1} (ts/\gamma)} -1 \over 1 + s} \right .
\nonumber\\
	&& \qquad \left . - {1 \over 4 st}
	(e^{4 \gamma \tan^{-1} (t/\gamma)} - 1 )
	(e^{4 \gamma \tan^{-1} (ts/\gamma)} - 1) \right ]
\nonumber\\
	&=& \int_0^1 {dt \over t^2} \int_0^1 {ds \over s}
	\left [ {e^{4 \gamma (\tan^{-1} (t/\gamma)
        + \tan^{-1} (ts/\gamma)} -1 \over 1 + s} \right .
\nonumber\\
        && \qquad \left . - {1 \over 1 + (ts/\gamma)^2}
	(e^{4 \gamma \tan^{-1} (t/\gamma)} - 1)
	e^{4 \gamma \tan^{-1} (ts/\gamma)} \right ]
\nonumber\\
	&& + {1 \over 4} \int_0^1 {dt \over t^3}
	(e^{4 \gamma \tan^{-1} (t/\gamma)} - 1)
	(e^{4 \gamma \tan^{-1} (t/\gamma)} - 1 - 4 t) \,,
\end{eqnarray}
where we have done the $s$ integral by parts for the second term.
We then do the $t$ integral by parts and observe that the $s$ integral
may be completed in a closed form to obtain
\begin{eqnarray}
    G_2(\gamma) &=& - \int_0^1 {ds \over s} \left [
	{e^{4 \gamma (\tan^{-1} \gamma^{-1}
	+ \tan^{-1} (s/\gamma)} - 1 \over 1 + s} \right .
\nonumber\\
	&& \qquad  \left . - {1 \over 1 + (s/\gamma)^2}
	e^{4 \gamma \tan^{-1} (s/\gamma)}
	( e^{4 \gamma \tan^{-1} \gamma^{-1}} - 1) \right ]
\nonumber\\
	&& + 4 \int_0^1 {dt \over t} \int_0^1 {ds \over s}
	\Biggr \{ {t^2 s/\gamma^2 \over [1 + (t/\gamma)^2]
	[1 + (st/\gamma)^2 ] } e^{4 \gamma ( \tan^{-1} (t/\gamma)
	+ \tan^{-1} (ts/\gamma)}
\nonumber\\
	&& \qquad - {s \over [1 + (ts/\gamma)^2]^2}
	(e^{4 \gamma \tan^{-1} (t/\gamma)} - 1)
	e^{4 \gamma \tan^{-1} (ts /\gamma)}
\nonumber\\
	&& \qquad + {1 \over 2}{ts^2 /\gamma
	\over [1 + (ts / \gamma)^2]^2}
	(e^{4 \gamma \tan^{-1} (t/\gamma)} - 1)
	e^{4 \gamma \tan^{-1} (ts/\gamma)} \Biggr \}
\nonumber\\
	&& + {1 \over 4} \int_0^1 {dt \over t^3}
	(e^{4 \gamma \tan^{-1} (t/\gamma)} - 1)
	(e^{4 \gamma \tan^{-1} (t/\gamma)} - 1 - 4 t)
\nonumber\\
	&=& - {1 \over 8} \left (1 + {1 \over \gamma^2} \right )
	(e^{4 \gamma \tan^{-1} \gamma^{-1}} - 1) + 2
	- \int_0^1 {dt \over t} {1 \over 1 + t}
	(e^{4 \gamma \tan^{-1} (t/\gamma)} - 1)
\nonumber\\
	&& + {1 \over 4} \left (1 + {1 \over \gamma^2} \right )
	\int_0^1 {dt \over (t + 1)^2} e^{4 \gamma \tan^{-1} (t/\gamma)}
	(e^{4 \gamma \tan^{-1} \gamma^{-1}} - 1) \,,
\end{eqnarray}
where we have chosen $t$ as the one parameter integrals variable
uniformly and have appropriately performed the $t$ integral by parts
several times.
Thus, $M (k)$ can be written in terms of one
parameter integrals:
\begin{eqnarray}
    M (k)\!\tab=\tab {\rm R\!y}^2 a_0^3
	{4 \pi \gamma \over 1{-}e^{-4 \pi \gamma}}
	\Biggr \{{-}{1 \over 4} \left (1{-}
	e^{- 4 \gamma \tan^{-1} \gamma^{-1}} \right )
	\left (1{-}{3 (1{+}5 \gamma^2) \over 1{+}\gamma^2}
	e^{-4 \gamma \tan^{-1} \gamma^{-1}} \right )
\nonumber\\
	\tab\tab\!{+}{16 \gamma^4 \over (1{+}\gamma^2)^2}
	e^{-8 \gamma \tan^{-1} \! \gamma^{-1}}
	{+}{1{+}3 \gamma^2 \over 1{+}\gamma^2} \!
	\Biggr [\!\left (1{-}e^{{-}4 \gamma \tan^{-1}\!\gamma^{-1}}
	\right )  \int_0^1 \! {dt \over (1{+}t)^2}
	e^{4 \gamma \left (\!\tan^{-1} (t/\gamma){-}
	\tan^{-1}\!\gamma^{-1} \! \right )}
\nonumber\\
	\tab\tab \left . \qquad \quad - {4 \gamma^2 \over 1{+}\gamma^2}
	e^{-8 \gamma \tan^{-1} \gamma^{-1}}
	\int_0^1 {dt \over t (t{+}1)}
	\left (e^{4 \gamma \tan^{-1} (t/ \gamma)}
	{-} 1 \right ) \right ] \Biggr \} \,.
\label{Mresult}
\end{eqnarray}

We now study the asymptotic behavior of $M(k)$.
For small $\gamma$, we shall only keep the terms up to order $O (\gamma^3)$.
This may be done by using the following asymptotic expansions
of the two integrals involved in Eq.~(\ref{Mresult}):
\begin{eqnarray}
    \int_0^1 {dt \over (1+t)^2} e^{4 \gamma \tan^{-1} (t/\gamma)
	- 4 \gamma \tan^{-1} \gamma^{-1}}
	\simeq {1 \over 2} + 4 \gamma^2 \ln 2 \gamma
\\
    \int_0^1 {dt \over t(1+t)} \left (
	e^{4 \gamma \tan^{-1} (t/\gamma)} - 1 \right )
	\simeq - 2 \pi \gamma \ln 2 \gamma \,.
\end{eqnarray}
These results can be derived by doing the $t$ integral by parts
and expanding the exponential into a power series.
Putting them into the expression for $M (k)$ gives
\begin{equation}
    M (k) \simeq \left [2 \pi \gamma{-}(4{+}\pi^2) \gamma^2
		{+}16 \pi \gamma^3 \ln \gamma
		{+}(20{+}16 \ln 2) \pi \gamma^3 \right ]
		{\rm R\!y}^2 a_0^3
		\quad {\rm as}\; \gamma \to 0 \,.
\label{Msmallasy}
\end{equation}
For $\gamma$ being large, it is not hard to see from the
expression~(\ref{Mresult}) that the leading behavior of
$M (k)$ is linear in $\gamma$. The proportion constant
is evaluated numerically to be $0.360$. Thus,
\begin{equation}
    M (k) \simeq 0.360 \gamma \qquad {\rm as} \;
	\gamma \to \infty \,.
\label{Mlargeasy}
\end{equation}

\begin {figure}[htbp]
        \epsfysize 5in
        \leavevmode {\hfill \epsfbox {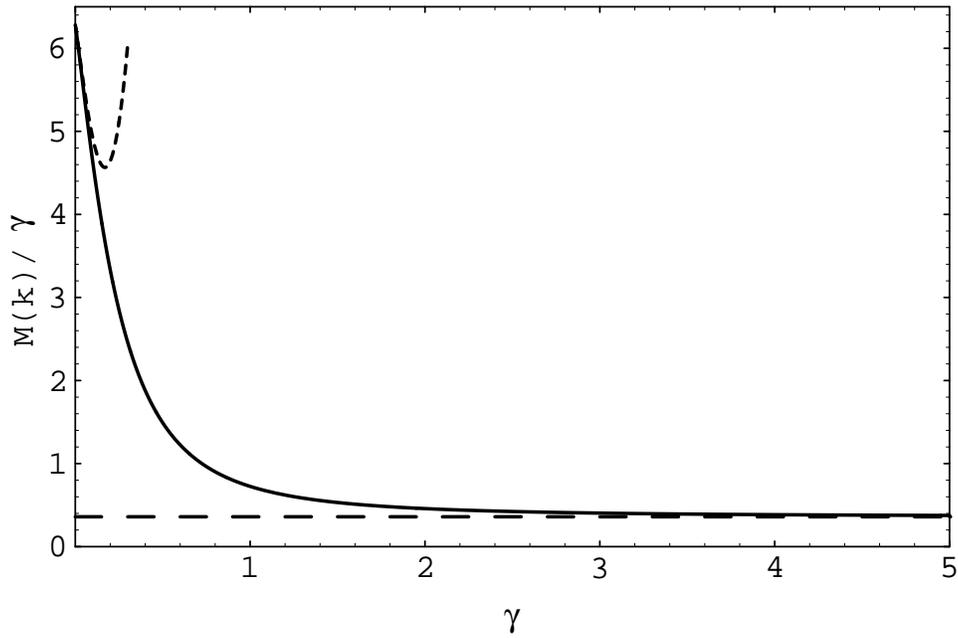} \hfill}
    \caption
        {%
        \advance\baselineskip by -8pt
        Numerical curve for $M(k)/\gamma$ as a function of $\gamma$.
	The long-dashed horizontal line is the asymptotic limit for
	large $\gamma$ given in Eq.~(\protect{\ref{Mlargeasy}}). The
	short-dashed curve at small value of $\gamma$ is the approximation
	given in Eq.~(\protect{\ref{Msmallasy}}).
	The vertical axis is in units of ${\rm R\!y}^2 a_0^3$
	while the horizontal axis is dimensionless.
        }%
    \label {fig:fourteen}
\end {figure}

\begin {figure}[htbp]
        \epsfysize 5in
        \leavevmode {\hfill \epsfbox {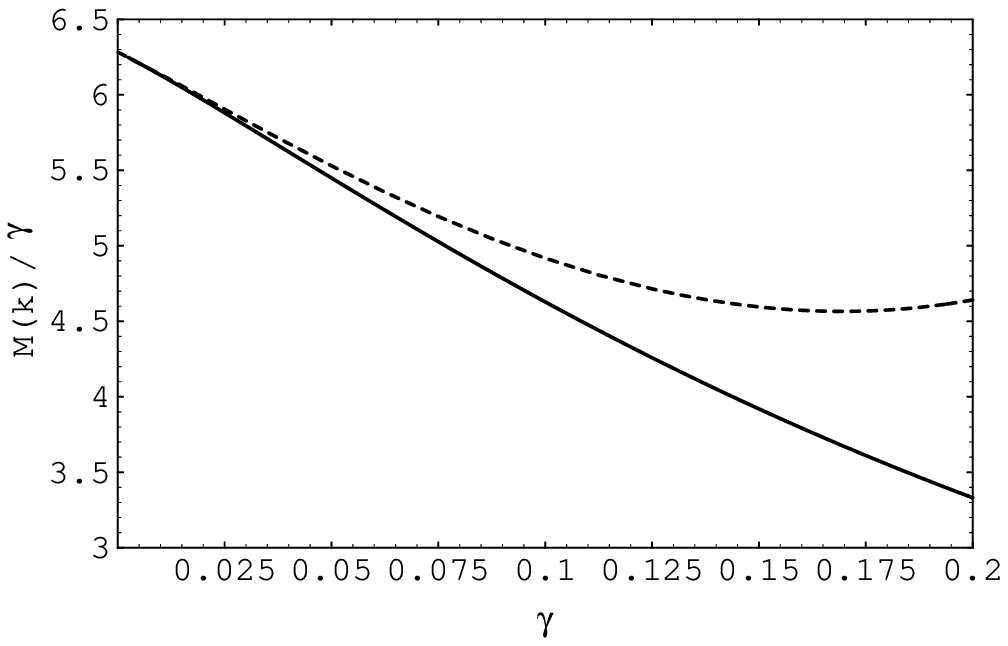} \hfill}
    \caption
        {%
        \advance\baselineskip by -8pt
        Numerical curve
        for $M(k)/\gamma$ (solid curve) as a function of $\gamma$
	compared to the small $\gamma$ expansion (dashed curve) given in
	Eq.~(\protect{\ref{Msmallasy}}).
	The vertical axis has the unit ${\rm R\!y}^2 a_0^3$
	and horizontal axis is dimensionless.
        }%
    \label {fig:fifteen}
\end {figure}

We now display the numerical curves of $M(k)/\gamma$ and its
asymptotic behavior as functions of
$\gamma$ in Fig.~\ref{fig:fourteen} and Fig.~\ref{fig:fifteen}.

\section{Details of the exchange terms}
We first examine the third term $T_3$ in Eq.~(\ref{calTresult2}).
For this term $T_3$, only the ``vacuum state''
$| 0 \rangle$ (i.e. the tritium nucleus) contributes to the intermediate
states when inserting a complete set of states just before the operator
$N_{\bf p} (0)$. Hence,
\begin{eqnarray}
    T_3 &=& \langle f | \psi^{\dagger} (0) | 0 \rangle
	{1 \over E - E_i} \langle 0 | N_{\bf p} (0)
	 | i \rangle
\nonumber\\
	&=& - \phi_f^* (0) {1 \over E - E_i}
	\langle {\bf p} |
	\left \{ {e^2 \over 4 \pi r} + v(r)
	\right \} | i \rangle \,,
\label{difference}
\end{eqnarray}
where $\langle {\bf p} |$ and $| i \rangle$ are the electron states
of the final beta ray and the initial tritium atom.
The Schr\"odinger equations for these energy eigenstates can be used
to rewrite
\begin{equation}
    T_3 = - \phi_f^* (0) \langle {\bf p} | i \rangle \,.
\label{exchange}
\end{equation}
This is an exchange term which represents the amplitude
for the electron produced by the weak interaction shaking the bound
electron of the original tritium out and being bound by the final
helium nucleus itself. We shall not evaluate $T_3$ because it is
canceled by a piece of the exchange term $T_{\rm e}$ which we now turn to
discuss.\footnote{It is not hard to see that $T_3$ is of order $\eta^4$
by examining the expression~(\ref{difference}) written in terms of
wave functions.
Since, for any reasonable comparison potential, the integral
\begin{displaymath}
    \int (d^3 {\bf r}) \phi_{\bf p}^* ({\bf r})
        \left \{ {e^2 \over 4 \pi r} + v(r)
        \right \}
\end{displaymath}
converges, and $1/p$ is much less than the Bohr radius, we can
replace the wave function $\phi_i ({\bf r})$
by its value at ${\bf r} = 0$ to obtain the leading order contribution.
Noting that $\phi_f^* (0)$ and $\phi_i (0)$ are of order $(\alpha m)^{3/2}$,
and approximating the integral displayed above
by the Fourier transform of the Coulomb potential
evaluated at momentum ${\bf p}$, yields
\begin{displaymath}
    T_3 \sim (\alpha m)^{3/2} \left ( m / p^2 \right )
	\left ( e^2 / p^2 \right )
	(\alpha m)^{3/2} \sim \eta^4 \,.
\end{displaymath}
}

The exchange part $T_{\rm e}$ in $T_2$ of Eq.~(\ref{decomp})
involves the ket $| i, {\bf r}_2{=}{\bf 0} \rangle$
For the evaluation of this exchange term, it is convenient to
interchange the roles of $r_1$ and $r_2$ in Eqs.~(\ref{H1def}),
(\ref{H2def}) and (\ref{interdef}) and define
\begin{eqnarray}
    {\tilde H}_1 &=& {{\bf p}_1^2 \over 2m} - {e^2 \over 4 \pi r_1} \,,
\\
    {\tilde H}_2 &=& {{\bf p}_2^2 \over 2m} - {2 e^2 \over 4 \pi r_2} \,,
\\
    {\tilde H}_I &=& {e^2 \over 4 \pi |{\bf r}_1 - {\bf r}_2|}
	- {e^2 \over 4 \pi r_1} \,.
\label{tlinterdef}
\end{eqnarray}
This enables us to expand $T_{\rm e}$ in powers of ${\tilde H}_I$ as
was done before [{\it c.f.} Eq.~(\ref{dirlead})].
The leading term in $T_{\rm e}$, which we now denote as $T_{\rm e}^0$, is
\begin{equation}
    T_{\rm e}^0 = \int (d^3 {\bf r}_1)
	\phi^*_{\bf p}({\bf r}_1) \phi_i ({\bf r}_1)
	\langle f | \left [
        {e^2 \over 4 \pi |{\bf r}_1 - {\bf r}_2|}
        - {2 e^2 \over 4 \pi r_1} - v(r_1) \right ]
        {1 \over {\tilde H}_2 + E_i - E - E_f - i \epsilon}
	|{\bf r}_2{=}{\bf 0} \rangle \,,
\label{exlead}
\end{equation}
upon using the fact that $|i \rangle$ is an eigenstate of
${\tilde H}_1$.
Noting that $\langle f |$ is an eigenstate of ${\tilde H}_2$,
we may write
\begin{eqnarray}
    T_{\rm e}^0 \tab=\tab  \! \int (d^3 {\bf r}_1)
        \phi^*_{\bf p}({\bf r}_1) \phi_i ({\bf r}_1) \langle f |
	\left ( {e^2 \over 4\pi |{\bf r}_1{-}{\bf r}_2|}
	- {e^2 \over 4\pi r_1}
	\right ) {1 \over {\tilde H}_2{+}E_i{-}E{-}E_f{-}i \epsilon}
        |{\bf r}_2{=}{\bf 0} \rangle
\nonumber\\
	\tab\tab + {1 \over E - E_i} \phi_f^* (0)
	\int (d^3 {\bf r}_1) \phi^*_{\bf p}({\bf r}_1) \phi_i ({\bf r}_1)
	\left [ {e^2 \over 4 \pi r_1} + v(r_1) \right ] \,.
\label{gulp}
\end{eqnarray}
The second term here exactly cancels $T_3$ in view of
Eq.~(\ref{difference}).

To evaluate the sum $  T_{\rm e}^0 + T_3 $,
the first line in Eq.~(\ref{gulp}), we follow the method of Appendix B
in writing the denominator as the integral of an exponential, use the
Heisenberg equation of motion with $\langle f | (\tilde H_2 - E_f) =
0$, and express the Coulomb Green's functions as Fourier integrals, to
obtain
\begin{eqnarray}
    T_{\rm e}^0 + T_3 \tab=\tab e^2  \! \int (d^3 {\bf r}_1)
        \phi^*_{\bf p}({\bf r}_1) \phi_i ({\bf r}_1) \int_0^\infty dt
	\, e^{ i (E - E_i) t }
\nonumber\\
	\tab\tab \int { (d^3{\bf k}) \over (2\pi)^3 } { 1 \over k^2 }
	e^{ i {\bf k} \cdot {\bf r}_1 } \langle f | \left(
	e^{ -i {\bf k} \cdot {\bf r}_2(t) } - 1 \right)
	|{\bf r}_2{=}{\bf 0} \rangle \,.
\end{eqnarray}
In leading order, which is equivalent to the $ p \to \infty $ limit,
we may replace the outgoing beta electron wave function
$ \phi^*_{\bf p}({\bf r}_1) $ by the free plane wave $e^{ -i {\bf p}
\cdot {\bf r}_1 } $, neglect $ E_i $ relative to $ E = p^2 /2m $, and
replace $ {\bf r}_2(t) $ by the free particle motion,
\begin{equation}
    {\bf r}_2(t) = {\bf r}_2 + {\bf p} t / m \,.
\end{equation}
Ordering the resulting exponential gives
\begin{equation}
    e^{ -i {\bf k} \cdot {\bf r}_2(t) } \simeq
	e^{ -i k^2 t / 2m } e^{ -i t {\bf k} \cdot {\bf p} / m }
	e^{ -i {\bf k} \cdot {\bf r}_2 } \,.
\end{equation}
Since the momentum operator generates a spatial translation, we now
arrive at
\begin{eqnarray}
T_{\rm e}^0 + T_3 \tab\simeq\tab i e^2  \! \int (d^3 {\bf r}_1)
	\int { (d^3{\bf k}) \over (2\pi)^3 } { 1 \over k^2 }
	e^{ -i ( {\bf p} - {\bf k} ) \cdot {\bf r}_1 }
	\phi_i ({\bf r}_1)
\nonumber\\
	\tab\tab
	\int_0^\infty dt \, e^{ i p^2 t / 2m } \left\{ e^{-i k^2 t / 2m }
	\langle f|{\bf r}_2 = {\bf k} t / m \rangle -
	\langle f|{\bf r}_2{=}{\bf 0} \rangle \right\} \,.
\end{eqnarray}
To exhibit the leading order contribution, we change variables by
writing
\begin{equation}
    {\bf k} = { m \over t } {\bf r} \,,
\end{equation}
and
\begin{equation}
   t = { m r \over p} \tau \,, \qquad
	{\bf r}_1 = {\bf r} + {1 \over p} {\bf u} \,.
\end{equation}
This change produces
\begin{eqnarray}
T_{\rm e}^0 + T_3 \tab\simeq\tab i 4\pi \eta^3 a_0^2 \! \int (d^3 {\bf r})
	{ 1 \over r^2 } \int_0^\infty { d\tau \over \tau }
	e^{ -i ( {\bf p} \cdot {\bf r} - pr / \tau ) }
\nonumber\\
	\tab\tab
	e^{ ipr\tau/2 } \left\{ e^{ -ipr/2\tau } \phi_f^*({\bf r})
	- \phi_f^*({\bf 0}) \right\}
	\int { (d^3{\bf u}) \over (2\pi)^3 }
	\phi_i( {\bf r} + {\bf u} /p )
    e^{ -i ( \hat {\bf p} - \hat {\bf r}/\tau ) \cdot {\bf u} } \,.
\end{eqnarray}
In the $ p \to \infty $ limit,
$ \phi_i( {\bf r} + {\bf u} /p ) $ may be replaced by
$ \phi_i( {\bf r} ) $, and we encounter
\begin{eqnarray}
   \int { (d^3{\bf u}) \over (2\pi)^3 }
    e^{ -i ( \hat {\bf p} - \hat {\bf r}/\tau ) \cdot {\bf u} }
	\tab=\tab \delta^{(3)} ( \hat {\bf p} - \hat {\bf r}/\tau )
\nonumber\\
	\tab=\tab
	\delta (1 - \tau) \sum_{lm} Y_{lm}^*(\hat {\bf p})
		Y_{lm}(\hat {\bf r}) \,,
\end{eqnarray}
which produces
\begin{equation}
T_{\rm e}^0 + T_3 \simeq i 4\pi \eta^3 a_0^2 \! \int (d^3 {\bf r})
	{ 1 \over r^2 } \sum_{l'm'} Y_{l'm'}^*(\hat {\bf p})
		Y_{l'm'}(\hat {\bf r})
	\left\{ \phi_f^*({\bf r})
	- e^{ ipr } \phi_f^*({\bf 0}) \right\} \phi_i( {\bf r}) \,.
\end{equation}
The last term involving $ e^{ ipr } \phi_f^*({\bf 0}) $ vanishes in
the large $p$ limit by virtue of its infinitely rapid phase oscillation.
The angular part of the $ {\bf r} $ integration just picks out the $l$
and $m$ values of the final atomic wave function,
\begin{equation}
     \phi_f^*({\bf r}) = R_{fl}^* (r) Y^*_{lm}(\hat {\bf r}) \,.
\end{equation}
Accordingly, to the leading order, the exchange amplitude is given by
\begin{equation}
    T_{\rm e}^0 + T_3 \simeq i \, 4 \pi \eta^3 \, a_0^2 \,
	Y_{lm}^* ({\hat {\bf p}})
	\int_0^{\infty} d r R_{fl}^* (r) \phi_i (r) \,,
\label{exresult}
\end{equation}
which is of order $\eta^3$.

Although the amplitude~(\ref{exresult}) is of order $\eta^3$, it
contributes to the decay rate only through the order $\eta^4$.
For $l {\not =} 0$, this result is of order $\eta^2$ relative to the
leading direct result~(\ref{lresult1}) which is of order $\eta$.
Therefore the exchange effect gives a correction to the decay rate at
order $\eta^4$.
For case $l=0$, the leading exchange amplitude~(\ref{exresult})
is relatively imaginary.
Since the leading imaginary amplitude appears at the order $\eta$,
the exchange correction to the decay rate is only of order $\eta^4$
We note here that this contradicts with the results in the literature
\cite{Bahcall}, where the dominant exchange {\it amplitudes}
appear in order $\eta^4$.


\begin{references}

\bibitem{Wilkerson}
    J. F. Wilkerson, Nucl. Phys. B {\bf 31} ({\it Proc. Suppl.}) (1993), 32.

\bibitem{Robertson}
    R. G. H. Robertson {\it et al.},
	Phys. Rev. Lett. {\bf 67} (1991), 957.

\bibitem{Holzschuh}
    E. Holzschuh {\it et al.}, Phys. Lett. B {\bf 287} (1992), 381.

\bibitem{Kawakami}
    H. Kawakami {\it et al.}, Phys. Lett. B {\bf 256} (1991), 105.

\bibitem{Stoeffl}
    W. Stoeffl, Bull. Am. Phys. Soc. 37 (1992), 925;
	D. J. Decman and W. Stoeffl, presented at conf. {\em The Many
	Aspects of Neutrino Physics}, Fermilab, Nov. 1991 (unpublished);
	W. Stoeffl and D. J. Decman, Proceedings of XXVIIIth Rencontre De
	Moriond Workshop, Villars sur Ollon, Switzerland, Jan 30 - Feb 6,
	15 (1993); W. Stoeffl and D. J. Decman, {\em Anomalous Structure
	in the Beta Decay of Gaseous Molecular Tritium},
	UCRL-JC-115771/Preprint, (Phys. Rev. Lett., to be published).

\bibitem{Backe}
    H. Backe {\it et al.}, Phys. Scr. T {\bf 22} (1988), 98;
	A. Picard {\it et al.},
	Nucl. Instr. and Methods in Phys. Research {\bf B63} (1992), 345;
	Ch. Weinheimer, {\it et al.}, 
	Phys. Lett. B {\bf 300} (1993), 210.

\bibitem{Knapp}
    D. A. Knapp, Los Alamos Ph.~D. thesis, (1986).

\bibitem{Martin}
    R. L. Martin and J. S. Cohen,
	Phys. Lett. A {\bf 110} (1985), 95;
    O. Fackler, B. Jeziorski, W. Lolos, J. J. Monkhorst, and
	K. Szalewica, Phys. Rev. Lett. {\bf 55} (1985), 1388;
    H. Agren and V. Carravetta,
	Phys. Rev. A {\bf 38} (1988), 2707.

\bibitem{Wilkinson}
    D. H. Wilkinson, Nucl. Phys. A {\bf 526} (1991), 131;
	J. J. Simpson, Phys. Rev. D {\bf 23} (1981), 649.

\bibitem{RobertsonRev}
    R. G. H. Robertson and D. A. Knapp,
	Ann. Rev. Nucl. Part. Sci.  {\bf 38} (1988), 185.

\bibitem{Durand}
    L. Durand and J. L. Lopez, Phys. Lett. B {\bf 198} (1987), 249;
    J. L. Lopez and L. Durand, Phys. Rev. C {\bf 37} (1988), 535.

\bibitem{Strikman}
    E. G. Drukarev and M. I. Strikman,
	Zh. Eksp. Teor. Fiz. {\bf 91} (1986), 1160
	[Sov. Phys. --- JETP {\bf 64} (1986), 686].

\bibitem{Rose}
    M. E. Rose, Phys. Rev. {\bf 49} (1936), 727.

\bibitem{Bahcall}
    J. N. Bahcall, Phys. Rev. {\bf 129} (1963), 2683.

\bibitem{Brown}
    See, for example, L. S. Brown,
	``Quantum Field Theory,'' pp. 219-298,
	Cambridge University Press, 1992.

\bibitem{Gottfried}
    See, for example, K. Gottfried,
        ``Quantum Mechanics,'' section 17, W. A. Benjamin, Inc., New York,
        1966. Notice that our wave function differs
	from the wave function there by a factor $(2 \pi)^{3/2}$
	due to different normalization, and our radial wave
	function $R_{kl} (r)$ is denoted by $C_l (k;r)$ there.

\bibitem{Landau}
    For the integral representation of the confluent hypergeometric
	function, see, for example, L. D. Landau and E. M. Lifshitz,
	``Quantum Mechanics,'' 2nd ed., p. 602, Pergamon Press, 1965.
	The formulas we need are (d.8) and (d.9), p. 602.

\end{references}
\end{document}